\documentclass[fleqn,usenatbib]{mnras}

\usepackage{newtxtext,newtxmath,epsfig,arydshln} 

\usepackage[T1]{fontenc}
\usepackage{ae,aecompl}

\usepackage{graphicx}	
\usepackage{amsmath}	
\usepackage{amssymb}	
\usepackage[dvipsnames,svgnames,x11names]{xcolor}
\usepackage[multiple]{footmisc}
\usepackage{multirow}
\usepackage{tcolorbox}
\PassOptionsToPackage{linktocpage}{hyperref}
\usepackage{breakurl}

\title[The SF and BH activity connection in the local Universe]{The connection between star formation and supermassive Black Hole activity in the local Universe}

\author[O. Torbaniuk et al.]{O. Torbaniuk$^{1}$\thanks{E-mail: olena.torbaniuk@unina.it},
M. Paolillo$^{1,2,3}$,
F. Carrera$^{4}$,
S. Cavuoti$^{2,1,3}$,
C. Vignali$^{5,6}$,
\newauthor{G. Longo$^{1,3}$ and J. Aird$^{7,8}$}
\\
$^{1}$Department of Physics, University of Napoli Federico II, via Cinthia 9, 80126, Napoli, Italy\\
$^{2}$INAF — Osservatorio Astronomico di Capodimonte, via Moiariello 16, 80131, Napoli, Italy\\
$^{3}$INFN — Sezione di Napoli, via Cinthia 9, 80126, Napoli, Italy\\
$^{4}$Instituto de Física de Cantabria (CSIC-UC), Avenida de los Castros, 39005 Santander, Spain\\
$^{5}$Department of Physics and Astronomy, University of Bologna, via Piero Gobetti 93/2, I-40129 Bologna, Italy \\
$^{6}$INAF — Osservatorio di Astrofisica e Scienza dello Spazio di Bologna, via Piero Gobetti 93/3, I-40129 Bologna, Italy\\
$^{7}$Institute for Astronomy, University of Edinburgh, Royal Observatory, Edinburgh EH9 3HJ, UK\\
$^{8}$School of Physics \& Astronomy, University of Leicester, University Road, Leicester LE1 7RJ, UK
}

\date{Accepted 2021 June 18. Received 2021 June 8; in original form 2021 January 26}

\pubyear{2021}

\begin{document}
\label{firstpage}
\pagerange{\pageref{firstpage}--\pageref{lastpage}}
\maketitle

\begin{abstract}
We present a study of the active galactic nucleus (AGN) activity in the local Universe ($z < 0.33$) and its correlation with the host galaxy properties, derived from an Sloan Digital Sky Survey (SDSS DR8) sample with spectroscopic star-formation rate (SFR) and stellar mass ($\mathcal{M}_{\ast}$) determination. To quantify the level of AGN activity we used  X-ray information from the XMM-Newton Serendipitous Source Catalogue (3XMM DR8). Applying multiwavelength AGN selection criteria (optical BPT-diagrams, X-ray/optical ratio etc) we found that 24\% of the detected sources are  efficiently-accreting AGN with moderate-to-high X-ray luminosity, which are twice as likely to be hosted by star-forming galaxies than by quiescent ones.  The distribution of the specific Black Hole accretion rate (sBHAR, $\lambda_{\mathrm{sBHAR}}$) shows that nuclear activity in local, non-AGN dominated galaxies peaks at very low accretion rates ($-4 \lesssim \log\lambda_{\mathrm{sBHAR}} \lesssim -3$) in all stellar mass ranges. However, we observe systematically larger values of sBHAR for galaxies with active star-formation than for quiescent ones, as well as an increase of the mean $\lambda_{\mathrm{sBHAR}}$ with SFR for both star-forming and quiescent galaxies. These findings confirm the decreased level of AGN activity with cosmic time and are consistent with a scenario where both star-formation and AGN activity are fuelled by a common gas reservoir.

\end{abstract}

\begin{keywords}
galaxies: active -- galaxies: elliptical and lenticular, cD -- galaxies: spiral -- galaxies: star formation -- X-ray: galaxies -- accretion, accretion discs
\end{keywords}


\section{Introduction}

Active Galactic Nuclei (AGN) seem to play a significant role in the evolution of their host galaxies and appear to be in close relation with the internal star-formation processes. Such co-evolution between the host galaxy and the supermassive Black Hole (SMBH) at its centre has been suggested by several studies. For instance, it was shown that the SMBH mass is tightly correlated with host galaxy properties such as luminosity \citep{Marconi:03}, bulge mass \citep{Haring:04, McConnel:13, Kormendy:13}, velocity dispersion of stars in the galactic bulge \citep{Ferrarese:00, Gebhardt:00, Gultekin:09, deNicola:19}, total stellar mass \citep{Haring:04,Reines:15}. Also, observational studies of the accreting SMBHs over a wide range of redshifts (i.e. cosmic times) show that AGN evolution follows the luminosity-dependent scenario in which high-luminosity AGN reach the peak of activity earlier than low-luminosity AGN \citep{Hasinger:05, Bongiorno:07, Bongiorno:12, Ueda:14}. This evolutionary picture is similar to the so-called `cosmic downsizing' trend observed in star-forming galaxies \citep{Cowie:96} and support the presence of a link between the evolution of SMBHs and their host galaxies. Furthermore, the shapes of the total AGN accretion curve as a function of redshift and of the global star-formation rate (SFR) are identical, both reaching a peak at redshift $z \sim 1$--3 with rapid decline toward the local Universe \citep{Delvecchio:14, Madau:14, Aird:15}.

A strong indication of a co-evolution between galaxies and their SMBHs is provided by a connection between AGN accretion and star-formation processes in the host galaxy. In fact it was observed that moderate-to-high luminosity AGN predominantly lie in galaxies with galaxies with active star-formation \citep{Merloni:10, Rosario:13, Heinis:16, Aird:18, Stemo:20} which can be explained by the cold gas fuelling both AGN activity and star-formation in the galaxy. Since the AGN accretion and star-formation processes operate typically on different spatial scales the details of the mechanisms responsible for such gas supply are still poorly known. Hydrodynamical simulations and individual observations of the closest galaxies suggest a scenario where a major role in the gas transportation to the galaxy centre can be played by the large-scale gravitational torques produced by galaxy major merger, minor interaction or various disk instabilities \citep{Fathi:06,Hopkins:10,Fischer:15}. In addition, the star-formation processes can significantly affect the nuclear accretion reducing it by `stealing' the gas or increasing it by the turbulence injection in the circumnuclear region, i.e. the so-called stellar feedback \citep{Schartmann:09,Hopkins:16}.
On the contrary, the presence of an AGN can affect star-formation in the galaxy through AGN feedback. According to the AGN feedback paradigm there are two main modes of AGN output \citep{Heckman:14}. The first form is the so-called \textit{quasar} or \textit{radiative} mode is mostly associated with high luminosity AGN with energetic output in the form of the electromagnetic radiation generated by the efficient ($>1\%$ Eddington) accretion onto the SMBH. The second one, \textit{jet} or \textit{radio} mode, is associated with less powerful AGN with radiatively inefficient accretion ($\ll 1\%$ Eddington), which generates powerful jets. Hence, AGN interacts with the gas in the host galaxy by radiation pressure which induce powerful wind (quasar mode) or an outflow of relativistic particles (jet mode), which heat and/or blow the cold gas away from the galaxy thus suppressing the formation of stars (negative AGN feedback). Through such mechanisms AGN feedback can quench star-formation \citep{Croton:06, Rosario:13} and also significantly reduce the accretion onto the SMBH thus resulting in self-regulation of the nuclear activity \citep{Fabian:12}. In some cases, AGN feedback may interact with the host galaxy in opposite way triggering the star-formation by compressing dense clouds in the interstellar medium, i.e. positive AGN feedback, \citep{Ishibashi:12,Zubovas:13,Combes:17}.

To constrain such scenario we need to be able to separate the AGN and host galaxy emission. This issue is is not trivial, especially in the local Universe because the population of local AGN are predominantly low-luminosity sources making their identification complicated in the optical and infrared (IR) bands where the host galaxy emission is dominant. Being produced in the innermost regions of the Active Nuclei, X-ray emission is an good tracer of accretion processes, allowing to study the AGN population over a wide range of redshifts down to relatively low luminosities~\citep{Alexander:12}. Despite this, previous investigations have found only little correlation between nuclear X-ray luminosity and stellar formation. For instance, AGN with similar X-ray luminosity lie in galaxies with broad ranges of stellar mass and SFR \citep{Mullaney:12a, Aird:13, Rosario:13, Azadi:15}. On the other hand, galaxies with the same SFR can contain AGN with a broad range of accretion rates~\citep{Bongiorno:12, Aird:12, Azadi:15}. This absence of correlation between SFR and accretion rate for the individual objects was explained by the different variability timescales of these two processes \citep{Hickox:09}. In fact even if nuclear activity and stellar formation are connected at any time, the SF is relatively stable over $\sim100$\,Myr, while AGN luminosity may vary of orders of magnitude on very short time scale $\sim 10^{5}$\,yr. \citep{Bongiorno:12, Mullaney:12a, Aird:13, Hickox:14, Paolillo:17}. Therefore, to discover the relation between AGN activity and galaxy properties we need to study the average properties of large samples of objects. 

For instance, \citet{Chen:13} used the far-IR-selected galaxies from \textit{Herschel Space Observatory} and AGN selected by IR and X-ray criteria and found a linear relation between average SMBH accretion rate (BHAR) and SFR for galaxies across a wide range of SFR and with redshifts $0.8 < z < 2.5$. Additionally, a similar relation between SFR and average specific BHAR (and stellar mass) was found by \citet{Delvecchio:15} on the basis of a far-IR sample of star-forming galaxies and X-ray selected AGN at $z < 2.5$. These and other studies \citep{Yang:18,Aird:19, Stemo:20} also show that the SFR--sBHAR correlation becomes more significant at redshift $z > 0.8$. Additional studies \citep{Masoura:18, Aird:19} found that the average AGN X-ray luminosity can change depending on the host galaxy position relative to the main sequence (MS) of star-forming galaxies, possibly indicating an enhancement of star-forming processes due to the AGN when its host lies below the MS line and its quenching when the host galaxy lie above the MS. However, \citet{Rovilos:12} and \citet{Shimizu:15} found no evidence of this effect for AGN in the local Universe. Such relations between sBHAR/X-ray luminosity and SFR are consistent with a scenario in which AGN activity and SFR are connected over galaxy evolution timescales by the common cold gas supply, but the AGN contribution to galaxy quenching remains controversial.

The study of large statistical samples of galaxies allows us to build the sBHAR probability function, i.e the probability of a galaxy with a given property (stellar mass, SFR and morphological type) to host an AGN with a given sBHAR, and derive the level of AGN activity in the Universe. Several works have shown that the sBHAR probability function follows a power-law shape with an exponential cut-off at high accretion rates, flattening or even decreasing toward low sBHAR \citep{Aird:12, Bongiorno:12, Aird:18}. Furthermore, the more recent study by \citet{Aird:18} for different host galaxy types shows that quiescent galaxies have typically lower probability of hosting an AGN than star-forming galaxies pointing toward a lower fraction of the cold gas (i.e. lower sBHAR) in the quiescent galaxies.

Most previous studies have focused on investigating the AGN-host galaxy connection at intermediate/high redshift from $z \sim 0.25$ up to $z \approx 4.0$ \citep{Chen:13,Rosario:13,Delvecchio:15,Aird:18,Aird:19, Stemo:20}, while in the local Universe our knowledge is limited by the absence of wide-area surveys, especially in the X-ray band. However, studies of the local sBHAR--SFR relation are valuable as local galaxies predominantly contain low-to-moderate luminosity AGN (i.e. with low-efficient SMBH accretion) which are difficult to trace at high redshifts. Additionally, the quiescent galaxy population in the local Universe allows us to study deeply the possible mechanism of star-formation suppression and explore the alternative AGN fuelling processes in the environment with the low level of cold gas.

In this paper we study the correlation between star-formation and AGN activity in the local Universe using a homogeneous Sloan Digital Sky Survey (SDSS-DR8) optical galaxy sample with robust SFR (in the range $10^{-3}$ to $10^{2}\mathcal{M}_{\odot}$\,year$^{-1}$) and $\mathcal{M}_{\ast}$ estimates (from $10^{6}$ to $10^{12}\mathcal{M}_{\odot}$), in combination with X-ray data from from XMM-Netwon Serendipitous Source Catalogue (3XMM-DR8), in order to identify AGN and estimate their sBHAR. We further investigate the intrinsic sBHAR distribution in the local Universe taking into account the variable XMM-Newton sensitivity across the sky and the relation between average sBHAR and SFR for star-forming and quiescent galaxies with different stellar masses. 

In Section 2 we describe the catalogue of host galaxy properties derived from the SDSS optical galaxy sample and extract the corresponding X-ray data from the XMM-Newton Serendipitous Source Catalogue. Section 3 describes the approach used to estimate of the intrinsic X-ray AGN luminosity and the corrections needed to account for  the contribution of the host galaxy. Section 4 describes the AGN identification criteria in the optical and the X-ray bands. In Section 5 we present our measurement of the specific BH accretion rate $\lambda_{\mathrm{sBHAR}}$ (Section 5.1), its distribution as a function of stellar mass and galaxy properties as well as the correlation between SFR and $\lambda_{\mathrm{sBHAR}}$. In Section 6 we discuss the reliability of our results. A brief summary of our results and concluding remarks are presented in Section 7. Throughout this paper, we adopt a flat cosmology with $\Omega_{\Lambda} = 0.7$ and $H_0 = 70$\;km\;s$^{-1}$ Mpc$^{-1}$ and assume a stellar initial mass (IMF) from \citet{Chabrier:03}.


\section{Sample selection and classification}

\subsection{The SDSS data set}\label{sec:sdss-data} 

Our galaxy sample is based on the \textit{galSpec} catalogue of galaxy properties\footnote{\url{https://www.sdss.org/dr12/spectro/galaxy\_mpajhu/}} which was produced by the MPA–JHU group as the subsample from the main galaxy catalogue of the 8th Data Release of the Sloan Digital Sky Survey (SDSS DR8). 
The stellar masses in the catalogue are obtained through Bayesian fitting of the SDSS \textit{ugriz} photometry to a grid of models. The details of the stellar masses determination are described in \citet{Kauffmann:03a, Tremonti:04}. 

The estimate of the star-formation rate (SFR) were done in two different ways described in \citet{Brinchmann:04}. In short, the object selection was based on the BPT criteria \citep{Baldwin:81}, according to which the objects with high signal-to-noise for four specific emission lines were classified as `star-forming galaxies', `AGN' and `composite'. The BPT criteria cannot be applied if the objects have low S/N value (S/N $< 3$) for at least one of the required emission line, therefore such objects were classified as low S/N star-forming galaxies and AGN. The objects with weak emission lines or no line at all were marked as unclassified (see \citet{Brinchmann:04} for details). The values of SFRs for SFGs and low S/N SFGs were determined using the H${\alpha}$ emission line luminosity. However, such SFR estimates can be underestimated due to the absorption by the galaxy dust. 
Therefore, the SFR estimates based on H${\alpha}$ luminosity are corrected for dust extinction on the basis of the Balmer decrement, D4000 \citep{Kauffmann:03c}. In addition, the H${\alpha}$ line is not an accurate indicator of SFR for all galaxies. Firstly, some galaxies do not have H${\alpha}$ line in their spectra or it can be blended with close emission lines (as N\,{\sc ii}\,$\lambda6583$). Secondly, according to the BPT-diagram classification, a major fraction of galaxies in \textit{galSpec} catalogue have an active nucleus and therefore the H${\alpha}$ line is due to both star formation and accretion processes. In these cases the SFRs were inferred from empirical relation between SFR and D4000  \citep{Kauffmann:03c}. All SFR measures are corrected for the fiber aperture following the approach proposed by \citet{Salim:07}. The stellar masses $\mathcal{M}_{\ast}$ and SFRs are computed by assuming a Kroupa IMF \citep{Kroupa:01} and are adjusted to a Chabrier IMF \citep{Chabrier:03}.

The \textit{galSpec} catalogue contains reliable spectroscopic information for about 1.5 million galaxies with redshift $z \leq 0.33$. We selected objects with reliable spectroscopic parameters (i.e. with \texttt{RELIABLE\,!=\,0}) and redshift (i.e. with \texttt{zWarning = 0}; see \citealt{Aihara:11}, for further details). Additionally, \textit{galSpec} catalogue may have multiple spectroscopic observations for an individual galaxy. Some of these objects were intentionally re-observed, as part of a different program or survey, or as part of a repeated plate observation. Therefore we rejected such duplicates and choose only objects which are `primary' in the sample using the \texttt{mode} flag (for `primary' objects \texttt{mode = 1}). The values of SFR and $\mathcal{M}_{\ast}$ for duplicate observations are very similar, with the majority of the objects in agreement within a few percent. In any case, objects with low quality SDSS photometry in \textit{r} and \textit{i} bands were excluded from our sample due to the need of accurate \textit{r-} and \textit{i-}band magnitudes for AGN selection described in Section\,\ref{sec:x-agn-select}. For this purpose, we used the basic photometric processing flags recommended on SDSS website\footnote{\url{https://www.sdss.org/dr12/algorithms/photo_flags_recommend/}} to clean the sample from objects with deblending (\texttt{PEAKCENTER\,!=\,0, NOTCHECKED\,!=\,0, DEBLEND\_NOPEAK\,!=\,0}) and interpolation problems (\texttt{PSF\_FLUX\_INTERP\,!=\,0, BAD\_COUNT\_ERROR\,!=\,0} and \texttt{INTERP\_CENTER\,!=\,0}). The magnitudes and photometric flags for our sample were obtained by the online service \texttt{CasJobs} SDSS SkyServer. After excluding all objects described above our final SDSS galaxy sample consists of 703\,422 sources. For clarity, a summary of all samples defined here and throughout the rest of this work is presented in Table\,\ref{tab:samples}.

\begin{table*}
 \caption{Summary of the different optical and X-ray samples defined in this work.}
 \label{tab:samples}
 \begin{tabular}{clcl}
  \hline
  \# & Samples & $N$ & Comments \\
  \hline
  1 & SDSS galaxy sample & 703\,422 & The optical sample of galaxies compiled from SDSS {\it galSpec} catalogue (Sec.\,\ref{sec:sdss-data})\\
  \hline
  2 & SDSS sample in the 3XMM footprint & 40\,914 & The number of SDSS galaxies that fall within the {\it XMM-Newton} footprint (Sec.\,\ref{sec:xmm-data}). \\
  \hline
  3 & 3XMM counterparts for the SDSS galaxy & 3742 & The crossmatch between SDSS galaxy sample (\#1) and 3XMM DR8 catalogue \\
  & sample &  & (Sec.\,\ref{sec:xmm-data}) \\
  \hline
  4 & 3XMM-SDSS sample & 1953 & Sample \#3 after filtering-out sources with unreliable photometry, extended sources\\
  & & & and sources with multiple X-ray detections (Sec.\,\ref{sec:xmm-data}) \\
  \hdashline
  4.1 & \hspace{0.3cm} out of which: optically-selected AGN & 661 & Sources in the 3XMM-SDSS sample (\#4) classified as AGN by at least one of \\
  &  &  & the optical BPT diagrams (Sec.\,\ref{sec:bpt-agn})  \\
  \hdashline
  4.2 & \hspace{0.3cm} out of which: `classical' X-ray AGN & 469 & Sources in the 3XMM-SDSS sample (\#4) classified as AGN according to the X-ray \\
  &  & & criteria defined in Sec.\,\ref{sec:x-agn-select}  \\
 \hline
    5 & {\it bona-fide} X-ray AGN sample & 1628 & The final X-ray AGN sample obtained from sample \#4 after the $L_{\mathrm{X}}$ correction for \\
    & & & the host-galaxy contribution (Section\,\ref{sec:corr_sfg_etg}) \\
    \hdashline
  5.1 & \hspace{0.3cm} out of which: optically-selected AGN & 494 & Sources in the {\it bona-fide} X-ray AGN sample (\#5) classified as AGN by at least one \\
  &  &  & of the optical BPT diagrams (Sec.\,\ref{sec:bpt-agn}) \\
\hdashline
  5.2 & \hspace{0.3cm} out of which: `classical' X-ray AGN & 454 & Sources in the {\it bona-fide} X-ray AGN sample (\#5) classified as AGN ccording to \\
  &  &  & the X-ray criteria defined in Sec.\,\ref{sec:x-agn-select} \\
  \hline
    5.3$^{\ast}$ & {\it bona-fide} X-ray AGN sample with {\tt DET\_ML} > 6 & 570 & The number of objects in the X-ray AGN sample (\#5) with detection likelihood   \\
    & & & {\tt DET\_ML} > 6 in the hard band; used only for the study of the sBHAR distribution \\
  & & & in Sec.\,\ref{sec:fupl}\\
  \hline
 \end{tabular}
\end{table*}

\subsection{The galaxy distribution on the SFR--$\mathcal{M}_{\ast}$ diagram}\label{sec:sdss-sfr-mass}

\begin{figure}
\centering
\includegraphics[width=0.99\linewidth]{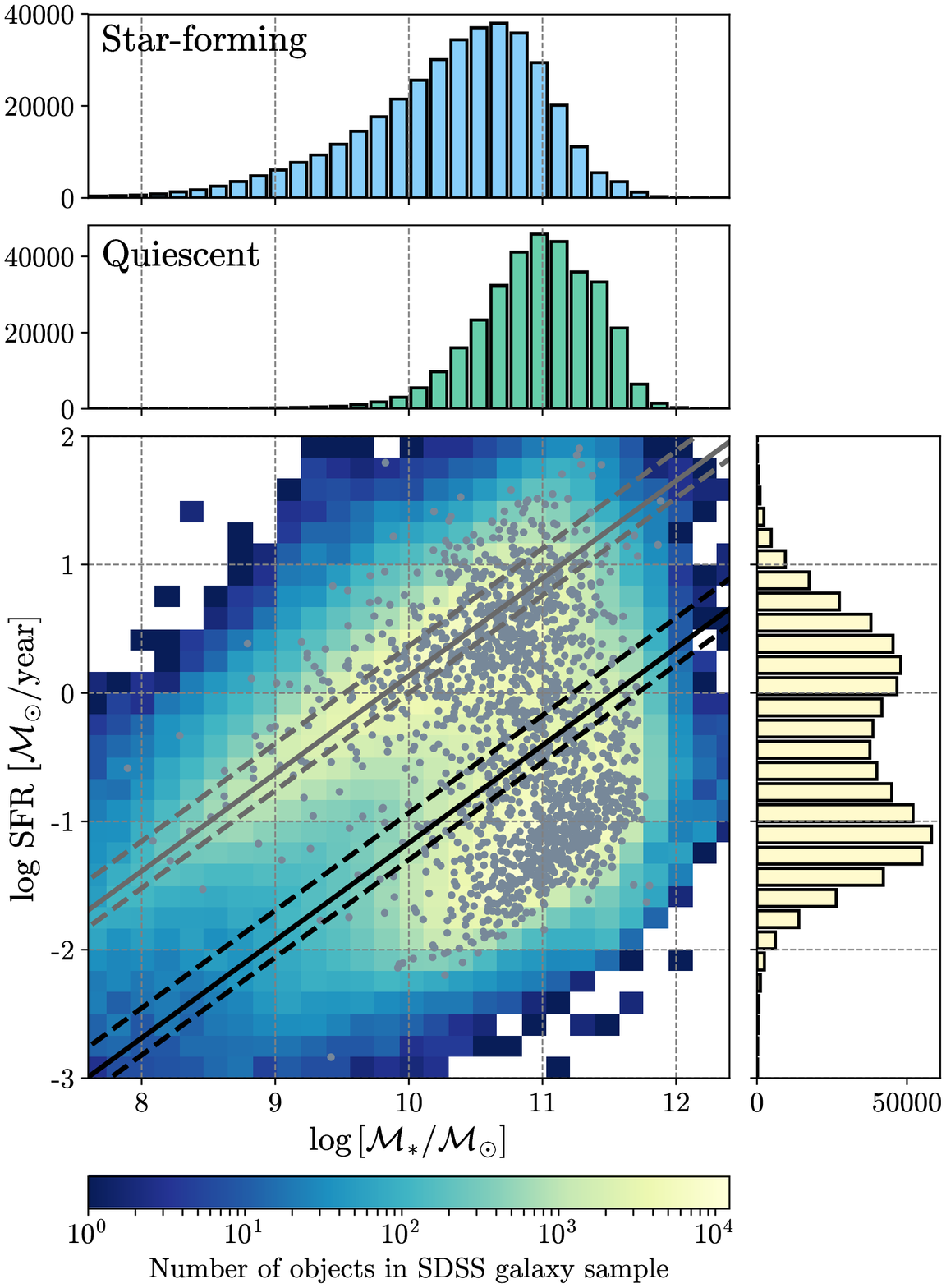}
\vspace*{-1.5ex}
\caption{The distribution of star-formation rate vs. stellar mass for our final SDSS galaxy sample. The histograms represent the SFR and $\mathcal{M}_{\ast}$ distributions. The right and top histograms represent the distribution of our sample in SFR and $\mathcal{M}_{\ast}$ The individual objects with X-ray emission found in 3XMM-DR8 catalogue are shown by grey circles. The grey lines show the main sequence (MS) of star-forming galaxies defined by Eq.\,\eqref{eq:sfg-pass-line}. All objects were divided into star-forming and quiescent galaxies by the cut 1.3\,dex below the MS of SFGs (black lines). The bottom dashed, solid and top dashed lines correspond to the lowest, mean and highest redshift in our sample ($z = 0.00, 0.11, 0.33$), respectively.}
\label{fig:sfr-mass-all}
\end{figure}

The distribution of galaxies on the SFR--$\mathcal{M}_{\ast}$ plain demonstrates that most galaxies can be separated into two main populations. First, the `star-forming' galaxies (SFGs) with steady processes of new stars formation and generally late-type morphologies with significant disc components (e.g. \citealt{Faber:07, Noeske:07, Blanton:09, Aird:17}), and second, the `quiescent' galaxies with passively evolving stellar populations and early-type morphologies. The distribution of our sample on SFR--$\mathcal{M}_{\ast}$ plane is shown in Fig.\,\ref{fig:sfr-mass-all}. 

We classify galaxies in our sample as star-forming or quiescent based on their SFRs relative to the evolving ‘star-forming main sequence’ setting the threshold between the two classes 1.3 dex below the main sequence defined by \citet{Aird:17}, and given by (Fig.\ref{fig:sfr-mass-all}):
\begin{multline}
\log\mathrm{SFR}_{\mathrm{cut}}(z)\,[\mathcal{M}_{\odot}\mathrm{year}^{-1}] =\\
= -8.9 + 0.76\log\,\mathcal{M}_{\ast}\,/\mathcal{M}_{\odot} + 2.95\log(1+z).
\label{eq:sfg-pass-line}
\end{multline}
Galaxies that fall below this cut were classified as passive or quiescent while those above the line as star-forming. Note that the relation in Equation\,\eqref{eq:sfg-pass-line} is redshift-dependent so that for the classification we used the redshift of each individual object.

In our SDSS galaxy sample 376\,938 sources are classified as star-forming (53.6\%) and 326\;484 as quiescent galaxies (46.4\%), respectively. 

\subsection{The 3XMM data set}\label{sec:xmm-data}

To quantify the accretion onto the central SMBH we used data from the XMM-Newton Serendipitous Source Catalogue\footnote{\url{http://xmmssc.irap.omp.eu/Catalogue/3XMM-DR8/3XMM_DR8.html}} (3XMM DR8, \citealt{Rosen:16})\footnote{We do not use the latest 4XMM-DR9 due to lack of sensitivity maps, which are required for sBHAR distribution reconstruction in Section~\ref{sec:fupl}.}. The comparison of our optical sample with 3XMM footprint revealed that only 40\,914 objects fall in the area of the sky observed by \textit{XMM-Newton}. The crossmatch of 3XMM DR8 with our optical sample (with a matching radius of $5^{\prime\prime}$) gives 3742 X-ray counterparts. Firstly, we rejected all objects with the detection flag\footnote{\url{https://xmmssc-www.star.le.ac.uk/Catalogue/3XMM-DR4/col\_flags.html}} {\sc sum\_flag > 3}, which indicates problems with the detection such as sources with a low coverage on the detector, sources in problematic areas near a bright source etc. Secondly, we selected objects with zero extension parameter to avoid including spatially extended objects (such as hot gas regions or galaxy clusters). In Section\,\ref{sec:extended} we evaluated the effect of including the extended sources on our results.

The XMM catalogue contains the observations from three cameras PN, MOS1 and MOS2\footnote{\url{https://www.cosmos.esa.int/web/xmm-newton/technical-details-epic}}. For our study we selected detections from the most sensitive PN camera. For objects with multiple observations we choose the one with the highest exposure time. When the data from PN camera were missing we used those from MOS1 or MOS2 cameras. This choice is motivated by the requirement to use robust flux upper limits in Section\,\ref{sec:fupl}.
After all these constrains the final 3XMM-SDSS sample (Table \ref{tab:samples}) contains 1953 objects (991 star-forming and 962 quiescent galaxies); their distribution on SFR--$\mathcal{M}_{\ast}$ plane is shown in Fig.\,\ref{fig:sfr-mass-all} by grey circles.

\section{Identification of AGN}

In order to identify AGN and study the properties of their host galaxies we need to reveal the presence of nuclear non-stellar emission. Since our primary sample is based on an optical SDSS `galaxy' catalogue, which by definition excludes quasar and other sources strongly dominated by nuclear emission, the majority of our sources are type 2 AGN. This type of AGN are partially obscured and provide the opportunity to observe emission not only from the nuclear region, but also from the host galaxy. Given the presence of both nuclear and star-formation emission in the galaxy we used a combination of multiwavelength techniques to identify `classical' AGN.

\subsection{Optical selection: BPT-diagrams}\label{sec:bpt-agn}

A common technique to distinguish AGN from normal star-forming galaxies is the BPT diagram discovered by \citet{Baldwin:81} and improved by \citet{Veilleux:87}. It is based on the comparison of the flux ratios of two pairs of strong emission lines with different level of ionisation.  Since these ratios are almost completely insensitive to reddening or to errors in the spectrophotometry, it is useful for separating AGN and star-forming galaxies due to the different processes producing these emission lines in their spectra. 

MPA-JHU SDSS catalogue already has the BPT classification flag ({\sc BPTCLASS} inside \textit{galSpecLines} file) based on [O\;{\sc iii}] $\lambda$5007/H$\beta$ versus the ratio [N\;{\sc ii}] $\lambda$6583/H$\alpha$ ratios (hereinafter BPT-[N\;{\sc ii}]) described in detail on \citet{Brinchmann:04} paper. The galaxies were classified as star-forming galaxies (SFGs), AGN or composite objects that have contribution from both AGN and star-formation. Additionally, \citet{Brinchmann:04} defined two additional classes: SFGs and AGN with low S/N for at least one of the emission lines required for the BPT classification. The residual objects with no emission lines at all were marked as unclassified. 
To verify the BPT classification already presented in the MPA-JHU SDSS catalogue we used the empirical criteria by \citet{Kauffmann:03c} to separate the `pure' SFGs from composite objects and \citet{Kewley:06} to separate the latter from AGN. Since the application of the BPT selection criteria requires the objects with S/N $> 3$ for all four emission lines, we selected 1260 objects (64.5\%) from our 3XMM-SDSS sample which satisfied the S/N condition for emission lines required for BPT-[N\;{\sc ii}] diagram. Additionally, we refine the BPT classification using the two other diagnostic diagrams proposed by \citet{Kewley:06} that involving the ratios [O\;{\sc i}] $\lambda$6300/H$\alpha$ (BPT-[O\;{\sc i}]) and [S\;{\sc ii}] $\lambda$6717/H$\alpha$ (BPT-[S\;{\sc ii}]) to distinguish the two classes of narrow-line AGN: Seyfert 2 and low-ionisation nuclear emission line objects (LINER). In this case, we selected 1165 objects (59.7\% of our 3XMM-SDSS sample) and 947 (48.5\%) objects in the BPT-[S\,\;{\sc ii}] and [O\,\;{\sc i}], respectively.

\begin{figure*}
    \centering
    \includegraphics[width=0.32\linewidth]{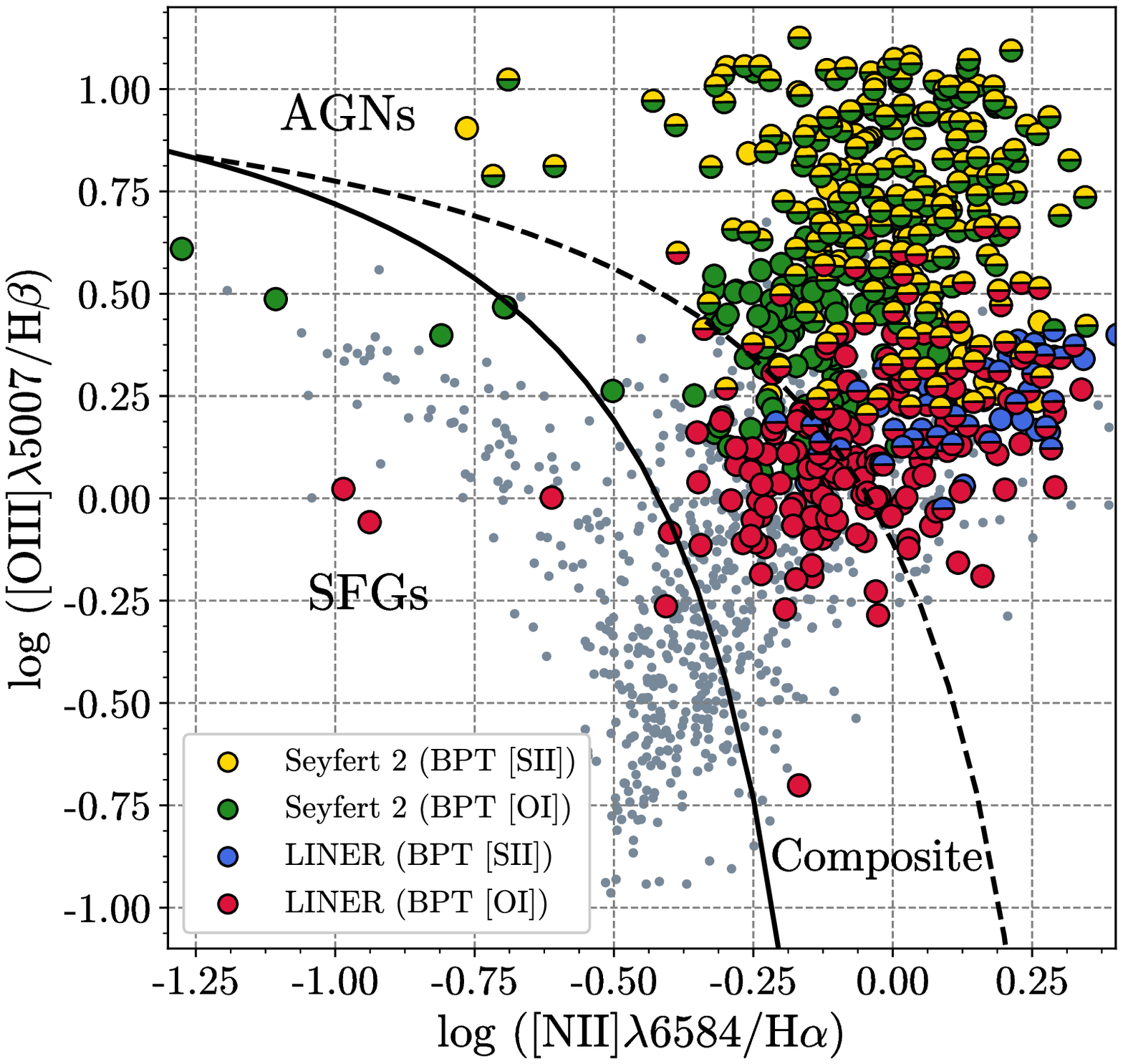}
    \includegraphics[width=0.32\linewidth]{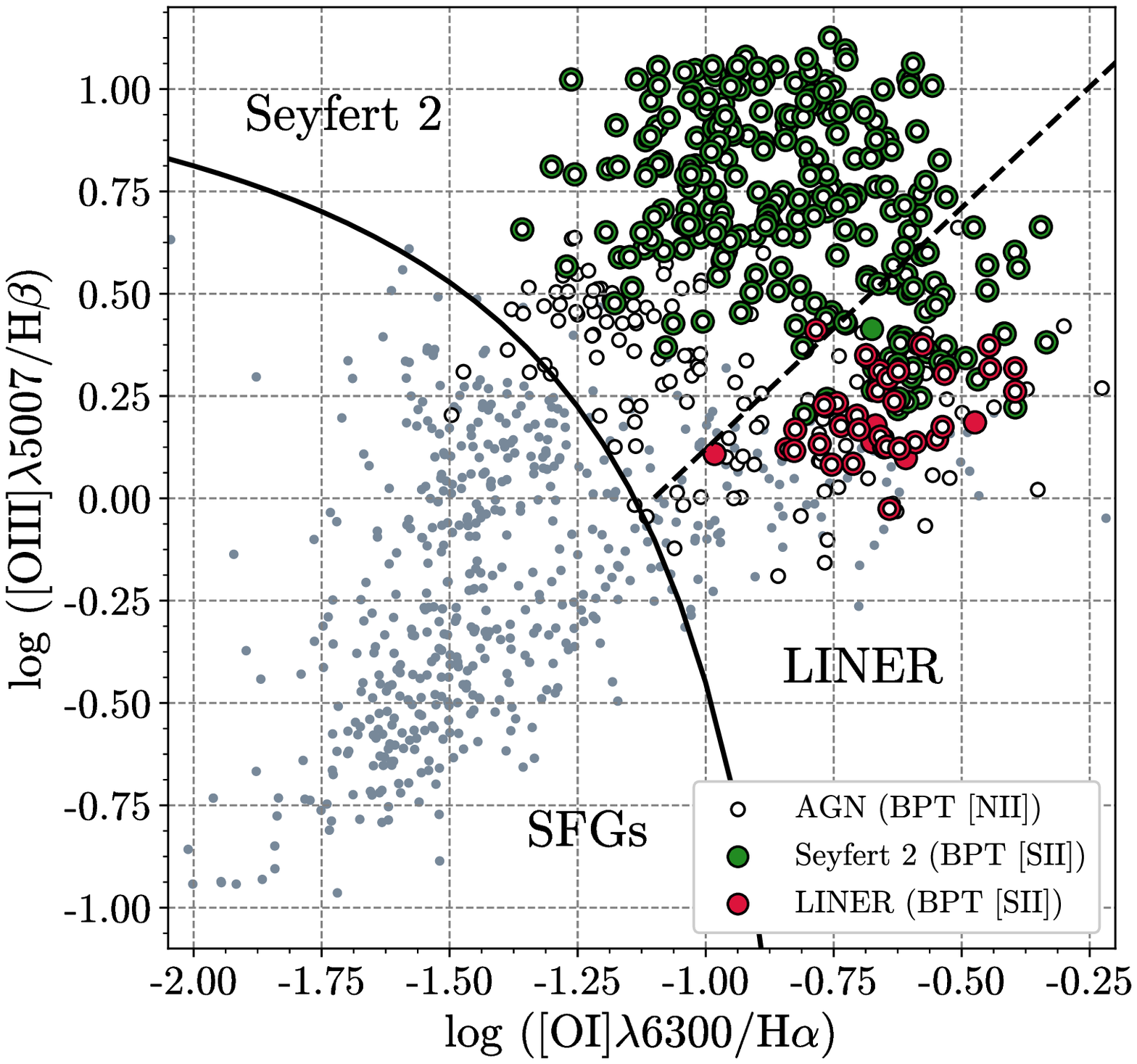}
    \includegraphics[width=0.32\linewidth]{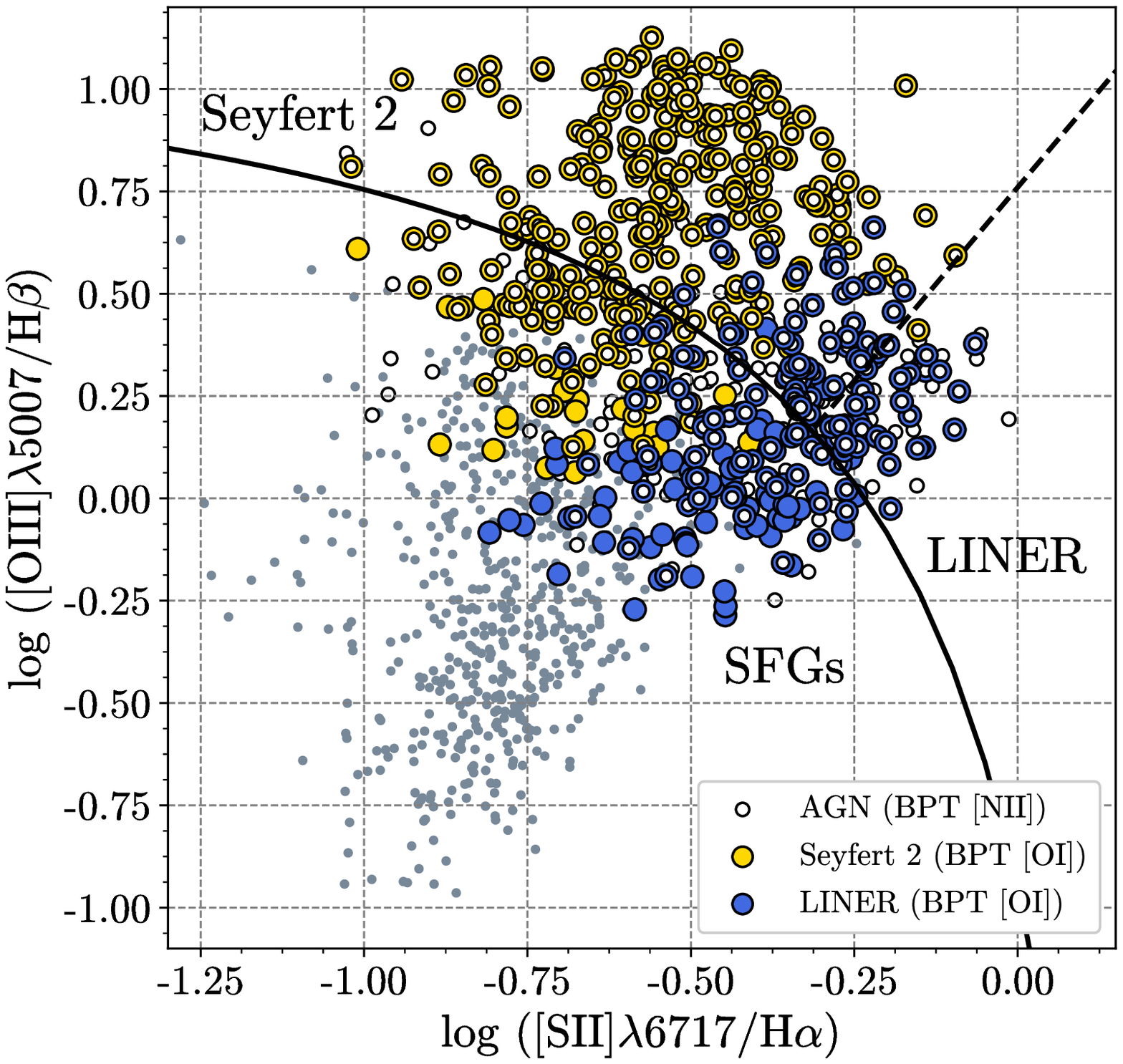}
    \vspace*{-1.5ex}
    \caption{The BPT-diagrams for the 3XMM-SDSS sample. \textit{Right:} [O\;{\sc iii}] $\lambda$5007/H$\beta$ vs [N\;{\sc ii}] $\lambda$6583/H$\alpha$ flux ratios diagram. The solid line represents the empirical criterion defined by \citet{Kauffmann:03a} for separating star-forming and composite galaxies. The dashed line is the criterion by \citet{Kewley:06} for AGN identification. The color circles show AGN identified by two other diagnostic criteria based on the [S\;{\sc ii}] $\lambda$6717 and [O\;{\sc i}] $\lambda$6300 lines. \textit{Center:} [O\;{\sc iii}] $\lambda$5007/H$\beta$ vs [O\;{\sc i}] $\lambda$6300/H$\alpha$ flux ratios. The criteria of \citet{Kewley:06} for SFG/AGN and Seyfert 2/LINER separation are represented by the solid and dashed lines, respectively. The colour circles show AGN identified by the two other criteria based on the [N\;{\sc ii}] $\lambda$6583 and [S\;{\sc ii}] $\lambda$6717 lines. \textit{Right:} [O\;{\sc iii}] $\lambda$5007/H$\beta$ vs [S\;{\sc ii}] $\lambda$6717/H$\alpha$ flux ratios. The colour circles show AGN identified by the two other diagnostic criteria on the basis of [N\;{\sc ii}] $\lambda$6583 and [O\;{\sc i}] $\lambda$6300 lines.}
    \label{fig:bpt}
\end{figure*}

The results obtained by BPT-[N\;{\sc ii}] coincide with the BPT flag obtained by \citet{Brinchmann:04} and it is presented on left panel of Fig.\,\ref{fig:bpt}. The BPT-[O\;{\sc i}] and BPT-[S\;{\sc ii}] diagrams are shown in Fig.\,\ref{fig:bpt} (centre and right panel). 

\begin{figure}
\centering
 \includegraphics[width=0.93\linewidth]{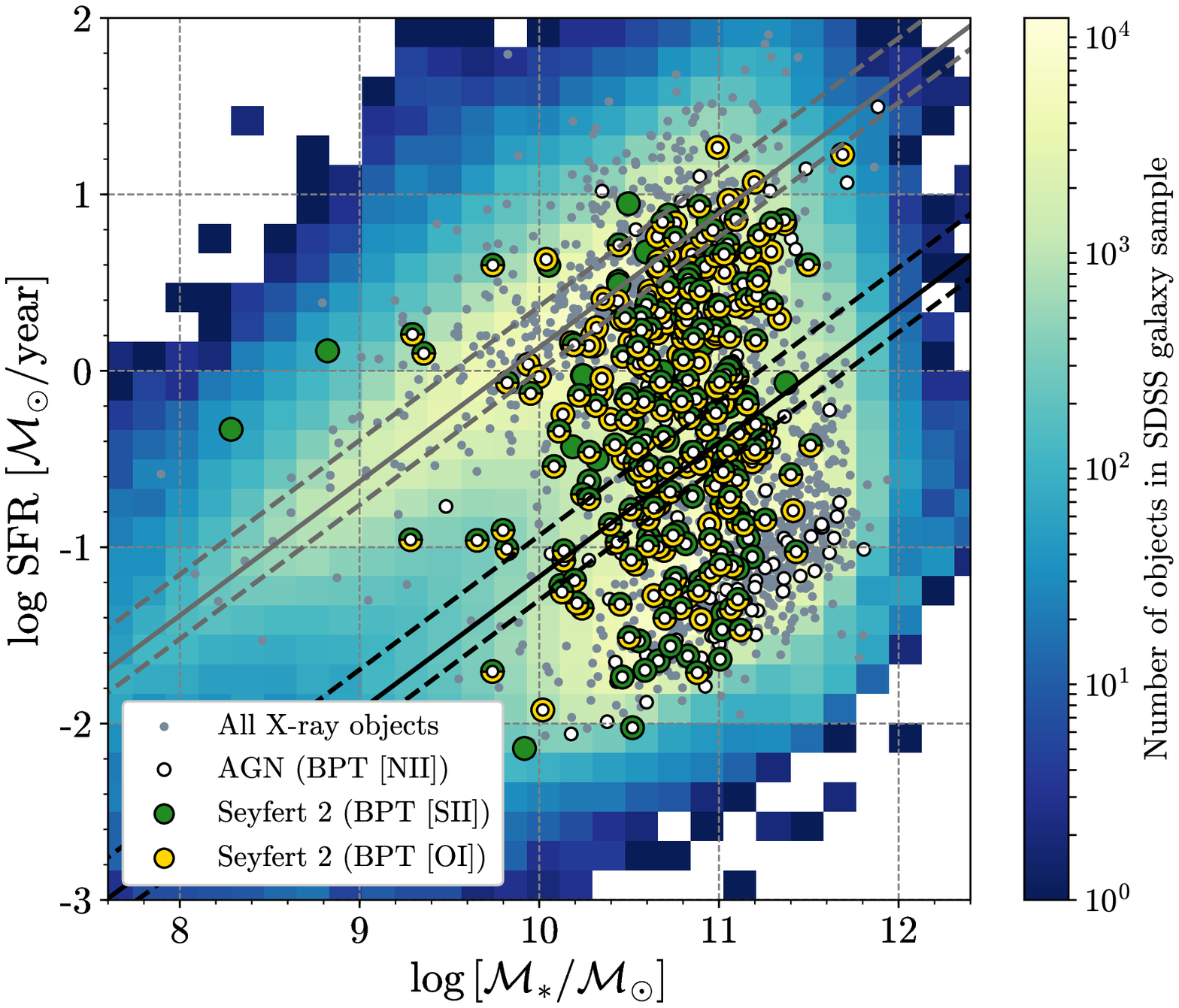}\\
 \includegraphics[width=0.93\linewidth]{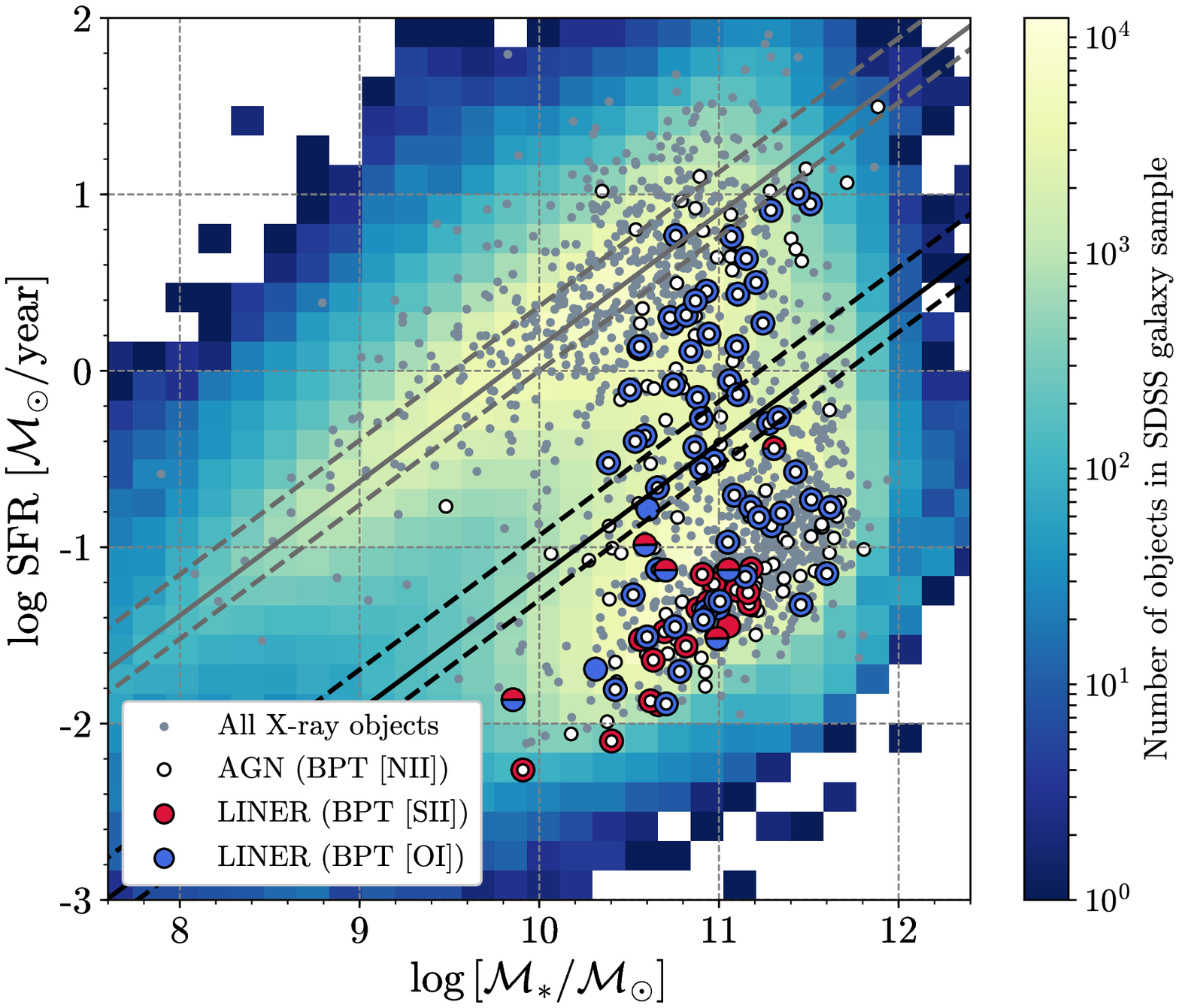}
 \vspace*{-1.5ex}
 \caption{\textit{Top:} The distribution of star-formation rate vs. stellar mass as in Fig.\,\ref{fig:sfr-mass-all} for the sample of optically identified AGN and Seyfert 2 according to the BPT criteria presented in Fig.\,\ref{fig:bpt}. \textit{Bottom:} The same distribution for LINERs. The black and grey lines are the same as in Fig.\,\ref{fig:sfr-mass-all}.
 }
 \label{fig:sfr-mass-bpt}
\end{figure}

We found that 553 objects were classified as AGN in the BPT-[N\,\;{\sc ii}] diagram, 318 as Seyfert 2 and 189 as LINER at least by one of BPT-[S\,\;{\sc ii}] or [O\,\;{\sc i}] diagrams and 50 objects were classified as Seyfert 2 by one diagram and LINER by another. The total number of AGN classified by three BPT diagrams (hereinafter optically-selected AGN) is summarised in Table\,\ref{tab:samples}. The distribution of these objects on the star-formation rate--stellar masses diagrams (see Fig.\,\ref{fig:sfr-mass-bpt}) shows that 15.2\% of sources in our 3XMM-SDSS sample, classified as AGN due to BPT-[N\,\;{\sc ii}], are located in SFGs and 13.2\% in quiescent galaxies. Simultaneously, Seyfert 2 identified by BPT-[S\,\;{\sc ii}] and/or BPT-[O\,\;{\sc i}] are mainly in SFGs (11.7\%) and only 4.6\% in quiescent galaxies. On the contrary, a higher percentage of LINER is located in quiescent galaxies (5.7\%) while only 4.0\% in SFGs. These results are consistent with the paradigm that LINERs are located mostly in host galaxies with little star formation and older stellar population \citep{Kauffmann:03c, Ho:08, Heckman:14} and may show the presence of a strong jet \citep{Falcke:04}. 

The BPT-diagram is powerful for AGN identification in optical band, but it has several significant limitations.  As it was mentioned before the application of BPT method requires high-S/N detection of 4 emission lines. {For example, objects with strong [O\,{\sc iii}]\,$\lambda5007$ line which is a reliable tracer of AGN activity can not be classified without the presence of less intense lines such as H$\beta$, [N\,{\sc ii}] etc.} Additionally, some AGN have weak or no emission lines due to high obscuration by the circumnuclear and galactic dust. Furthermore, some SFGs may contain a low-luminosity AGN, which may not be identified since the emission of the star-forming processes will dominate the spectrum.

\subsection{X-ray AGN selection criteria}\label{sec:x-agn-select}

As it will be discussed in Section\,\ref{sec:corr_sfg_etg} the contribution of the host galaxy to the total X-ray emission is generally smaller than in the optical, which makes X-ray detection a robust technique for AGN identification \citep{Brandt:05}.

We classified an X-ray source as an AGN if it satisfied at least one of the following three criteria: \textit{(1)} an intrinsic luminosity $L_{\mathrm{X,int}} \geq 3 \cdot 10^{42}$\;erg\;s$^{-1}$ in the hard band { defined in Section\,\ref{sec:int-lum}}; \textit{(2)} X-ray-to-optical flux ratio of $\log\;(f_{\mathrm{X}}/f_{\mathrm{opt}}) > -1$ and \textit{(3)} X-ray-to-IR flux ratio of $\log\;(f_{\mathrm{X}}/f_{\mathrm{Ks}}) > -1.2$. The last two criteria, X-ray/optical and X-ray/IR ratios, were calculated in the form: 
\begin{equation}
\log\,(f_{\mathrm{X}}/f_{{j}}) = \log f_{\mathrm{X}}+\frac{\mathrm{mag}_{j}}{2.5} + C_{j},   
\end{equation}
where $f_{\mathrm{X}}$ is the hard-band detected flux, $f_{j}$ is $f_{\mathrm{opt}}$ in the SDSS \textit{r-} or \textit{i-}band fluxes or $f_{\mathrm{Ks}}$ in the 2MASS $K_{\mathrm{S}}$-band flux; mag$_{j}$ is the magnitude in \textit{r-}, \textit{i-} or $K_{\mathrm{S}}$-band, $C_{j}$ is calibration constant determined from band parameters as described in \citet{Ananna:17}. 

\begin{figure*}
\centering
 \includegraphics[width=0.32\linewidth]{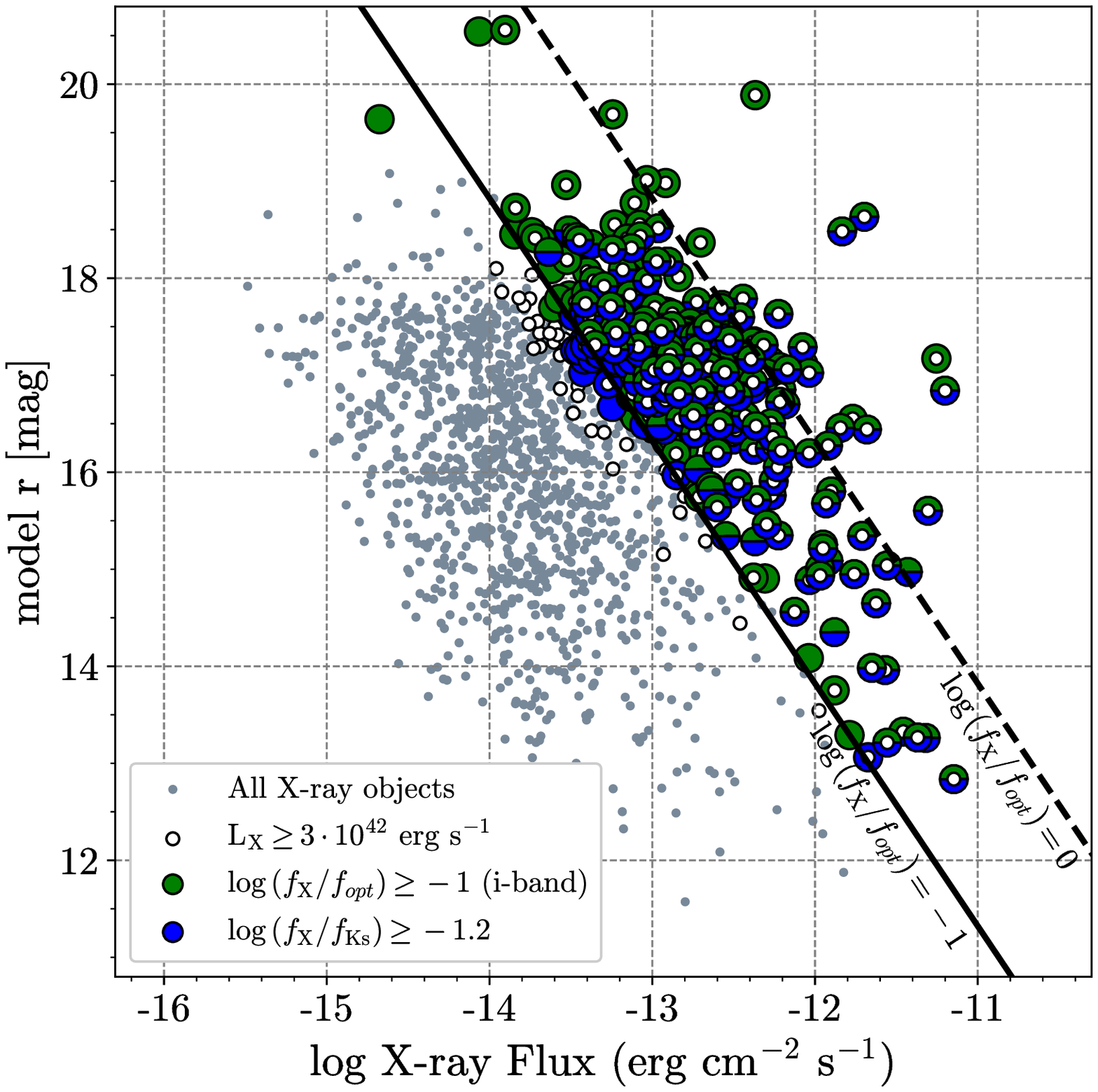}
 \includegraphics[width=0.32\linewidth]{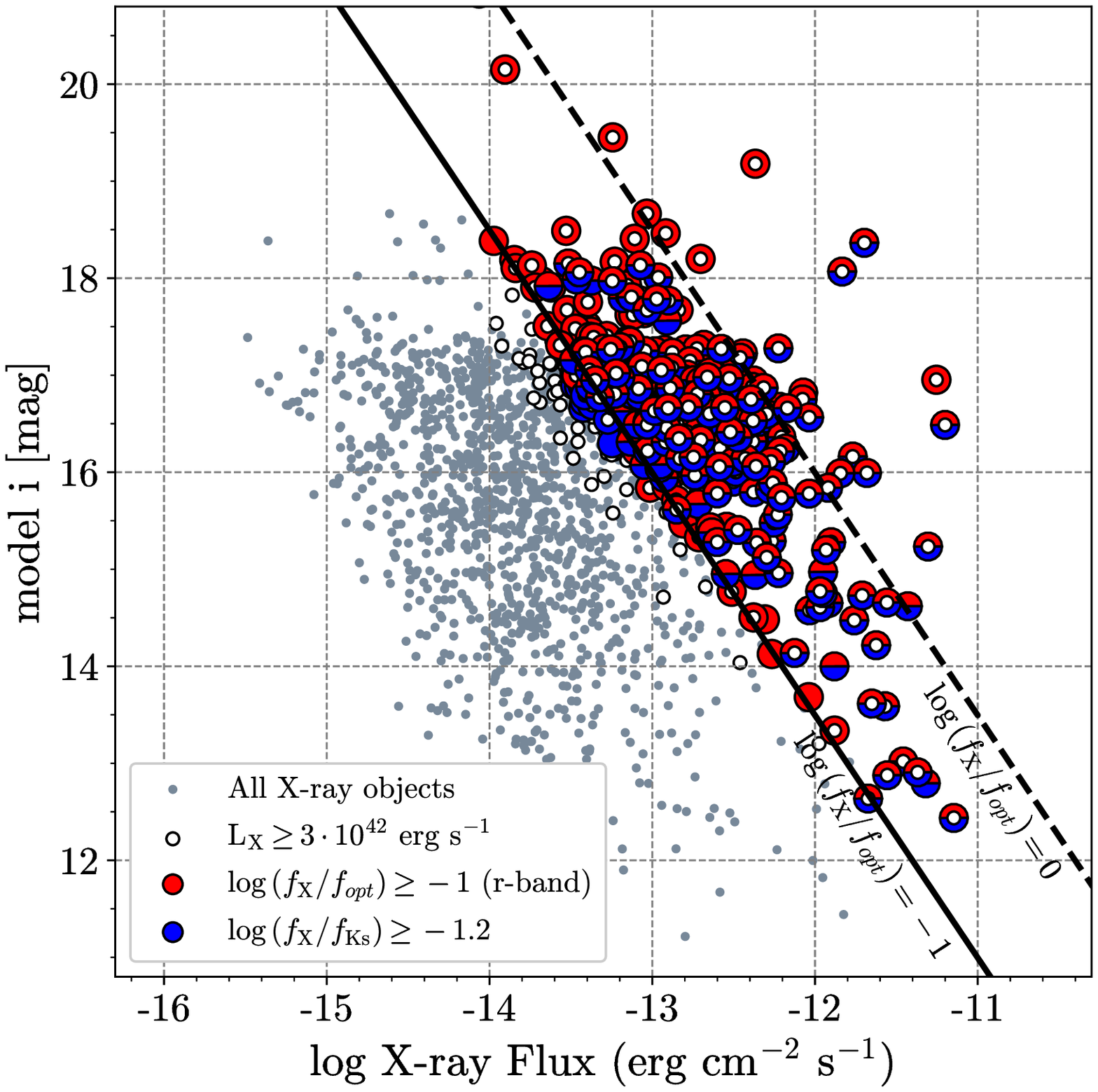}
 \includegraphics[width=0.32\linewidth]{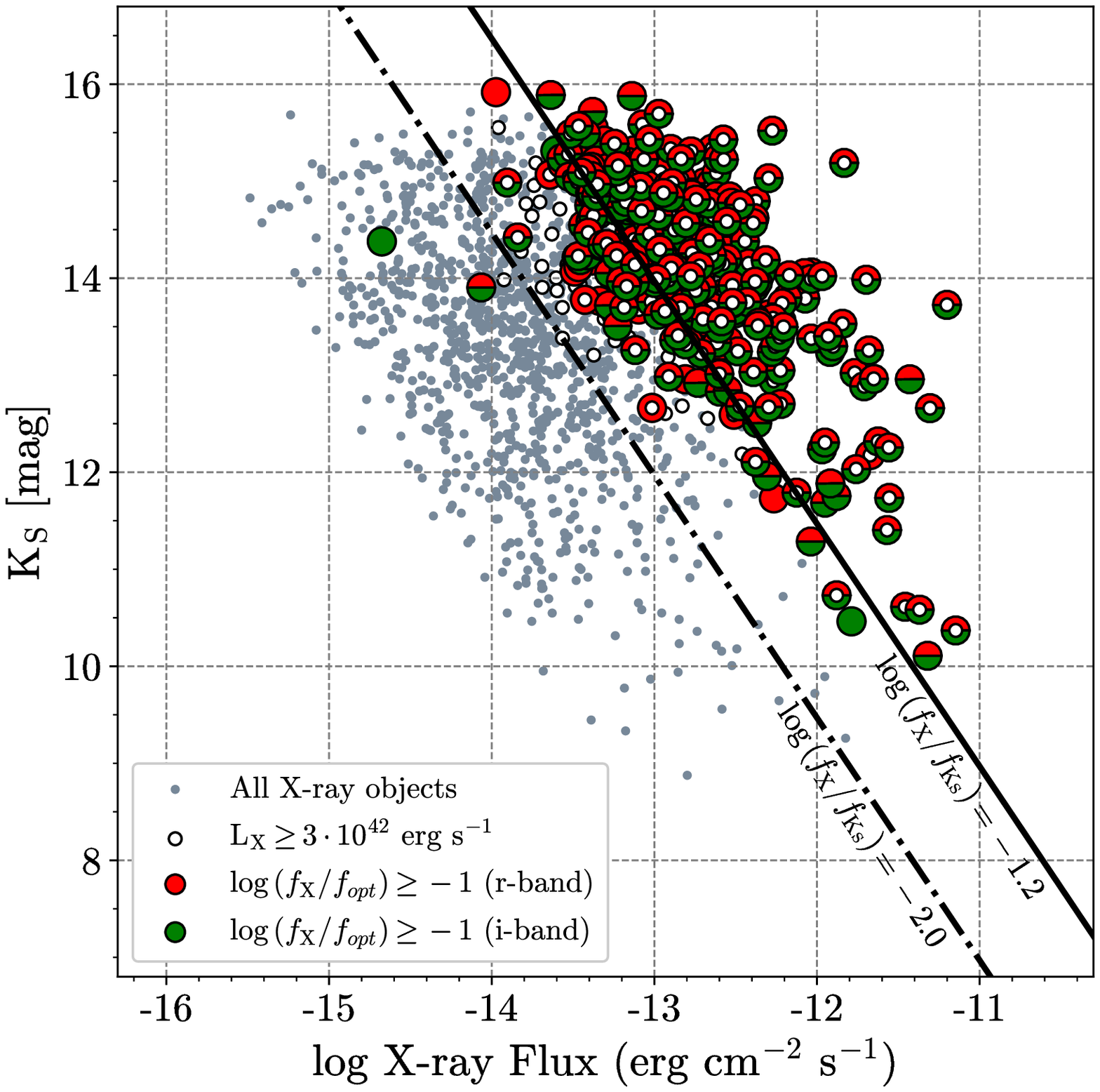}
 \vspace*{-1.5ex}
 \caption{\textit{Left and centre:} The X-ray flux in hard band vs. optical SDSS \textit{r-}band and \textit{i-}band magnitude for sources in 3XMM-SDSS sample (grey circles). White circles indicate AGN having $L_{\mathrm{X,int}} \geq 3 \cdot 10^{42}$\;erg\;s$^{-1}$ in Fig.\,\ref{fig:lum-z}. AGN selected by the X-ray/optical flux ratios in \textit{r-} and \textit{i-}band are represented by red and green circles, by X-ray/IR ratio by blue circles. Diagonal lines indicate constant flux ratios between the SDSS \textit{r-} and \textit{i-}band and X-ray hard band. \textit{Right:}  The X-ray flux in hard band vs. 2MASS infrared $K_{\mathrm{S}}$-band magnitude. The colours of circles are the same as on left and central panel. Diagonal lines indicate constant flux ratios between the 2MASS $K_{\mathrm{S}}$-band and X-ray hard band.}
 \label{fig:x-agn-ratio}
\end{figure*}

\begin{figure}
\centering
\includegraphics[width=0.85\linewidth]{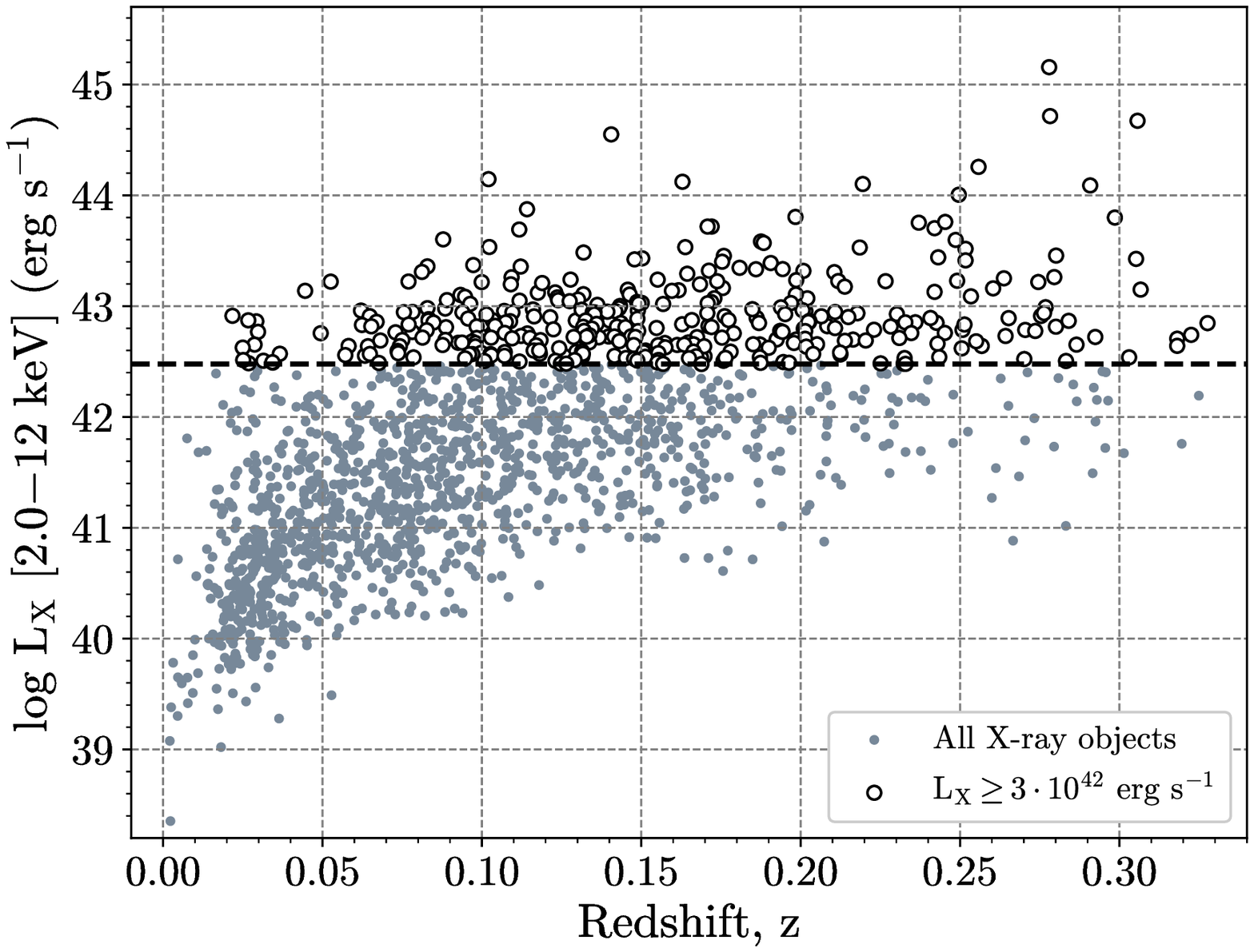}
\vspace*{-1.5ex}
\caption{The X-ray luminosity vs. redshift for the objects in 3XMM-SDSS sample. The horizontal dashed line indicates $L_{\mathrm{X,int}} \geq 3 \cdot 10^{42}$\;erg\;s$^{-1}$ criteria utilised to classify AGN (black circles).}
\label{fig:lum-z}
\vfill
\centering
\includegraphics[width=0.9\linewidth]{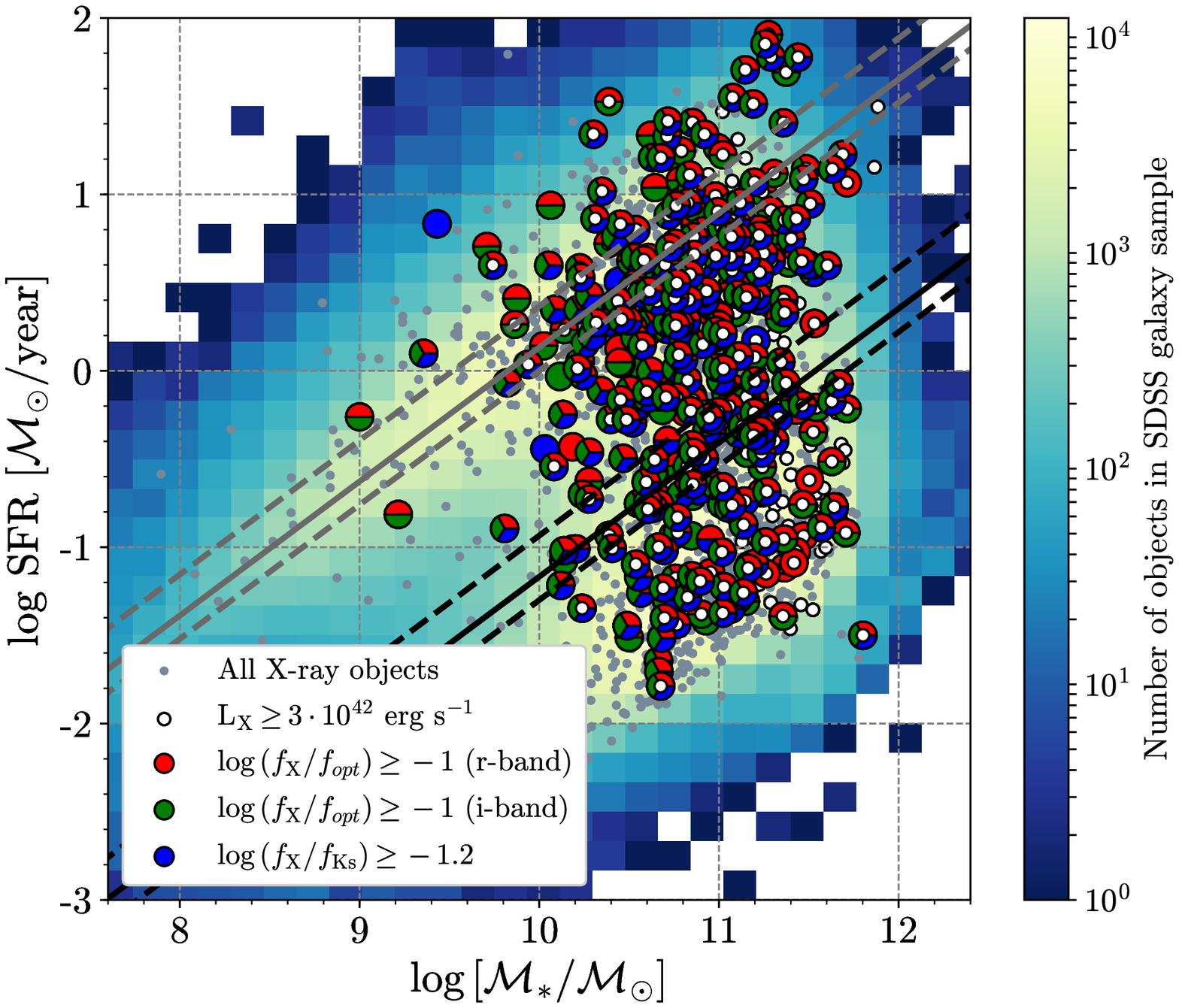}
\vspace*{-1.5ex}
\caption{The distribution of star-formation rate vs. stellar mass for 3XMM-SDSS sample in hard band (grey circles). The gradient from blue to yellow shows the 2D histogram of the density distribution of galaxies in our optical SDSS sample. All symbols represent the AGN selected by the X-ray criteria described in Section\,\ref{sec:x-agn-select}. The black and grey lines are the same as in Fig.\,\ref{fig:sfr-mass-all}.}
\label{fig:sfr-m}
\end{figure}

In total, we identified 469 AGN (24.0\% of our 3XMM-SDSS sample), 362 were selected by the X-ray luminosity threshold, 397 and 372 by X-ray-to-optical flux ratio in \textit{r-} and \textit{i-}bands and 281 by X-ray/IR flux ratio (see summary in Table\,\ref{tab:samples}). For simplicity hereafter we will refer to AGN selected by such X-ray criteria as `classical' X-ray AGN. Only 212 objects are classified as AGN by all criteria. The redshift distribution of AGN selected by the $L_{\mathrm{X}}$ criterion are shown in Fig.\,\ref{fig:lum-z}, the X-ray/optical flux ratios in SDSS \textit{r-} and \textit{i-}band (left and central panels) and X-ray/IR flux ratio are presented in Fig.\,\ref{fig:x-agn-ratio} (right panel).

The distribution of our sample on the SFR vs stellar mass diagrams in the hard band is shown in Fig.\,\ref{fig:sfr-m}. AGN selected with the above X-ray criteria show the tendency to occupy predominantly the star-forming main-sequence (67.6\% of all X-ray selected AGN), while only 32.4\% of AGN were found in quiescent galaxies. Such preference of AGN to be hosted by star-forming galaxies was also found by \citet{Mullaney:12a, Mendez:13, Rosario:13, Shimizu:15, Delvecchio:15, Aird:18, Stemo:20} for AGN selected by different IR and X-ray criteria. At the same time, the percentage of star-forming and quiescent galaxies hosting an AGN in our 3XMM-SDSS sample are 32.0\% and 15.8\%, respectively.

\section{Calculation of the intrinsic X-ray luminosity}\label{sec:int-lum}

We computed the rest-frame X-ray luminosities in the hard band (2.0--12\,keV) after applying a K-correction; following \citet{Luo:17}, we assumed a photon index $\Gamma = 1.4$ appropriate for a moderately obscured AGN spectrum with the absorption column density 
$\log{N_H/\mathrm{cm}^{-2}}\simeq 22.5$ (also see \citealt{Tozzi:06, Liu:17}).

The assessment of the level of AGN activity from the X-ray emission requires to determine the contribution of the host galaxy. Galaxy populations have different contributions to their X-ray emission; for example, X-ray radiation in star-forming galaxies is mainly due to X-ray binaries. Low mass X-ray binaries (LMXBs) are associated to the old stellar population in bulges of spirals, while high-mass X-ray binaries (HMXBs) are associated to younger stars and are concentrated preferentially in the disk of spirals galaxies. On the other hand, quiescent galaxies with elliptical morphology have only one type of X-ray binaries, i.e LMXBs, but the fraction of X-rays emitted by the hot gas can be significant or even dominate the total emission \citep{Fabbiano:89, Kim:13}. Hence it is clear that the X-ray luminosity has to be corrected by taking into account the different types of contributions. The correction for our 3XMM-SDSS sample is made separately for the two populations of galaxies (star-forming and quiescent) defined in Section\,\ref{sec:sdss-sfr-mass}. Also, we decided to use only the hard X-ray band (2.0--12\,keV) since it allows to minimise the contribution from hot gas, SN remnants and other soft X-ray components.


\subsection{Subtraction of X-ray emission due to star formation}\label{sec:corr_sfg_etg}

\begin{figure*}
\includegraphics[width=0.99\linewidth]{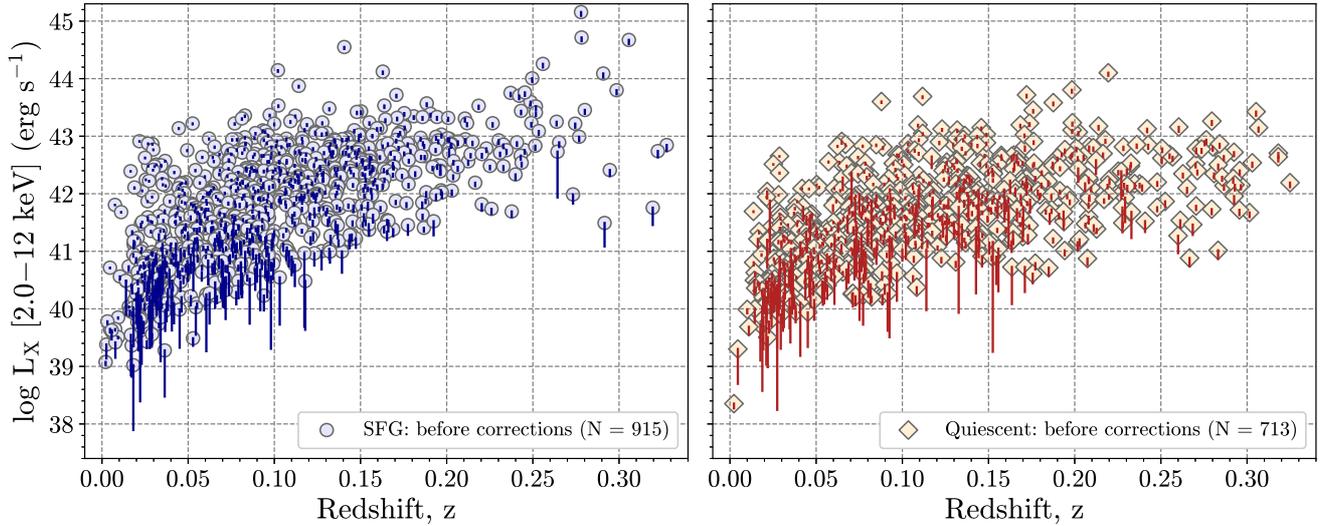}
 \vspace*{-1.5ex}
\caption{The final nuclear X-ray luminosity vs redshift distribution of our X-ray AGN sample. The uncorrected L$_\mathrm{X}$ values for SFGs and quiescent galaxies are presented as solid circles and diamonds, respectively. The change in L$_\mathrm{X}$ after corrections for each object is shown by a solid line.}
\label{fig:Lum-corr}
\end{figure*}

Different independent analyses show that X-ray emission from SFGs correlates directly with SFR \citep{Ranalli:03, Mineo:12, Vattakunnel:12, Fragos:13, Symeonidis:14, Lehmer:16}. We calculated the expected X-ray luminosities of SF galaxies using the scaling relation between $L_{\mathrm{X,SF}}$ and SFR, stellar masses $\mathcal{M}_{\ast}$ and redshift $z$ of galaxies from \citet{Lehmer:16} in the following form:
\begin{multline}
    L_{\mathrm{X, hard}}\,[\text{erg\;s}^{-1}] = \alpha(1+z)^{\gamma}\mathcal{M}_{\ast}\,/\mathcal{M}_{\odot}+ \\
    +\beta(1+z)^{\delta}\mathrm{SFR}\,[\mathcal{M}_{\odot}\;\text{year}^{-1}],
    \label{eq:sfg-lehmer}
\end{multline}
where $\log\alpha=29.37\pm0.17$, $\log\beta=39.28\pm0.05$, $\gamma = 2.03\pm0.06$ and $\delta = 1.31\pm 0.13$ for hard band. The X-ray luminosity vs redshift distribution for SFGs, before and after correcting for the contribution from XRBs, is shown in Fig.\,\ref{fig:Lum-corr}. 

Since the X-ray luminosity of quiescent galaxies (ETGs) is mainly due to LXMBs and hot gas \citep{Boroson:11, Kim:13, Civano:14} we have to apply a different type of correction. We use the relation between luminosity of the galaxy in the $K$-band and $L_\mathrm{X}$ of different components of quiescent galaxies. Using 30 normal early-type galaxies observed by \textit{Chandra} \citet{Boroson:11} found that the X-ray luminosity due to LMXBs correlates with K-band luminosity as 
\begin{equation}
    L_{\mathrm{X}}\,[\text{erg s}^{-1}] = 10^{29.0\pm0.176}\cdot L_{\mathrm{K}}\,[L_{\mathrm{K}\odot}].
    \label{eq:etg-lmxb}
\end{equation}

Other types of stellar sources that can radiate X-rays and therefore provide a contribution to the total X-ray luminosity of the galaxy are coronally active binaries (ABs) and cataclysmic variables (CVs). For their study \citet{Boroson:11} used the \textit{Chandra} observation of M31 and M32 galaxies in hard band (2--10\;keV) as their proximity allows us to resolve the individual X-ray sources inside the galaxies. They found a similar relation between X-ray and $K$-band luminosity:
\begin{equation}
   L_{\mathrm{X}}\,[\text{erg s}^{-1}] = 4.5^{+0.8}_{-0.6} \cdot 10^{27} \cdot L_{\mathrm{K}}\,[L_{\mathrm{K}\odot}].
    \label{eq:etg-ab-cv}
\end{equation}

In addition, to evaluate the contribution of the hot gas we used the relation between the X-ray emission and the $K$-band luminosity in the form $L_{\mathrm{X}} \sim L_{\mathrm{K}}^{\alpha}$ with exponential slope $\alpha = 4.5$ from \citet{Civano:14}.

We used $K_{S}$ magnitudes from the 2MASS\footnote{\url{https://old.ipac.caltech.edu/2mass/}} Point Source Catalogue (PSC) and Extended Source Catalogue (XSC) \citep{Skrutskie:06}. Firstly, we crossmatched our 3XMM-SDSS sample with the XSC as the galaxies in our sample are located at low redshift and should have extended shapes in 2MASS. Most of the objects in the PSC are stars of the Milky Way, but the catalogue also contains a significant number of unresolved, more distant galaxies; therefore we made additional crossmatch between the PSC and our sample.
We found that 1457 objects (74.6\% of our 3XMM-SDSS sample) have counterparts in XSC and 1743 (89.2\%) in PSC. 1323 objects in 3XMM-SDSS sample were found in both catalogues because PSC contains entries and point source-processed flux measurements for virtually all extended sources in the XSC. We refined our selection rejecting objects with low S/N, contamination or blending according to the quality flags described in the Explanatory Supplement to the 2MASS\footnote{\url{https://old.ipac.caltech.edu/2mass/releases/allsky/doc/explsup.html}} and reduced the number of sources to 1809 (1330 extended and 479 point-like sources).

We calculate the $K_S$-band luminosity in units of solar luminosity for 921 quiescent galaxies using the equation from \citet{Civano:14}:
\begin{equation}
    L_{K_{S}}\,[L_{\odot}] = 10^{-(K_{S}-K_{\odot})/2.5}\cdot(1+z)^{\alpha-1}\cdot(D_{L}\,[\text{pc}]/10)^{2}
    \label{eq:k-mag-lum}
\end{equation}
where $K_{S}$ is the magnitude from the 2MASS catalogue, $z$ is the redshift, and $D_{L}$ is the luminosity distance in parsecs and $K_{\odot} = 3.33$\;mag is the magnitude of Sun in $K$-band. To evaluate rest-frame $K$-band luminosities, we assumed a spectral shape of the type $f_{\nu} \varpropto \nu^{\alpha}$, where $\alpha = -(J-K_{S}/\log(\nu_{J}/\nu_{K_{S}}))$, where $J-K_{S}$ is colour taken from the 2MASS catalogue.

We determined the X-ray luminosity for AB+CV (from Eq.\;\ref{eq:etg-ab-cv}), LMXBs (Eq.\;\ref{eq:etg-lmxb}) and hot gas components (with the slope $\alpha = 4.5$) and subtracted the luminosities from our intrinsic X-ray luminosity. The redshift distribution of X-ray luminosity in hard band for quiescent galaxies after corrections is shown in Fig.\,\ref{fig:Lum-corr} (right panel). 

Our final {bona-fide} AGN sample (hereinafter X-ray AGN sample) contains 1628 objects with positive residual X-ray luminosity after correction for X-ray emission from binaries and hot gas (83.4\% from the 3XMM-SDSS sample before correction), 915 of which are SFGs and 713 are quiescent ETGs. All following results have been obtained based on this X-ray AGN sample. Note however that, in order to study the sBHAR distribution we were forced to further filter our X-ray AGN sample due to a absence of flux upper limits estimates for part of the source (see details in Section\,\ref{sec:fupl} and Table\,\ref{tab:samples}).

\section{The specific Black Hole accretion rate}\label{sec:bhar}

To investigate the AGN activity we calculate the specific Black Hole accretion rate ($\lambda_{\mathrm{sBHAR}}$), 
defined as: 
\begin{equation}
    \lambda_{\mathrm{sBHAR}} \propto \frac{k_{\mathrm{bol}}\,{L}_{\mathrm{X,hard}\,}[\mathrm{erg\,s}^{-1}]}{\mathcal{M}_{\ast}\,/{\mathcal{M}_{\odot}}} 
    \label{eq:spec-bhar}
\end{equation}
where $k_{\mathrm{bol}}$ is a bolometric correction factor for the hard band and ${L}_{\mathrm{X,hard}}$ is the 2.0--12\,keV X-ray luminosity. 
Although the bolometric correction factor is dependent on the luminosity \citep{Marconi:04, Lusso:12, Bongiorno:16}, here we adopted an average bolometric correction of $k_{\mathrm{bol}} = 25$ since the other systematics discussed below dominate the final uncertainty. 
$\lambda_{\mathrm{sBHAR}}$ is an observationally defined quantity, tracing the level of AGN activity per unit stellar mass of the host galaxy. The proportionality constant is conventionally set to be  $(0.002\times 1.3\cdot10^{38} [\mathrm{erg\,s}^{-1}])^{-1}$ so that $\lambda_{\mathrm{sBHAR}}$ is comparable to the Eddington ratio $\lambda_{\mathrm{Edd}} \propto \mathrm{L}_{\mathrm{X}}/\mathcal{M}_{\mathrm{BH}}$ in the assumption that the Black Hole mass scales with the host galaxy stellar mass as $\mathcal{M}_{\mathrm{BH}} = 0.002\,\mathcal{M}_{\ast}/\mathcal{M}_{\odot}$ \citep{Haring:04}. This assumption is however ambiguous since ${M}_{\ast}$ is not an unbias tracer of $\mathcal{M}_{\mathrm{BH}}$. Here we retain this definition for an easier comparison with previous works \citep[e.g.][]{Aird:12, Aird:18} but we also report in the plots $\lambda_{\mathrm{sBHAR}}$ in ${L}_{\mathrm{X}}/\mathcal{M}_{\ast}$ units in accordance to, e.g,  \citet{Georgakakis:14,Bongiorno:16}. The connection between $\lambda_{\mathrm{sBHAR}}$ and  $\lambda_{\mathrm{Edd}}$  will be discussed further in Section\,\ref{sec:sbhar-vs-edd}).

\subsection{The sBHAR distribution as a function of stellar mass}\label{sec:fupl}

\begin{figure}
\centering
\includegraphics[width=0.9\linewidth]{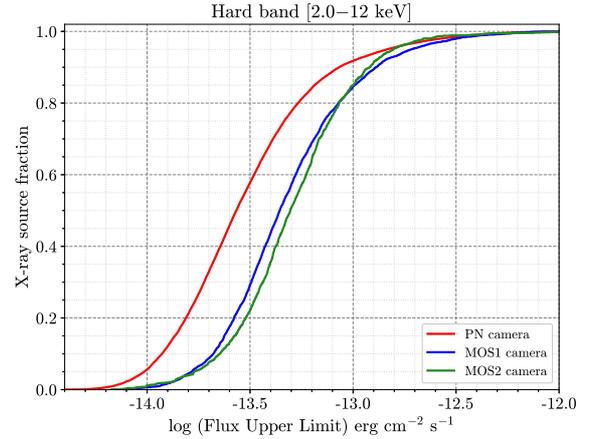}
\vspace*{-1.5ex}
\caption{The cumulative histogram of flux upper-limit in hard band [2.0--12\;keV] for XMM cameras from the XMM FLIX service.}
\label{fig:cumul-hard-flix}
\end{figure}

\begin{figure*}
\centering
\includegraphics[width=0.98\linewidth]{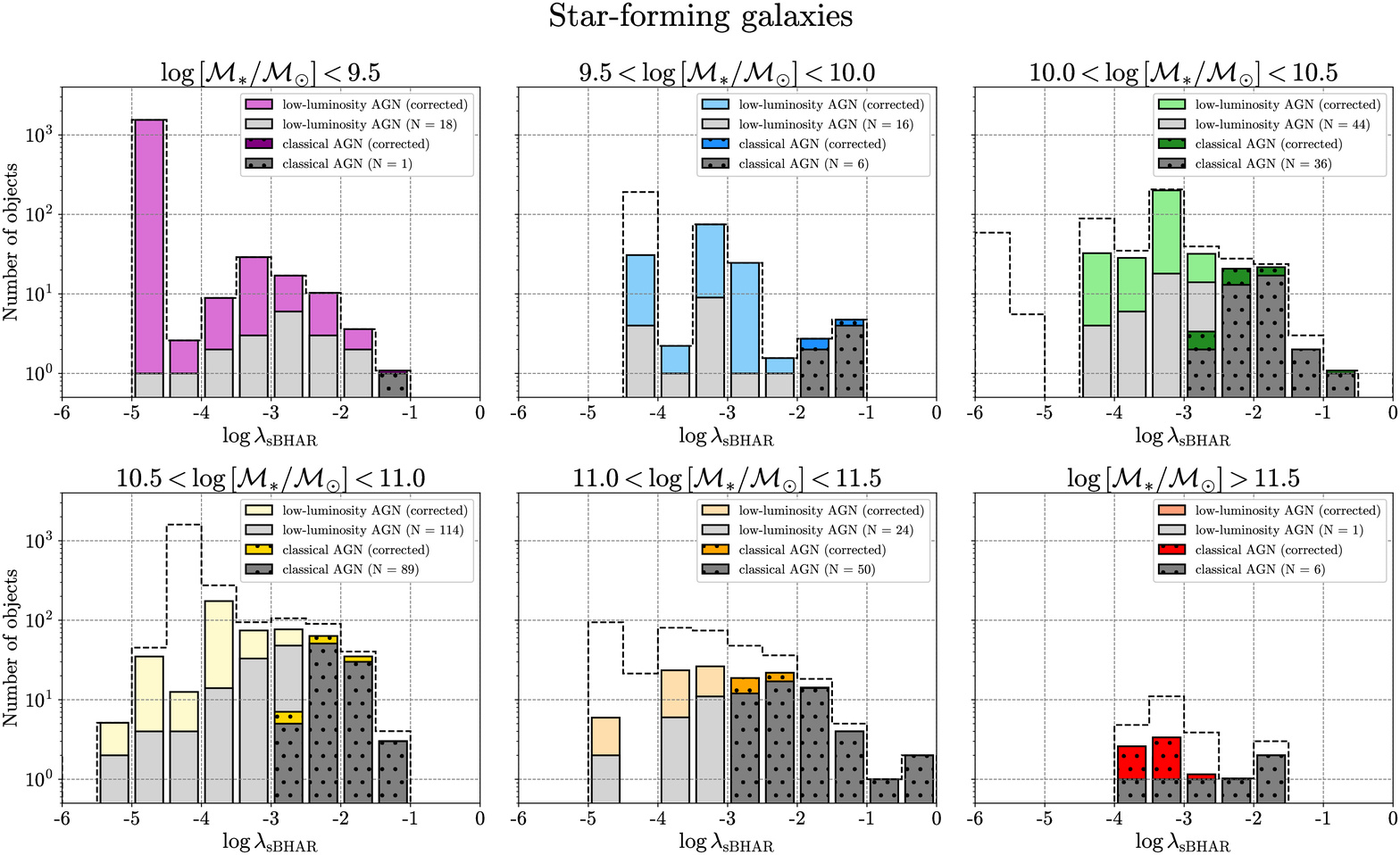}
\vspace*{-1.5ex}
\caption{The distribution of the specific Black Hole accretion rate (sBHAR) for AGN in \textit{star-forming galaxies} in six $\log\,[\mathcal{M}_{\ast}/\mathcal{M}_{\odot}]$ ranges. The `classical' AGN (Section\,\ref{sec:x-agn-select}) are shown by the dotted area and darker colour. The original data, before correcting for the 3XMM survey sensitivity, are shown by grey colours. The black dashed line represents the total distribution of sources in the X-ray AGN sample without SF/quiescent galaxy separation. }
\label{fig:histo-bhar-sfg}
\end{figure*}

\begin{figure*}
\centering
\includegraphics[width=0.98\linewidth]{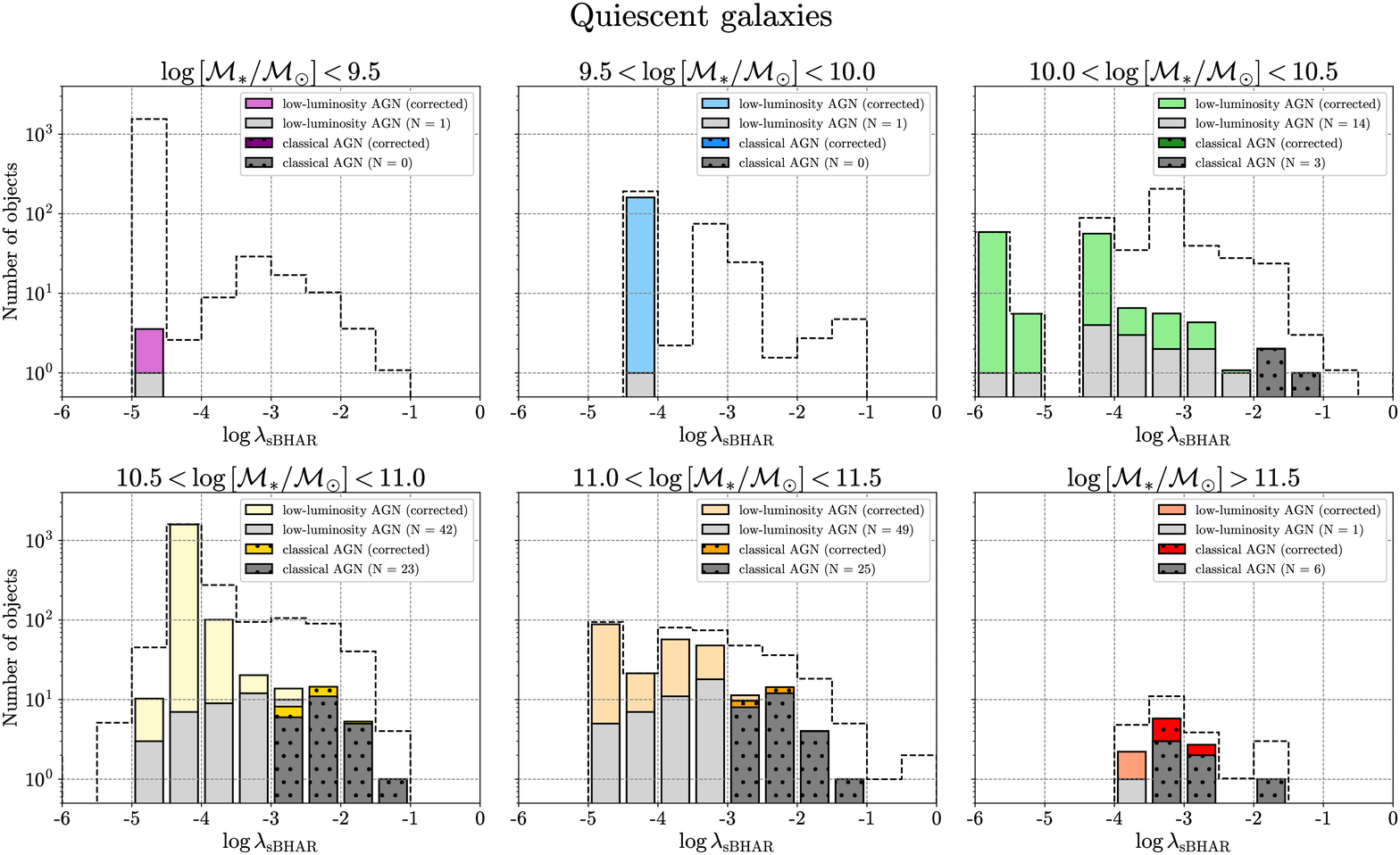}
\vspace*{-1.5ex}
\caption{The same distribution of sBHAR as Fig.\,\ref{fig:histo-bhar-sfg} for \textit{quiescent galaxies}.}
\label{fig:histo-bhar-pass}
\end{figure*}

To study the sBHAR in the local Universe we plot first the observed sBHAR distribution for star-forming and quiescent galaxies. As the sensitivity of the X-ray observations covering our SDSS galaxy sample varies across the sky (due to different exposure time, off-axis angle, detector) we have to correct our X-ray AGN sample for the fraction of missed sources as a function of flux. For this purpose, we used the count/flux upper limit service for XMM data, XMM FLIX\footnote{\url{https://www.ledas.ac.uk/flix/flix.html}}~\citep{Carrera:07}. The 3XMM-DR8 catalogue includes sources with EPIC detection likelihood \texttt{DET\_ML} > 6 in the full band (0.2--12\,keV); however a large number of these sources have a lower detection probability in the hard band (2.0--12\,keV) because of the AGN spectral shape coupled with the lower sensitivity of the \textit{XMM-Newton} detectors at higher energies. Since FLIX allows to calculate the flux upper limits only for the values of the likelihood detection higher than 6 ($3\sigma$), we were forced to reduce our final sample from 1628 objects to 570 objects choosing only those objects with detection likelihood \texttt{DET\_ML} > 6 in the hard band (grey area in Fig.\,\ref{fig:histo-bhar-sfg} and \ref{fig:histo-bhar-pass}). We then collected the values of the flux upper limit, which corresponds to a $3\sigma$ detection threshold of the 3XMM survey, at the position of each source in our optical sample falling in the 3XMM footprint and compiled the cumulative curves shown in Fig.\,\ref{fig:cumul-hard-flix} which describe the likelihood of detecting the X-ray counterpart of our galaxies at each flux level. The cumulative curves were applied as statistic weight to scale the number of objects in sBHAR distribution histograms in Fig.\,\ref{fig:histo-bhar-sfg} and \ref{fig:histo-bhar-pass} to correct for the variable sensitivity across the sky. 

The corrected $\lambda_{\mathrm{sBHAR}}$ distribution shows that `classical' AGN have higher accretion rates ($\log\,\lambda_{\mathrm{sBHAR}} \geq -3$) than the rest of the accreting SMBH population in all ranges of stellar masses. This result is in agreement with previous works \citep{Mullaney:12b, Chen:13, Mendez:13, Delvecchio:15, Stemo:20}. Such limit on sBHAR for X-ray selected `classical' AGN can be caused by the fact that the selection criteria based on the X-ray flux/luminosity were calibrated to detect moderate and high luminosity AGN, and do not work properly for low-luminosity AGN. 

The shape of the completeness-corrected $\lambda_{\mathrm{sBHAR}}$ distribution is approximately consistent with a power-law flattening at low accretion rates between $-3 \lesssim \log\lambda_{\mathrm{sBHAR}} \lesssim -2$ for all stellar mass ranges. Although star-forming galaxies show a slightly higher values of sBHAR peaking at $\log\lambda_{\mathrm{sBHAR}} \approx-3$ (Fig.\,\ref{fig:histo-bhar-sfg}) than the quiescent ones peaking at $\log\lambda_{\mathrm{sBHAR}} \approx -4$ (Fig.\,\ref{fig:histo-bhar-pass}). On the other hand, the $\lambda_{\mathrm{sBHAR}}$ distribution shows a lack of objects at high ($\log\,\lambda_{\mathrm{sBHAR}} > -2$) sBHAR due to the lack of bright SMBH accreting up to the Eddington limit usually found in AGN studies based on flux-limited surveys extending to high redshifts. In fact, as discussed before, our initial sample was selected from optical galaxies with an estimate of the intrinsic SFR derived from the optical spectrum, and thus, by definition, non-AGN dominated systems; furthermore we are limited to low-redshift sources and thus the presence of bright AGN and quasars is suppressed by the strong evolution of the AGN luminosity function with cosmic time. 

\subsection{The sBHAR and $L_{\mathrm{X}}$ correlation with stellar mass and SFR}\label{sec:bhar-sfr-m}

\begin{figure*}
\centering
\includegraphics[width=0.47\linewidth]{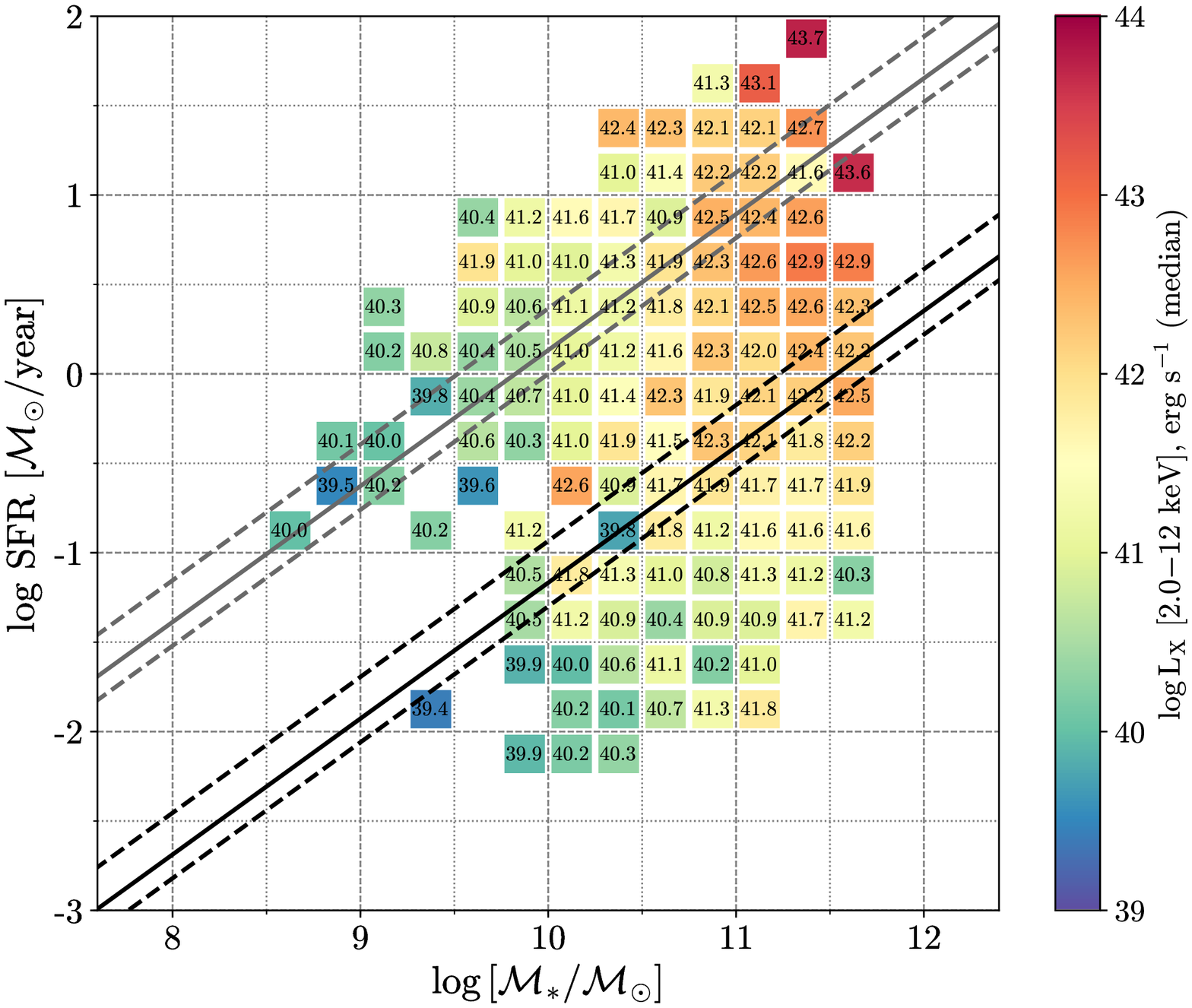}
\includegraphics[width=0.49\linewidth]{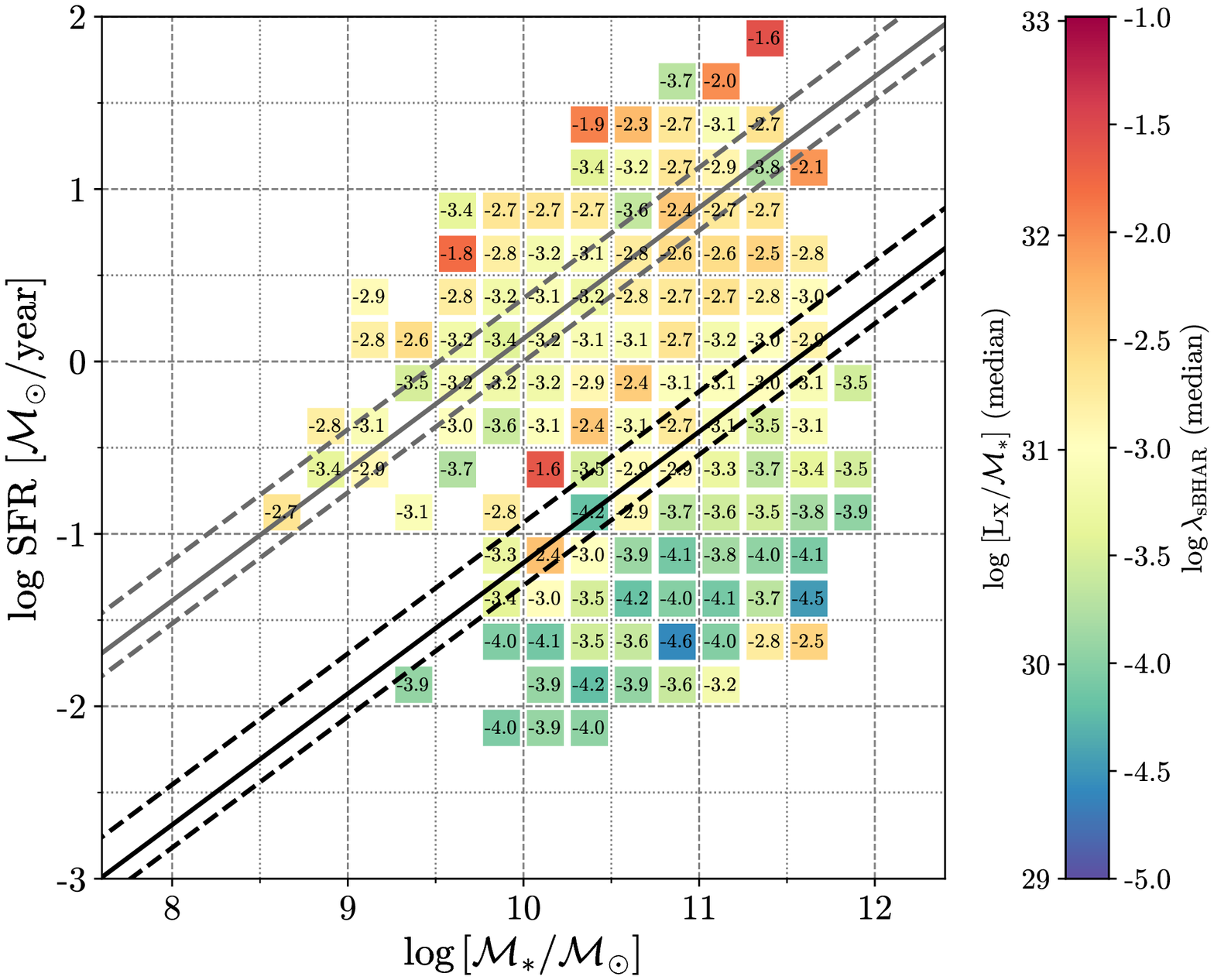}
\vspace*{-1.5ex}
\caption{The distribution of X-ray luminosity (\textit{left}) and the specific BH accretion rate $\lambda_{\mathrm{sBHAR}}$ (\textit{right}) on SFR--$\mathcal{M}_{\ast}$ plane. The actual median value of $\lambda_{\mathrm{sBHAR}}$ (X-ray luminosity) for each bin of SFR and $\mathcal{M}_{\ast}$ is written inside the square. The black and grey lines are the same as in Fig.\,\ref{fig:sfr-mass-all}. Number of points in both diagrams ranges from 60 in the central part to 2-3 in the edges.}
\label{fig:median-bhar-sfr-m}
\end{figure*}

To infer the dependence of AGN activity on galaxy properties, we divided our X-ray AGN sample in bins of SFR and $\mathcal{M}_{\ast}$ and calculated the median $\lambda_{\mathrm{sBHAR}}$ and $L_{\mathrm{X}}$ in each bin; we point out that negative values (after the correction for the host galaxy contamination) are included in the calculation to avoid biasing the result. The distribution of $\lambda_{\mathrm{sBHAR}}$ (and $L_{\mathrm{X}}$) on the SFR--$\mathcal{M}_{\ast}$ diagram is shown in Fig.\,\ref{fig:median-bhar-sfr-m}. The Figure shows that $L_{\mathrm{X}}$ increase with $\mathcal{M}_{\ast}$ for both star-forming and quiescent galaxies. To verify the statistical significance of this trend we applied the one-way \texttt{ANOVA} analysis for $L_{\mathrm{X}}$ values in six stellar mass bins for star-forming and quiescent galaxies separately. As a result we found a significant statistical difference in $L_{\mathrm{X}}$ between these mass bins for star-forming ($P$ value $= 0.001$) and quiescent galaxies ($P$ value $= 0.004$). This trend is consistent with the results of \citet{Mullaney:12b, Delvecchio:15, Stemo:20} according to which high-mass galaxies have the tendency to host AGN with much higher X-ray luminosity than less massive (dwarf) galaxies. 
At the same time the right panel of Fig.\,\ref{fig:median-bhar-sfr-m} shows that sBHAR for star-forming galaxies increases with stellar mass ($P$ value $= 0.008$), while quiescent galaxies with different stellar masses have on average the same values of sBHAR ($P$ value $= 0.351$). However, star-forming galaxies have systematically higher $\lambda_{\mathrm{sBHAR}}$ (and $L_{\mathrm{X}}$ respectively) at fixed $\mathcal{M}_{\ast}$ than quiescent ones; on average the median $\lambda_{\mathrm{sBHAR}}$ for SFGs varies from $-3.2$ to $-2.4$, while for quiescent galaxies from $-4.2$ to $-3.5$. The same result was found for optical, IR and X-ray selected samples in \citet{Rodighiero:15, Heinis:16}. 

\begin{figure*}
\centering
\includegraphics[width=0.98\linewidth]{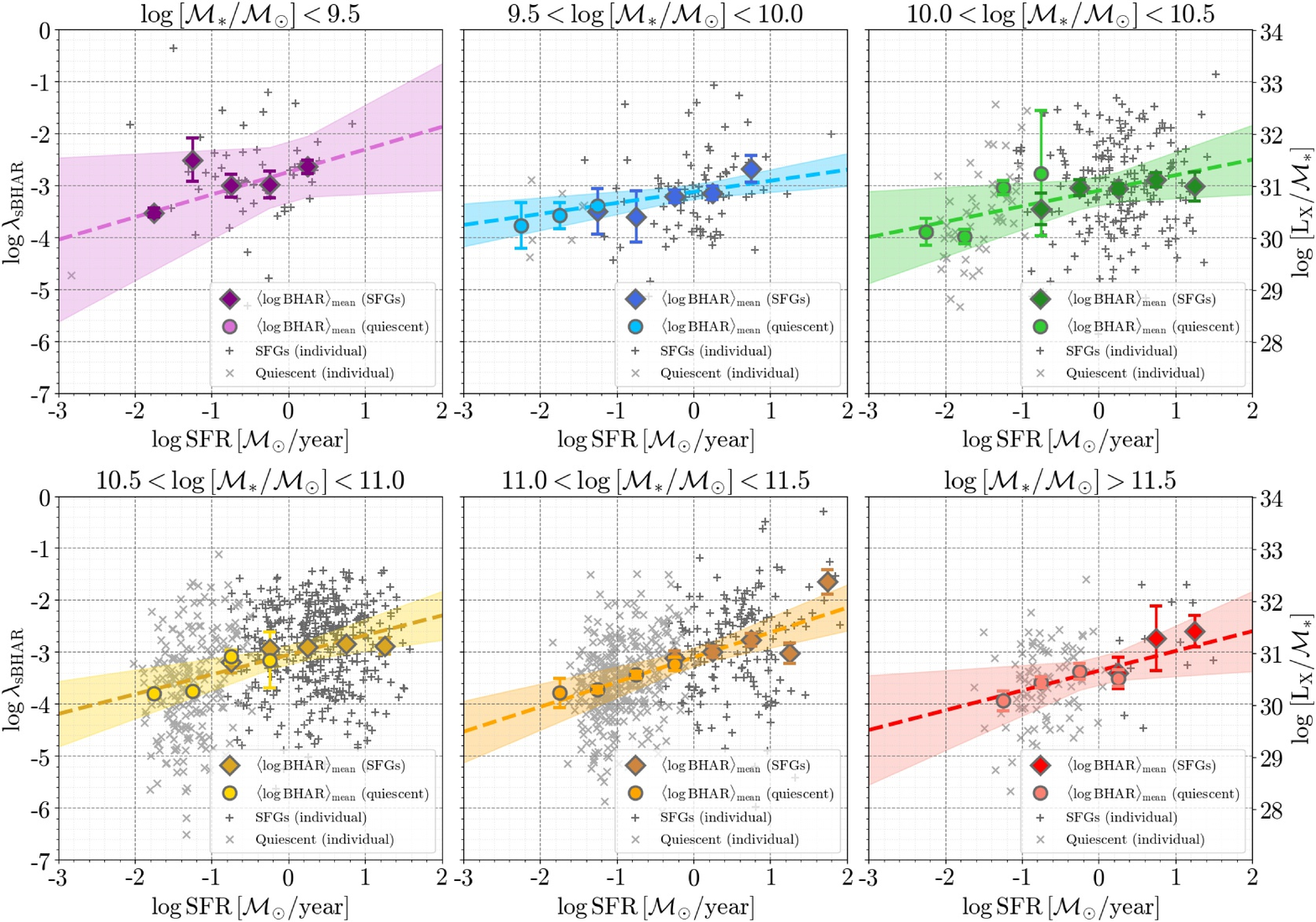}
\vspace*{-1.5ex}
\caption{The jackknife mean value of sBHAR vs SFR for star-forming (diamond) and quiescent galaxies (circles) for six stellar masses ranges. The individual objects from our X-ray AGN sample represented by grey crosses (SFGs) and pluses (quiescent). The errorbars were calculated as a variance of the jackknife mean. The dashed line shows the least-square linear best-fit with 95\% confidence interval. The best-fit and goodness-of-fit parameters are presented in Table\,\ref{tab:wls-fit}.}
\label{fig:bhar-sfr}
\end{figure*}

The correlation between $\langle\log\,\lambda_{\mathrm{sBHAR}}\rangle$ and $\log\,\mathrm{SFR}$ is presented separately for quiescent and star-forming galaxies in six stellar mass ranges in Figure\,\ref{fig:bhar-sfr}. For each $\mathcal{M}_{\ast}$ interval the mean $\langle\lambda_{\mathrm{sBHAR}}\rangle$ was calculated in 10 bins of SFR in the range $-3.0 < \log\,\mathrm{SFR} < 2.0$. The uncertainty of $\langle\lambda_{\mathrm{sBHAR}}\rangle$ was computed using jackknife resampling. The Figure confirms the lower level of accretion rate for quiescent galaxies in 5 stellar mass ranges, while for the lowest stellar masses, we cannot verify the existence of the same trend due to presence of only two quiescent galaxies in this mass range (see left top panel in Fig.\,\ref{fig:bhar-sfr}). To evaluate the statistical significance of the sBHAR-SFR correlation we applied a regression analysis and fitted our data using the least-squares approximation by linear function. The best-fit parameters are listed in Table\,\ref{tab:wls-fit}: in particular the $P$-values confirm that sBHAR is correlated with SFR at $>95$\% confidence ($P$-value $<0.05$) for all six stellar mass intervals. The best-fitting slope is close to the result found in \citet{Delvecchio:14} for the low redshift subsample, but is systematically flatter compared to high redshift samples in \citet{Chen:13, Delvecchio:15, Aird:19}. This would indicate that the $\langle\log\,\lambda_{\mathrm{sBHAR}}\rangle$--$\log\,\mathrm{SFR}$ relation is not linear and flatter at low SFR than at high SFR. On the other hand, the high-z studies do not sample well the low-SFR regime and thus a definitive conclusion is not straightforward.

\begin{figure}
\centering
\includegraphics[width=0.88\linewidth]{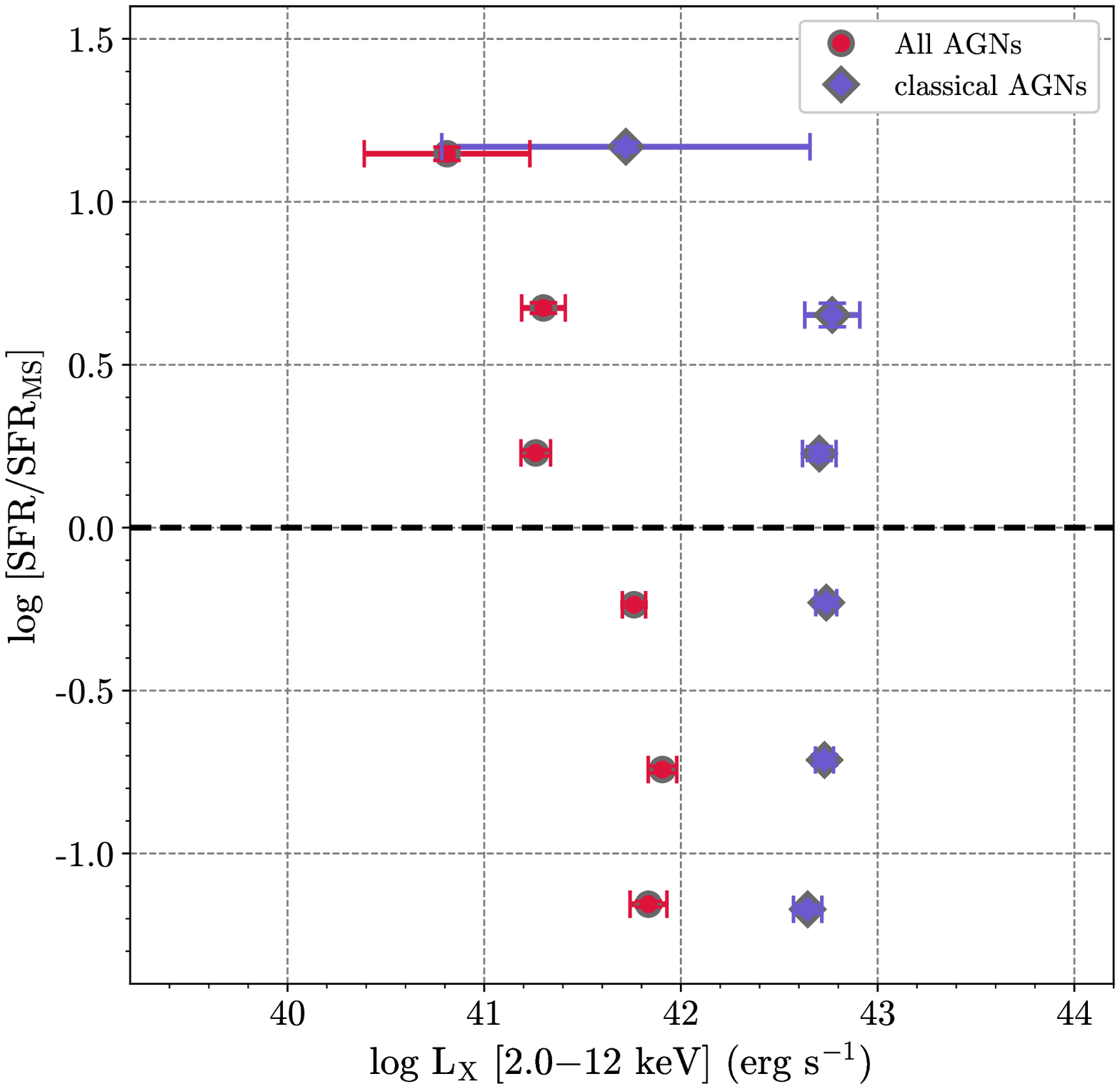}
\vspace*{-1.5ex}
\caption{The normalised SFR as a function of the mean $L_{\mathrm{X}}$ obtained by the jackknife resampling in $\log\,[\mathrm{SFG}/\mathrm{SFR}_{\mathrm{MS}}]$ bins for SFGs in X-ray AGN sample (circles) and for `classical' AGN selected by X-ray criteria in Section\,\ref{sec:x-agn-select} (diamonds). The errorbars were calculated as a variance of the jackknife mean. The dashed line is the position of the main sequence of star-forming galaxies (see definition in the text).}
\label{fig:ms-offset}
\vfill
\centering
\includegraphics[width=0.88\linewidth]{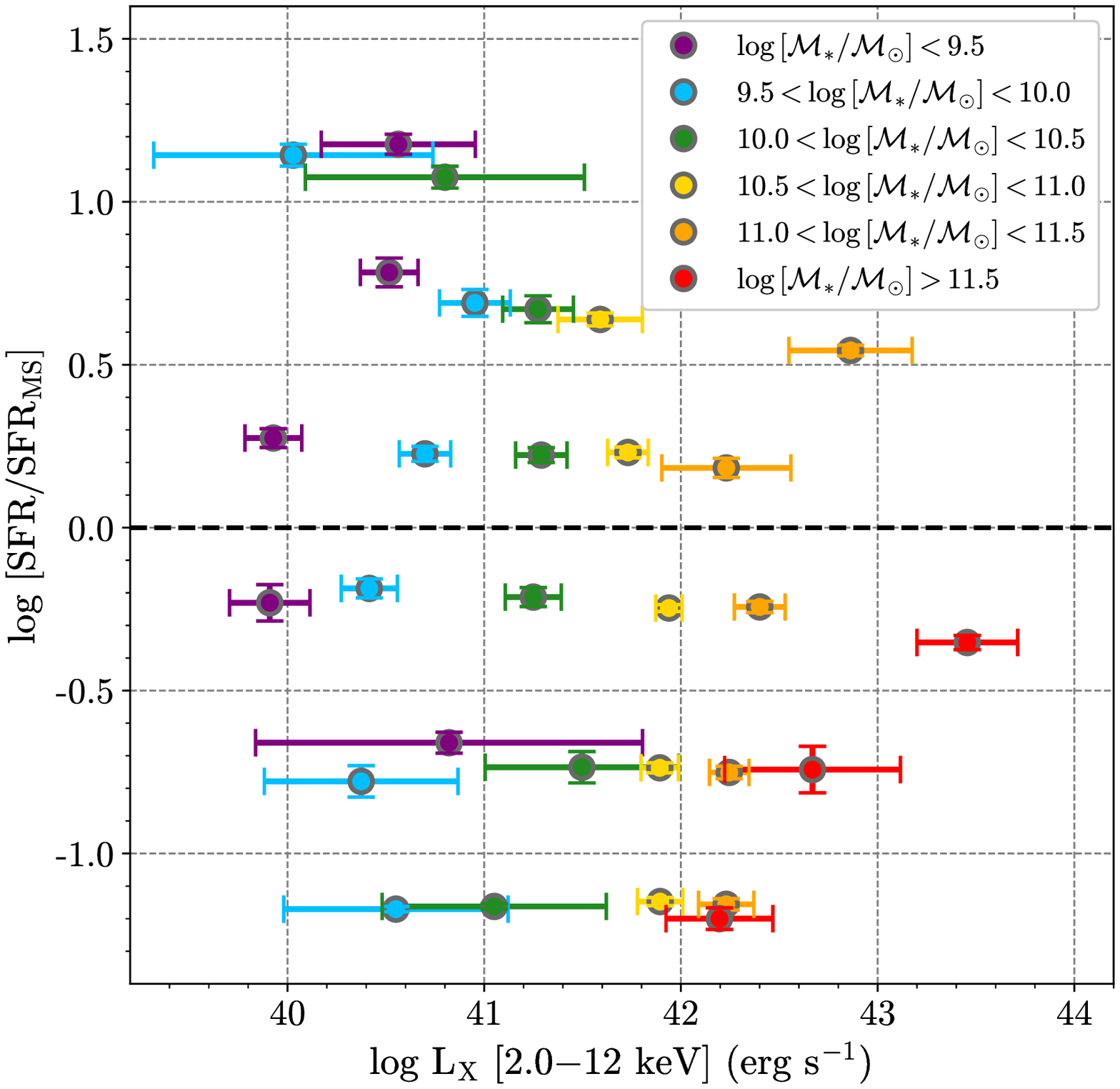}
\vspace*{-1.5ex}
\caption{The normalised SFR as a function of the mean $L_{\mathrm{X}}$ obtained by the jackknife resampling in $\log\,[\mathrm{SFG}/\mathrm{SFR}_{\mathrm{MS}}]$ bins for SFGs in X-ray AGN sample for six stellar mass bins. The errorbars were calculated in the same way as for Fig.\,\ref{fig:ms-offset}.}
\label{fig:ms-offset-mass}
\end{figure}

A number of studies suggest the existence of a connection between X-ray luminosity for star-forming galaxies and their location related to main sequence.  For instance, \citet{Masoura:18} found that $L_{\mathrm{X}}$ increases with SFR for galaxies below the MS and decrease with SFR above MS, and suggested that this trend can be explained by the enhancing/quenching of star-formation processes by AGN depending on the position of the host galaxy in relation to MS. To verify the existence of this effect in our X-ray AGN sample, we calculated the normalised SFR as the ratio of the SFR of each galaxy to the SFR of the main-sequence at the same mass. For each interval of the normalised SFR the mean $L_{\mathrm{X}}$ and its uncertainty were calculated using jackknife resampling in 6 bins of $\log\,[\mathrm{SFR}/\mathrm{SFR}_{\mathrm{MS}}]$ and presented in Fig.\,\ref{fig:ms-offset}.
At first sight it looks like the effect is confirmed, as we observe an average lower $L_{\mathrm{X}}$ at larger (normalized) SFR that is qualitatively consistent with the result obtained by \citet{Masoura:18}. However, analysing the $L_{\mathrm{X}}$--SFR relation separately for six stellar mass bins (Fig.\,\ref{fig:ms-offset-mass}) it seems that this effect is mainly due to the fact that at lower $\mathcal{M}_{\ast}$ we have only SFGs with typically lower $L_{\mathrm{X}}$ than at higher stellar mass, and thus could be driven due to the incompleteness effects. Furthermore no evidence of this effect is observed for `classical' AGN,  (see Fig.\,\ref{fig:ms-offset}) in agreement with the studies of \citet{Rovilos:12, Shimizu:15} for low-redshift samples.
 
\begin{table*}
 \caption{The best-fit parameters obtained from a linear relation between $\langle\log\,\lambda_{\mathrm{sBHAR}}\rangle$ and $\log\,\mathrm{SFR}$ for six stellar mass ranges (see Fig.\,\ref{fig:bhar-sfr}). The slope, intercept with their standard errors and all statistics parameters ($F$-statistic, $P$ value and $R^{2}$) were found from a least-square linear regression. In this work, we consider the confident level as $P$-value $<0.05$. $N$ is the number of points in each stellar mass bin.}
 \label{tab:wls-fit}
 \begin{tabular}{cccccccc}
  \hline
\# & Stellar mass range & slope & intercept & \textit{F}-statistic & $P$ value (\textit{F}-stat) & $R^{2}$ & $N$ \\
\hline
1 & $\log\,[\mathcal{M}_{\ast}/\mathcal{M}_{\odot}] < 9.5$            & $0.43\pm0.08$ & $-2.75\pm0.12$ & 26.66 & 0.0141 & 0.899 & 5 \\[1pt] 
2 & $9.5 < \log\,[\mathcal{M}_{\ast}/\mathcal{M}_{\odot}] < 10.0$     & $0.21\pm0.04$ & $-3.12\pm0.05$ & 31.56 & 0.0014 & 0.840 & 8 \\[1pt] 
3 & $10.0 < \log\,[\mathcal{M}_{\ast}/\mathcal{M}_{\odot}] < 10.5$    & $0.30\pm0.09$ & $-3.11\pm0.09$ & 11.09 & 0.0126 & 0.613 & 9 \\[1pt] 
4 & $10.5 < \log\,[\mathcal{M}_{\ast}/\mathcal{M}_{\odot}] < 11.0$    & $0.38\pm0.07$ & $-3.06\pm0.06$ & 30.90 & 0.0009 & 0.815 & 9 \\[1pt] 
5 & $11.0 < \log\,[\mathcal{M}_{\ast}/\mathcal{M}_{\odot}] < 11.5$    & $0.48\pm0.06$ & $-3.10\pm0.06$ & 56.49 & 0.0001 & 0.890 & 9 \\[1pt] 
6 & $\log\,[\mathcal{M}_{\ast}/\mathcal{M}_{\odot}] > 11.5$           & $0.38\pm0.10$ & $-3.36\pm0.07$ & 13.92 & 0.0136 & 0.736 & 7 \\[1pt] 
  \hline
 \end{tabular}
\end{table*}

\section{Sources of uncertainty in the specific Black Hole accretion rate determination}\label{sec: reliability}

The results discussed in the previous Section are affected by various factors and assumptions. To evaluate these effects on sBHAR determination and SFR--sBHAR relation we made several tests which are presented below.

\subsection{The reliability of SFR and $\mathcal{M}_{\ast}$}\label{sec:reliability-sfr-mass}

In order to test the reliability of our results we investigated the accuracy of our optical star-formation rates with those derived using different measurement methods. An analysis of the accuracy of SFR measurements based on H${\alpha}$ and D4000 from the \textit{galSpec} catalogue are presented by \citet{Popesso:19}. According to their work the values of SFR from H${\alpha}$ in \textit{galSpec} are in the good agreement with SFR obtained from far-IR luminosity from \textit{Wide-field Infrared Survey Explorer (WISE)} \citep{Salim:16} and \textit{Herschel} \citep{Elbaz:11}. However, the D4000-based SFR (which, we recall, was used by \citealt{Brinchmann:04} when line-based measurements were not possible, i.e. mainly for quiescent galaxies) reveals a systematic underestimate with respect to $L_{\mathrm{IR}}$-based SFR and an overestimate at low SFRs ($< 0.01 \mathcal{M}_{\odot}$yr$^{-1}$) compared to SED-fitting measurements \citep{Salim:16}. A correction to D4000-based SFR in \textit{galSpec} was proposed by \citet{Oemler:17} as a calibration parameter derived from UV+IR estimates with a correction based on the galaxy inclination and the NUV--g rest-frame colour (GALEX--SDSS filters). The combination of UV and IR bands was chosen in order to reproduce the fraction of galaxy radiation lost in optical band, e.g. the ionising UV emission from hot stars and mid-IR due to radiation absorbed by dust.

\begin{figure*}
\centering
\includegraphics[width=0.95\linewidth]{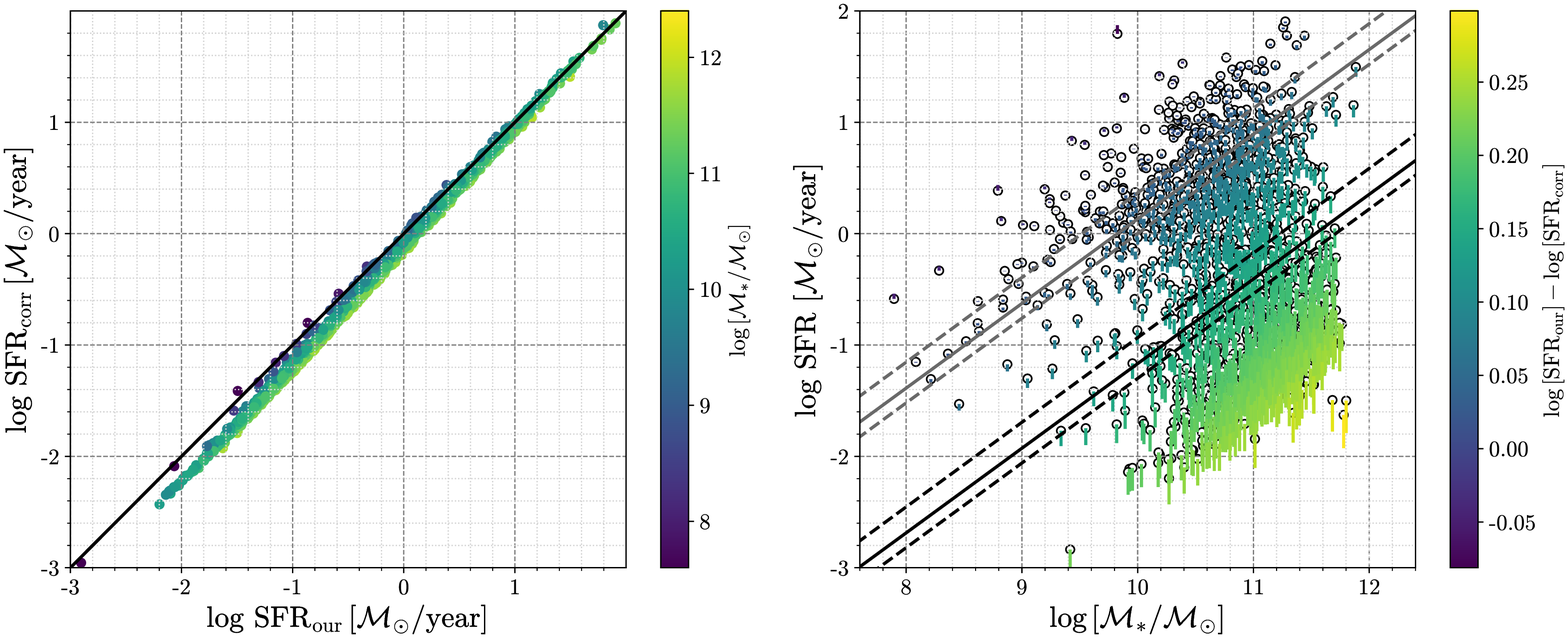}
\vspace*{-1.5ex}
\caption{\textit{Left:} The ratio of SFR for our X-ray AGN sample calculated by \citet{Brinchmann:04} and SFR corrected according to the criteria proposed by \citet{Oemler:17}. The colour shows the $\log\,[\mathcal{M}_{\ast}/\mathcal{M}_{\odot}]$ range in our sample. The 1:1 line is represented by black solid line. \textit{Right:} The distribution of star-formation rate vs. stellar mass for our X-ray AGN sample (black circles). The value of SFR correction is represented by colour. The black and grey lines are the same as in Fig.\,\ref{fig:sfr-mass-all}.}
\label{fig:sfr-corr}
\centering
\includegraphics[width=0.96\linewidth]{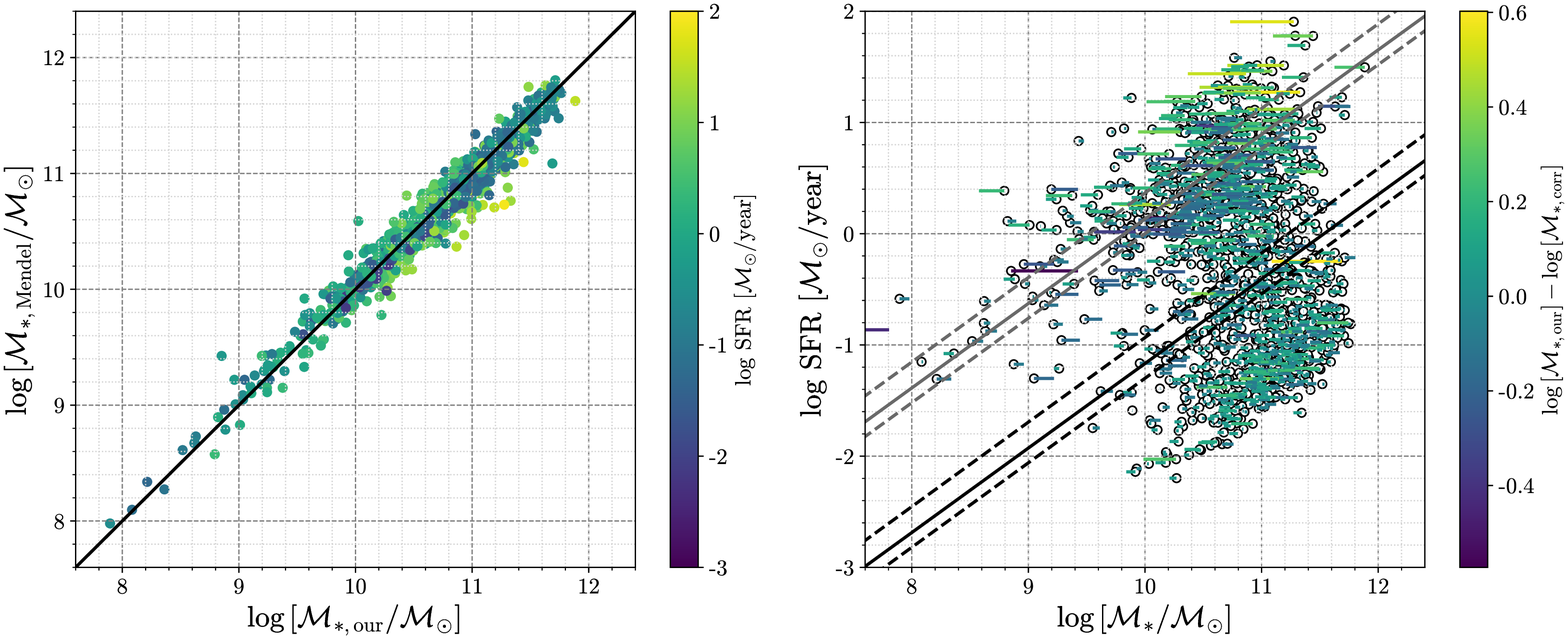}
\vspace*{-1.5ex}
\caption{\textit{Left:} The stellar mass for our X-ray AGN sample calculated by \citet{Brinchmann:04} vs. the stellar mass computed by \citet{Mendel:14}. The colour shows the distribution of $\log\,\mathrm{SFR}$ in our sample. The 1:1 line is represented by black solid line. \textit{Right:} The distribution of star-formation rate vs. stellar mass for our X-ray AGN sample (black circles). The value of $\log\,[\mathcal{M}_{\ast,\mathrm{our}}]-\log\,[\mathcal{M}_{\ast,\mathrm{Mendel}}]$ is represented by colour. The black and grey lines are the same as in Fig.\,\ref{fig:sfr-mass-all}.}
\label{fig:mass-corr}
\end{figure*}

We calculated a corrected estimate of SFR for our X-ray AGN sample using the relation $\log(\mathrm{sSFR})_{\mathrm{corr}} = 1.07\cdot\log(\mathrm{sSFR})_{\mathrm{our}}+0.64$ derived by \citet{Oemler:17}, where sSFR is the so-called specific star-formation rate defined as $\log(\mathrm{sSFR}) = \log(\mathrm{SFR}) - \log{\mathcal{M}_{\ast}}$. 
A comparison of the uncorrected and corrected SFR for our X-ray AGN sample is presented in Fig.\,\ref{fig:sfr-corr} (left panel) and the effect of the SFR correction on the SFR--$\mathcal{M}_{\ast}$ diagram is shown in the right panel of Fig.\,\ref{fig:sfr-corr}. It is clear that the SFR correction becomes significant only at the lowest SFRs, i.e. for quiescent galaxies with SFR$< 0.01 \mathcal{M}_{\odot}$yr$^{-1}$ (with a maximum difference of the $(\log(\mathrm{SFR})_{\mathrm{uncorr}}-\log(\mathrm{SFR})_{\mathrm{corr}}) \sim 0.3$) as was claimed also by \citet{Oemler:17,Popesso:19}. In our work, we used the SFR in two cases: as tracer of the X-ray emission from X-ray binaries in star-forming galaxies (see Section\,\ref{sec:corr_sfg_etg}) and in evaluating the correlation between SFR and specific Black Hole accretion rate (sBHAR, Section\,\ref{sec:bhar-sfr-m}). The maximum SFR correction $\Delta\mathrm{SFR} = 0.15$ derived for SFGs in our X-ray AGN sample does not change significantly the X-ray luminosity correction in Section\,\ref{sec:corr_sfg_etg} ($\Delta L_{\mathrm{XBs}} =  1.41$\,erg\,s$^{-1}$]) and therefore it does not affect significantly the sBHAR calculation. The SFR-$\lambda_{\mathrm{sBHAR}}$ can change due to the SFR correction for the most massive quiescent galaxies ($\mathcal{M}_{\ast} > 10^{11.5} \mathcal{M}_{\odot}$), but we have a small fraction of such object in our sample and therefore it does not change our final results either. 

To evaluate the accuracy of the stellar masses derived by \citet{Brinchmann:04} we used the catalogue of bulge, disk and total stellar masses for SDSS DR7 from \citet{Mendel:14}. These masses were estimated from SED fitting in \textit{u,g,r,i,z} SDSS bands similar to the mass estimate described in \citet{Kauffmann:03a}, but with significant differences in the stellar population synthesis (SPS) model grid. In fact \citet{Mendel:14} excluded bursty star-forming histories from their SPS model grid considering their rarity in the local Universe (<10\%) and the difficulty to identify them only by photometric data. In total we thus found counterparts for 1362 galaxies (83.3\% of our X-ray AGN sample) in the \citet{Mendel:14} catalogue.

\begin{figure}
\centering
\includegraphics[width=0.92\linewidth]{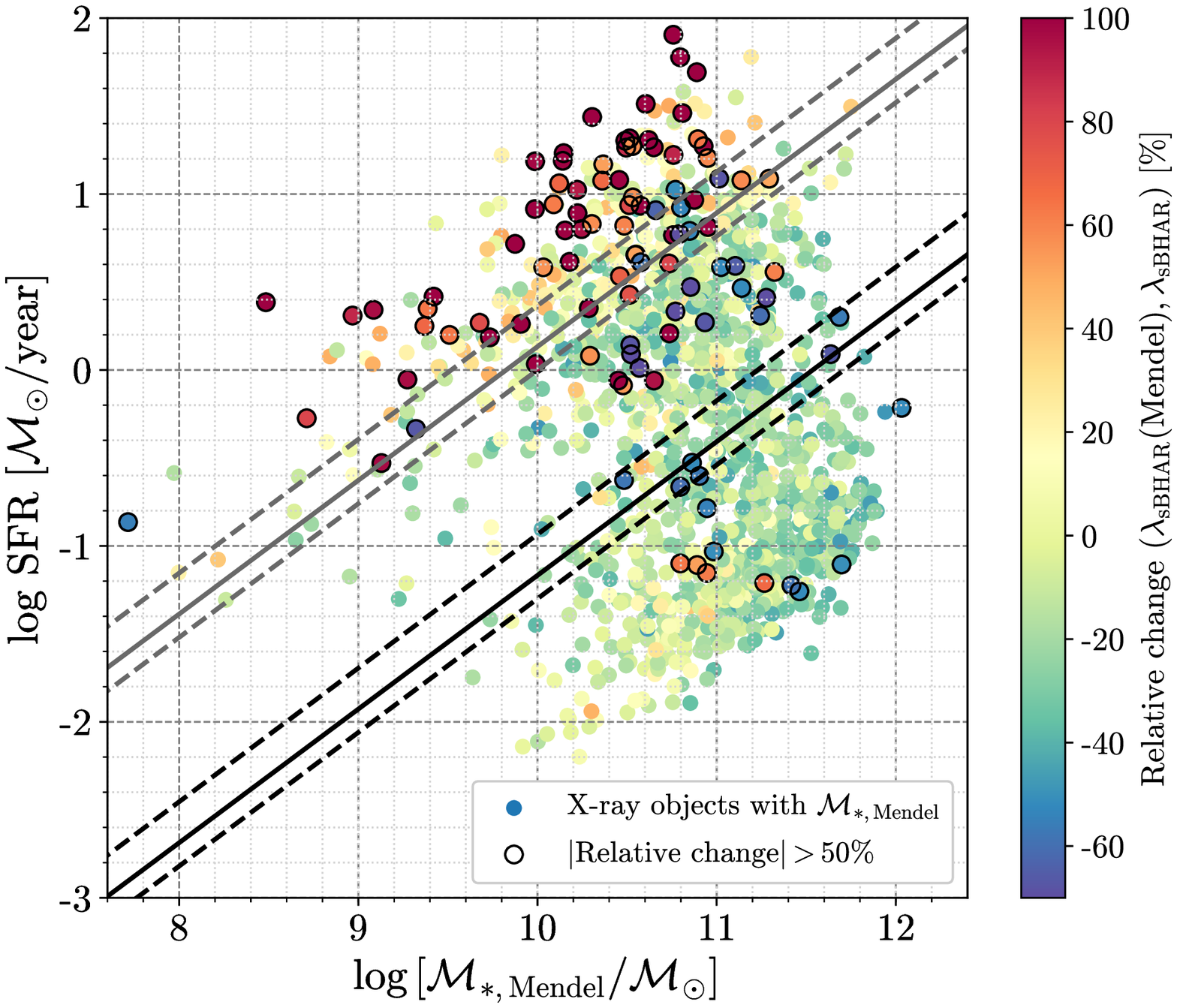}
\vspace*{-1.5ex}
\caption{The distribution of star-formation rate vs. stellar mass from Mendel catalogue for 1362 X-ray objects. The colour gradient shows the relative percentage change between sBHAR calculated on the basis of $\mathcal{M}_{\ast}$ from SDSS sample and $\mathcal{M}_{\ast}$ from Mendel catalogue. Black circles represent the objects with the absolute value of relative change more than 50\%.}
\label{fig:sfr-mass-mendel-rc}
\end{figure}

The  comparison between the stellar masses used in our work and those derived by \citet{Mendel:14} in the left panel of Fig.\,\ref{fig:mass-corr} shows that $\mathcal{M}_{\ast}$ and $\mathcal{M}_{\ast,\mathrm{Mendel}}$ are generally consistent for all objects in our X-ray AGN sample. However, galaxies with extreme SFR tend to have systematically lower values of $\mathcal{M}_{\ast,\mathrm{Mendel}}$ stellar masses compared to those used in this paper. The same difference is also visible in the SFR--$\mathcal{M}_{\ast}$ diagram (right panel of Fig.\,\ref{fig:mass-corr}) and it is most likely caused by the exclusion of bursty star-forming galaxies from SPS model grid used by \citet{Mendel:14}. 
The accuracy of stellar mass determination affects the X-ray correction for SFGs (see Eq.\,\eqref{eq:sfg-lehmer} in Section\,\ref{sec:corr_sfg_etg}) and sBHAR determination (Section\,\ref{sec:bhar}) as $\lambda_{\mathrm{sBHAR}} \propto \mathcal{M}_{\ast}^{-1}$. We calculated $\lambda_{\mathrm{sBHAR}}$ using the stellar mass from \citet{Mendel:14} catalogue and compared the obtained sBHAR with that one obtained in Section\,\ref{sec:bhar}. To evaluate the change of sBHAR calculated on the basis of different stellar mass we calculated the relative change in percentage unit. The sBHAR relative change distribution on SFR--$\mathcal{M}_{\ast}$ plane in Fig.\,\ref{fig:sfr-mass-mendel-rc} shows only 105 objects significantly changed sBHAR and their absolute values of relative change are more than 50\%. At the same time, these objects are located predominantly above MS of SFGs where it was discussed above such high number of changes can be caused by the underestimation of stellar mass for starburst galaxies in \citet{Mendel:14} catalogue.

According to all findings discussed in this section we decided not to use any correction for SFR and stellar mass as the existent uncertainties of SFR and stellar mass have insignificant effect on our final results.

\subsection{The extended X-ray sources and their effect on the sBHAR}\label{sec:extended}

\begin{figure}
\centering
\includegraphics[width=0.94\linewidth]{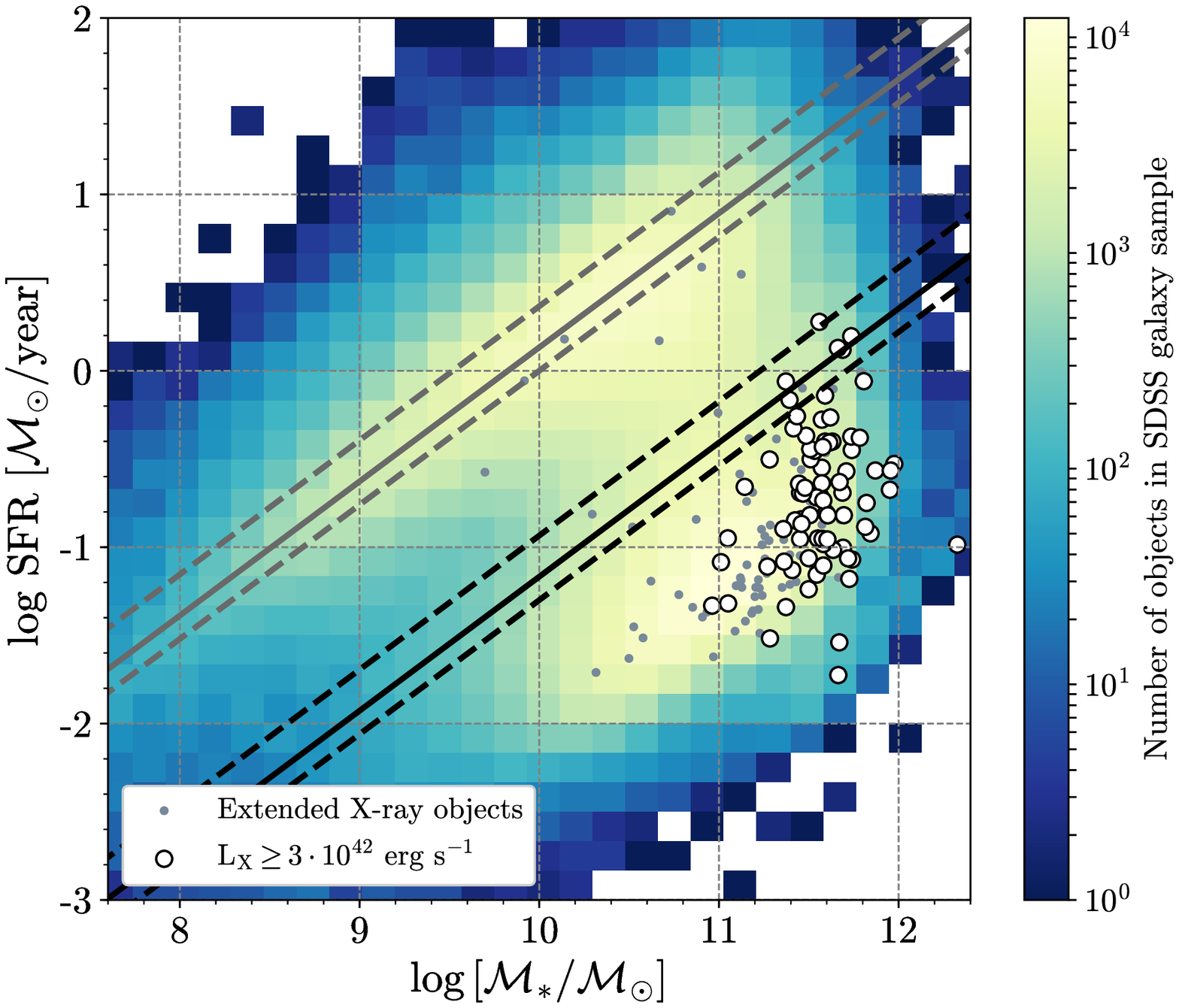}
\vspace*{-1.5ex}
\caption{The distribution of star-formation rate vs. stellar mass for extended X-ray sources (grey circles). The blue to yellow colorscale reflects the 2D density distribution of galaxies in our optical SDSS sample. Black circles represent the AGN selected by the X-ray luminosity criterion $L_{\mathrm{X}} \geq 3 \cdot 10^{42}$\;erg\;s$^{-1}$. The black and grey lines are the same as in Fig.\,\ref{fig:sfr-mass-all}.}
\label{fig:ext-sfr-mass}
\end{figure}

In Section\,\ref{sec:xmm-data} we selected the X-ray sources with zero extension parameter to avoid spatially extended objects, i.e. individual galaxies whose X-ray emission may be dominated by hot gas and LMXBs or which are part of massive galaxy clusters. As our primary SDSS sample contains source at low redshift there is a probability to have a small fraction of nearby objects with resolved X-ray cores (i.e non-zero extension) which harbour a faint accreting SMBH. The rejection of such objects affects the completeness of our X-ray AGN sample and may lead to an underestimation of the median sBHAR. To evaluate this effect we expanded the analysis to extended sources as well. We selected 159 extended objects following the same step as for non-extended sources (see Section\,\ref{sec:xmm-data}). The distribution of extended sources on the SFR--$\mathcal{M}_{\ast}$ diagram (Fig.\,\ref{fig:ext-sfr-mass}) shows that nearly 85\% of such objects are located in massive quiescent galaxies ($\geq 10^{11}\,\mathcal{M}_{\odot}$) and 45\% of these sources have $L_{\mathrm{X,int}} \geq 3 \cdot 10^{42}$\;erg\;s$^{-1}$. The visual inspection of SDSS images and spectra shows that 50\% of extended sources are represented by isolated elliptical galaxies, while 45\% are located in galaxy clusters and probably are central dominant (cD) galaxies. Some galaxies also have a nearby galaxy or a point-like source (likely an AGN/quasar) which can contribute to the detected X-ray emission. The large X-ray luminosity ($L_{\mathrm{X,int}} \geq 3 \cdot 10^{42}$\;erg\;s$^{-1}$) of some extended sources could suggest the presence of an AGN contributing to the total X-ray emission. However, according to the optical images half of these objects are located in galaxy clusters and could thus also be explained by emission from the hot intra-cluster medium. Only 8 sources in our extended sample are located in the star-forming galaxy region (Fig.\,\ref{fig:ext-sfr-mass}) they are all low-redshift ($z < 0.05$) spiral galaxies with large angular diameter. 

\begin{figure}
\centering
\includegraphics[width=0.96\linewidth]{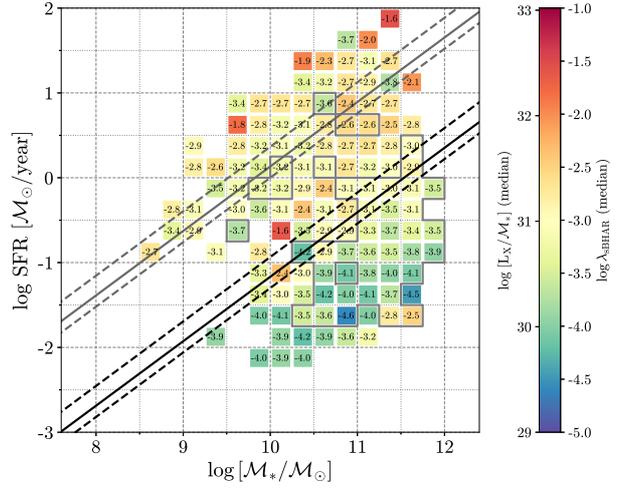}
\vspace*{-1.5ex}
\caption{The distribution of the specific BH accretion rate ($\lambda_{\mathrm{sBHAR}}$) for the combined sample of extended and non-extended X-ray sources. The SFR--$\mathcal{M}_{\ast}$ bins with some contribution from extended sources are encompassed by a grey solid line area. The black and grey lines are the same as in Fig.\,\ref{fig:sfr-mass-all}.}
\label{fig:ext-point-median-bhar}
\end{figure}

After applying the X-ray luminosity correction according to the criteria described in Section\,\ref{sec:corr_sfg_etg} we calculated the sBHAR for extended objects using Eq.\,\eqref{eq:spec-bhar} and combined them with our X-ray AGN sample. The distribution of the median $\lambda_{\mathrm{sBHAR}}$ on the SFR--$\mathcal{M}_{\ast}$ plane for the combined (extended and point-like) dataset (see Fig.\,\ref{fig:ext-point-median-bhar}) shows that the inclusion of extended sources to our X-ray AGN sample leads to an increase of $\lambda_{\mathrm{sBHAR}}$ for massive/quiescent galaxies compared to the result presented in  Fig.\,\ref{fig:median-bhar-sfr-m} (right panel). However the overall increase not significant, and varies from 1.2\% for $\log\,[\mathcal{M}_{\ast}/\mathcal{M}_{\odot}] = {11.0}$ to 2.1\% at $\log\,[\mathcal{M}_{\ast}/\mathcal{M}_{\odot}] = {11.5}$.

In conclusion, we decided to exclude extended objects from our main study to avoid overestimating the average accretion rate since for such objects the correction to the total X-ray luminosity due to the hot gas contribution (see Section\,\ref{sec:x-agn-select}) is likely underestimated.

\section{The specific Black Hole accretion rate versus Eddington ratio}\label{sec:sbhar-vs-edd}

The masses of SMBH in the local Universe can be measured with high accuracy using direct methods based on the kinematics of gas and stars around SMBH \citep{Shen:11,Kormendy:13}. However, these direct techniques require the long-time observation of individual galaxies thus making such mass determination for large number of objects a challenging effort. Also, the accuracy of such method depends on the spectral resolution, on the orientation and geometry of broad-line region, on obscuration etc \citep{Merloni:10, Shen:11, Reines:15}. To estimate the SMBH masses for larger samples and over wide redshift ranges the usual approach consists in using indirect methods where the SMBH mass is inferred from observable host-galaxy properties that correlate with the Black Hole mass (i.e scaling relation). For instance, several studies showed that the mass of supermassive Black Hole correlates with the velocity dispersion of stars in the galaxy bulge \citep{Ferrarese:00, Gultekin:09}, the bulge luminosity and mass \citep{Haring:04, Kormendy:13, McConnel:13} and the total stellar mass of the host galaxy \citep{Reines:15, Shankar:17, Shankar:20}. 

The knowledge of the host galaxy parameters allows us to use scaling relations as an indirect method for BH mass determination, but the accuracy of SMBH mass derived by such method is dependent on the uncertainties and biases of the underlying scaling relations. In fact the assumption that SMBH mass correlates with the total stellar mass of the host galaxies is the limiting factor in the use of the sBHAR as a tracer of the Eddington ratio in AGNs. \citet{Reines:15} showed that the BH-to-stellar mass relation for nearby galaxies varies greatly depending on morphological type and AGN activity. Local AGN host-galaxies with SMBH masses measured by reverberation mapping or virial methods show significantly lower values than quiescent galaxies with masses determined by dynamical methods. It was suggested that such difference in BH-to-stellar mass relation is caused by the host galaxy properties. For instance, local AGN tend to be located in late-type spiral galaxies with pseudobulges, while the early-type elliptical galaxies (i.e. spheroids or classical bulges) mainly host inactive BHs and have a tendency to follow the canonical BH-to-bulge mass relation \citep{Haring:04, Kormendy:13, McConnel:13}. This result was also confirmed by \citet{Shankar:17, Shankar:20} based on the sample of early and late-type galaxies with dynamical BH mass estimates derived by \citet{Savorgnan:16}. Furthermore, \citet{Bernardi:07, Shankar:16} showed that the BH-to-stellar mass relation can be systematically biased by the spatial resolution limits of current instruments or other observational effects. Since the gravitational sphere of influence of relatively less massive SMBH ($\mathcal{M}_{\mathrm{BH}} \lesssim 10^{7} \mathcal{M}_{\odot}$) cannot be resolved, the dynamical estimation of the SMBH mass becomes inaccurate and distorts the observed scaling relations. To reproduce the unbiased scaling relations \citet{Shankar:16} performed several Monte Carlo tests and found that the normalisation of the intrinsic (unbiased) BH-to-stellar mass scaling relation is lower by a factor $\sim50-100$ at small stellar masses ($\mathcal{M}_{\ast} \sim 10^{10}-10^{10.5} \mathcal{M}_{\odot}$) than the dynamical $\mathcal{M}_{\mathrm{BH}}$ estimates.
Moreover, this result is consistent with the SMBH masses for local AGN found by \citet{Reines:15} and hence, it can reconcile the observed difference between SMBH masses for local AGN host-galaxies and early-type galaxies. 

\begin{figure}
\centering
\includegraphics[width=0.98\linewidth]{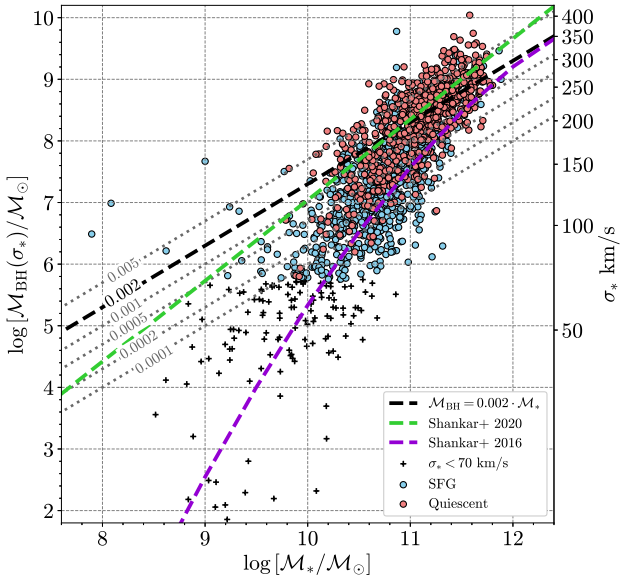}
\vspace*{-1.5ex}
\caption{The BH-to-stellar mass diagram for star-forming and quiescent galaxies in our X-ray AGN sample. The BH mass is calculated using the scaling relation between BH mass and stellar velocity dispersion derived by \citet{McConnel:13}. The objects with $<70$\,km/s are marked by the black crosses. The black dashed line show the mass scaling relation $\mathcal{M}_{\mathrm{BH}} = 0.002 \cdot \mathcal{M}_{\ast}$ used in $\lambda_{\mathrm{sBHAR}}$ determination in Section\,\ref{sec:bhar}. The grey dotted lines are correspond to the BH-to-stellar mass relation with a scaling coefficient in range from $0.0001$ to $0.005$. The green dashed line corresponds to the BH-to-stellar scaling relation from \citet{Shankar:20}, which is obtained from the sample with reliable dynamically-measured $\mathcal{M}_{\mathrm BH}$ in the centre of local quiescent galaxies. The violet dashed line is represent the intrinsic or unbiased BH-to-stellar mass correlation from \citet{Shankar:16} obtained by a series of the Monte Carlo simulations to correct the bias effect due to the observation limits.  }
\label{fig:masses-bh-stellar}
\end{figure}

To investigate the link between the sBHAR and the Eddington ratio in our sample, 
we estimated the SMBH masses using the canonical scaling relation between SMBH mass and stellar velocity dispersion ($\sigma_{\ast}$). For this we used the values of $\sigma_{\ast}$ from the SDSS {\it galSpec} catalogue and $\mathcal{M}_{\mathrm{BH}}$-$\sigma_{\ast}$ relation derived by \citet{McConnel:13}. The comparison of the derived SMBH masses with the total stellar masses (see Figure\,\ref{fig:masses-bh-stellar}) shows that for massive quiescent galaxies $\mathcal{M}_{\mathrm{BH}}$ scales with stellar mass by coefficient close to $0.002$, adopted in the definition of the sBHAR (Section\,\ref{sec:bhar}), while star-forming and low-mass galaxies
present a wide distribution of $\mathcal{M}_{\mathrm{BH}}$. 
Furthermore, stellar velocity dispersion measurements have large uncertainties for objects with stellar mass $\mathcal{M}_{\ast} \lesssim 10^{10.5} \mathcal{M}_{\odot}$, as the resolution of SDSS spectra allows to measure the velocity dispersion reliably only for $\sigma_{\ast} >70$\,km/s, which corresponds to $\mathcal{M}_{\mathrm{BH}}$ near $10^6 \mathcal{M}_{\odot}$. 

Based on this analysis we conclude that the assumption that the BH-to-stellar mass scaling factor is $\sim 0.002$ is roughly valid (within one dex) only for quiescent galaxies above $\mathcal{M}_{\ast}\gtrsim 10^{10} \mathcal{M}_{\odot}$; thus for star-forming galaxies the sBHAR could underestimate the Eddington ratio by more than one order of magnitude, especially at $\mathcal{M}_{\ast} \lesssim 10^{10.5} \mathcal{M}_{\odot}$. On the other hand this comparison strengthens the difference in AGN accretion between quiescent and star forming galaxies, since the latter tend to have on average Eddington ratios larger than their measured sBHAR.

\section{Discussions}\label{sec: discussions}

In Section\,\ref{sec:bhar} we described the relation between specific BH accretion rate (sBHAR) and SFR over wide range of total stellar mass for two main populations of galaxies, star-forming and quiescent. In this section, we discuss and interpret the obtained results in the context of current paradigm of AGN/host galaxy co-evolution. 

\subsection{AGN and their host galaxies in the local Universe}\label{sec:dic1-agn-popul-local}

Previous study examining the properties of AGN host galaxies show that X-ray AGN prefer to reside in gas-rich massive galaxies with active star formation \citep{Lutz:10, Mullaney:12a, Mendez:13, Rosario:13, Shimizu:15, Birchall:20, Stemo:20}. We extend these results to low redshift, finding that AGN selected by X-ray criteria (X-ray luminosity threshold, X-ray-to-optical ratio, see Section\,\ref{sec:x-agn-select}) reside mainly on the main sequence of star-forming galaxies (67.6\% of all X-ray selected AGN) as opposed to the quiescent galaxiy region (only 32.4\%) in Fig\,\ref{fig:sfr-m}. This result shows that the fraction of AGN in SFGs is a factor $\sim 2$ higher than in quiescent galaxies and it is consistent with the result obtained using different AGN selection criteria. For instance, AGN selected by variability in the optical band \citep{Heinis:16} or through IR and X-ray criteria \citep{Lutz:10, Rosario:13, Stemo:20} show that AGN are more likely hosted by galaxies with higher star-formation activity, younger stellar population and late-type morphologies over a wide redshift range. Additionally, \citet{Mullaney:12b} and \citet{Shimizu:15} studied in  detail the position of X-ray selected AGN on the SFR--$\mathcal{M}_{\ast}$ diagram and found that AGN preferentially reside in galaxies in transition from the main-sequence of SFG to quiescent galaxies; instead galaxies with a higher level of SFR on and above the MS (i.e. starburst) have a smaller probability of hosting an AGN. In our work, we found that 239 AGN hosted by SFGs (75.4\% of all AGN found in SFGs) are located below the MS, while only 78 AGN are above the MS (24.6\%). The larger fraction of AGN in the transition region from SFGs to quiescent seem to support the scenario in which high-efficiency accreting AGN play a role in the quenching of star formation via AGN feedback \citep{Croton:06,Fabian:12}.

We point out that only 24.0\% of all X-ray sources in our 3XMM-SDSS sample were classified as AGN according to X-ray criteria mentioned above. The analysis performed in Section\,\ref{sec:corr_sfg_etg} shows however that many galaxies that do not respect those criteria, present an excess X-ray emission indicating that they likely host low luminosity AGN. In fact, the empirical AGN selection criteria identify AGN which dominate the host galaxy in some part of the electromagnetic spectrum, and therefore tend to miss sources with low luminosity and/or inefficient accretion. The numbers of AGN selected without/with X-ray luminosity correction are presented in Table\,\ref{tab:agn-fraction}. The combination of `classical' and `low luminosity' AGN reveal a similar tendency of AGN to be hosted by SFGs as was shown above for `classical' AGN. The same result was obtained by several studies over a broad redshift range \citep{Mullaney:12a, Delvecchio:15, Aird:18} with the use of near- and far-IR and X-ray data corrected for star-formation processes in the host galaxies.

\begin{table*}
 \caption{Composition of the 3XMM-SDSS X-ray sample. The left columns ('X-ray sources') show the number of AGN and other sources selected according to the 'classical' X-ray criteria to SDSS sources detected in the 3XMM catalogue (Section\,\ref{sec:x-agn-select}); the right columns show the number of AGN in the X-ray AGN sample of sources which could be corrected for the host-galaxy contribution, thus identifying also 'low luminosity' AGN. The percentages refer to the respective sample sizes.}
 \label{tab:agn-fraction}
 \begin{tabular}{lcccccccccccccc}
  \hline
     & \multicolumn{5}{c}{X-ray sources ($N = 1953$)} & \, & \multicolumn{8}{c}{AGN after $L_x$ correction ($N = 1628$)} \\[1pt] 
    \cline{2-6}\cline{8-15}\\[-5pt]
     & \multicolumn{2}{c}{`classical' AGN} & \, & \multicolumn{2}{c}{other X-ray sources} & \, & \multicolumn{2}{c}{`classical' AGN} & \, & \multicolumn{2}{c}{`low luminosity' AGN} & \, & \multicolumn{2}{c}{all AGN} \\[1pt] 
    \cline{2-3}\cline{5-6}\cline{8-9}\cline{11-12}\cline{14-15}\\[-6pt]
    & $N$ & \% & & $N$ & \% & & $N$ & \% & & $N$ & \% & & $N$ & \% \\[1pt] 
    All & 469 & 24.0\% & & 1484 & 76.0\% & & 454 & 27.9\% & & 1174 & 72.1\% & & -- & -- \\[1pt] 
    Star-forming & 317 & 16.2\% & & 674 & 34.4\% & & 317 & 19.5\% & & 598 & 36.7\% & & 915 & 56.2\% \\[1pt] 
    Quiescent & 152 & 7.8\% & & 810 & 41.5\% & & 137 & 8.4\% & & 576 & 35.4\% & & 713 & 43.8\% \\[1pt] 
  \hline
 \end{tabular}
\end{table*}

Finally, we can derive the fraction of galaxies hosting AGN, based on our 3XMM-SDSS sample. As mentioned in Section\,\ref{sec:xmm-data}, 40\,914 galaxies (20\,462 star-forming and 20\,452 quiescent) of our SDSS sample fall within the 3XMM footprint. Of these only 4\% (1628 objects) have a residual X-ray emission after correcting for the host-galaxy emission, suggesting the presence of an AGN, and only 469 (1\%) of these are `classical' AGN. Specifically 4.5\% of star-forming galaxies in the local Universe host AGN (1.5\% `classical' AGN) and 3.5\% of quiescent galaxies host AGN (0.7\% `classical' AGN). 


\subsection{sBHAR distribution in the local Universe}\label{sec:dis2-bhar-distr}

In Section \ref{sec:fupl} we showed that the distribution of the specific BH accretion rate ($\lambda_{\mathrm{sBHAR}}$) for both types of galaxy population (see Fig.\,\ref{fig:histo-bhar-sfg} and\,\ref{fig:histo-bhar-pass}) has an approximately power-law shape with flattening at low accretion rates between $-3 \lesssim \log\lambda_{\mathrm{sBHAR}} \lesssim -2$ for all stellar mass ranges indicating the prevalence of low-efficiency accretion in the local Universe. This trend is consistent with the studies of the sBHAR probability function which shows a power-law shape with an exponential cut-off at high sBHAR and flattering toward low sBHAR \citep{Aird:12,Bongiorno:12, Georgakakis:14, Aird:18}. On the one hand, `classical' AGN hosted by both star-forming and quiescent galaxies show moderate-to-high specific accretion rates ($\log\,\lambda_{\mathrm{sBHAR}} \geq -3$). The same result was found in the study of \citet{Birchall:20} for the sample of dwarf galaxies. Also, \citet{Mendez:13} showed that AGN selected by IR and X-ray criteria show high specific BH accretion rates, while X-ray sources identified as AGN only by IR have lower specific BH accretion rates. This result is in agreement with the fact that low-redshift galaxies are known to have lower nuclear activity than their high-redshift counterparts, based on evolutionary studies of AGN \citep{Boyle:98, Ueda:03,Hasinger:05, Ho:08}. On the other hand, to estimate SFR, we rely on an optical sample that by definition excludes bright AGN (Seyfert 1 and quasars), where the AGN continuum dominates the host galaxy emission. This explains in part the lack of sources with specific accretion rates $\log\,\lambda_{\mathrm{sBHAR}} \geq -1$ which are the found in higher redshift samples containing a larger fraction of quasars and Type 1 AGN. {We examined the SDSS DR8 quasar catalogue and found 587 objects at $z < 0.33$ with the X-ray detection in hard band in 3XMM-DR8 catalogue. Assuming the average BH mass $\sim 10^{8}\mathcal{M}_{\odot}$ we calculated the $\lambda_{\mathrm{sBHAR}}$ by Eq.\,\eqref{eq:spec-bhar} and found that these objects have predominantly high specific accretion rates with a peak at ($\log\,\lambda_{\mathrm{sBHAR}} = -1$).}

The small fraction of objects with high sBHAR could be related to the small gas supply in local galaxies as well as to the low merger rate, that could trigger AGN activity by feeding the central BH \citep{Kauffmann:00, Hopkins:08}. This result is consistent with \citet{Aird:18} which found a  decline in the number of sources with $\log\,\lambda_{\mathrm{sBHAR}} \approx 0.1$ in low redshift X-ray AGN ($z \leq 0.5$). Furthermore, quiescent galaxies show a systematically lower sBHAR values for all stellar masses (see Figures\,\ref{fig:histo-bhar-sfg} and\,\ref{fig:histo-bhar-pass}). This would imply that AGN in quiescent galaxies are fuelled by a much lower gas fraction and can not sustain the same phase of SMBH accretion as AGN in star-forming galaxies with the same stellar mass \citep{Rosario:13, Goulding:14}. 
Several studies show that 22\% of early-type galaxies in local Universe contain a significant fraction of molecular gas \citep{Young:11}, 40\% and 73\% show the presence of neutral Hydrogen \citep{Serra:12} and ionised gas \citep{Davis:11} respectively and weak SF activity \citep{Crocker:11}. Also, \citet{Thom:12} found that the halo of early-type galaxies at z < 1 contains a significant fraction of cold gas which can possibly feed star-formation and AGN activity. Hence quiescent galaxies, usually-associated with elliptical galaxies, may contain a sufficient reservoir of cold gas to sustain low efficiency SMBH accretion and generate the low-luminosity AGN that we observe. \citet{Kauffmann:09} have also proposed an alternative scenario where the primary gas supply from stellar mass-loss in the galaxy can constantly provide gas with low angular momentum and feed a low-luminosity (i.e low-efficiently) AGN. However, such scenario predicts a decrease in sBHAR at higher stellar masses as the mass-loss rate of the older stellar population in the most massive galaxies is lower, while our result in Fig.\,\ref{fig:median-bhar-sfr-m} and \ref{fig:bhar-sfr} seem in contradiction with this prediction, revealing an increase of $\lambda_{\mathrm{sBHAR}}$ with stellar mass for star-forming galaxies, while quiescent ones have on the average the same $\lambda_{\mathrm{sBHAR}}$ for all range of stellar masses.

Additionally, the lower level of AGN activity might be the signature of different accretion mode. Several theoretical models \citep{Churazov:05, Best:12} claim that the radiatively inefficient SMBH accretion can be produced by so-called advection-dominated accretion flows (ADAF) mode \citep{Narayan:97} or jet-dominated mode with mechanism of Bondi accretion of hot gas \citep{Allen:06,Hardcastle:07}. 


\subsection{The connection between star-formation and BH activity}\label{sec:dic3-sfg-bhar}

Several observational studies show that the SFR density and the AGN activity evolve in similar patterns with cosmic time, both peaking at redshift $z \sim 2-3$ and declining sharply as we move toward the present time \citep{Madau:14}, indicating that the evolution of galaxy growth and their central SMBH proceeds in a coherent way. However, the BHAR density (BHAD) has a slightly faster decay since $z \sim 2$ down to $z \sim 0$ compared to the SFR density (SFRD) since BHAD $\propto$ SFRD$^{1.4\pm0.2}$ \citep{Aird:10, Delvecchio:14, Ueda:14}. 
In the present work, we used the median value of sBHAR to trace the average level of accretion in the local Universe (see Section\,\ref{sec:bhar-sfr-m}) and to study the sBHAR and SFR correlation. We found that star-forming galaxies possess a larger median sBHAR with respect to quiescent galaxies at fixed $\mathcal{M}_{\ast}$. A similar difference of $\log\,\lambda_{\mathrm{sBHAR}}$ for the two different galaxy populations was also presented in \citet{Delvecchio:15, Rodighiero:15, Aird:18} and can be explained, as discussed in the previous Section, by a scenario where both star-formation and AGN activity are triggered by fuelling from a common cold gas reservoir \citep{Alexander:12}. 

Additionally, we examined the correlation between mean sBHAR and SFR in more detail and found that quiescent galaxies have not only lower level of sBHAR in comparison with SFGs, but in general there seems to be a continuous trend of increasing sBHAR with SFR (see Fig.\,\ref{fig:bhar-sfr}) for all ranges of stellar mass. We found a significant correlation between the average SFR and the specific BH accretion rate for all stellar mass ranges. The linear regression analysis suggests a flatter relation (see the column with slope values in Table\,\ref{tab:wls-fit}) for our local sample compared to other studies obtained over a wide range of both stellar mass and redshift (up to $z \sim 2.5$) \citep{Chen:13, Delvecchio:15, Aird:19}. A similarly flatter relation was observed by \citet{Delvecchio:15} for their low redshift subsample and can be likely explained by the fact that the local Universe has a smaller fraction of high-luminosity AGN and powerful quasars. 

The existence of a correlation between SFR and sBHAR supports a scenario where both AGN activity and star-formation processes are triggered and fuelled by a common gas supply. However, a number of studies also suggest an alternative scenario where AGN can enhance or quench the star-formation due to the feedback processes \citep{Fabian:12, Ishibashi:12, Heinis:16, Bluck:20}. \citet{Masoura:18} compared the star-formation rate for galaxies of similar mass with/without AGN for wide redshift range and found that the AGN luminosity (i.e activity) depends on the location of the host galaxy relative to the main sequence (MS) of star-formation; they interpret this result as evidence that the AGN quenches star-formation when the galaxy is located above MS and enhances it when galaxy is below the MS. 
We followed the same approach and examined the connection between X-ray luminosity for star-forming galaxies and its location related to the MS of SFGs (see Section\,\ref{sec:bhar-sfr-m}) and found a mild decrease of AGN X-ray luminosity with increasing SFR (normalised to SFR of MS), similar to the result by \citet{Masoura:18}, but with a slight shift in $\log[\mathrm{SFR}/\mathrm{SFR}_{\mathrm{MS}}]$. 

However when splitting the sample according to the host-galaxy mass we found that the trend disappears (Fig.\ref{fig:ms-offset}), and is likely a systematic effect due to the lack of massive X-ray luminous galaxies with large SFR. We point out however that \citet{Masoura:18} sample larger X-ray luminosities than we do, as well as higher redshifts. In fact, \citet{Rovilos:12} found a correlation between X-ray luminosity and so-called starburstiness (i.e. the ratio of the specific SFR of the source over the main-sequence value at the given redshift) only for sources with redshift $z > 1$, while there is no correlation at lower redshift. This result is in agreement with the study of \citet{Shimizu:15} which demonstrated that X-ray luminosity of the local AGN with $z < 0.05$ (selected by the ultra-hard X-ray emission in \textit{Swift}/BAT catalogue) does not show any relation with the increasing of distance from MS of star-forming galaxies (i.e. change of SFR). Such difference between the results in local Universe and at high redshift can indicate that AGN participated in quenching of SFG strongly in the past where they were more powerful, while in the local Universe the average AGN output is not sufficient to affect directly star-formation processes. On the other hand, we cannot exclude either that the absence of $L_{\mathrm{X}}$--$\mathrm{SFR/SFR}_{\mathrm{MS}}$ relation can be due to the fact that both \citet{Rovilos:12} and \citet{Shimizu:15} used a sample of bright X-ray AGN, since our `classical' AGN also reveal a lack of correlation (see diamond points in Fig.\,\ref{fig:ms-offset}). A conclusive test would thus require to compare consistent samples over similar mass, SFR and redshift ranges.

\section{Summary and conclusions}

In this paper we analysed the intrinsic distribution of SMBH accretion in the local Universe. The parent sample is extracted from the SDSS galaxy catalogue produced by the MPA–JHU group, containing  spectroscopic SFR and $\mathcal{M}_{\ast}$ estimates. Using X-ray detections from the 3XMM-DR8 we measured the average sBHAR in these galaxies and investigated the relation between sBHAR and SFR for star-forming and quiescent galaxy over a wide range of stellar masses. Our main conclusions are the following:

\begin{enumerate}
    \item 'classical' AGN with moderate to high efficiency accretion ($\log\,\lambda_{\mathrm{sBHAR}} > -3$) only represent 24\% of our X-ray detected sample (1\% of the whole galaxy population) and are twice as likely to occupy galaxies with active star-forming processes (i.e. high SFR) than quiescent systems;
     
    \item Overall 5\% of our galaxy sample hosts an accreting SMBH. The majority of these objects accrete at very low specific rates, revealing a power-law sBHAR distribution with flattering at $\log\,\lambda_{\mathrm{sBHAR}} \lesssim -2$ for star-forming galaxies; quiescent galaxies show a tendency to accrete at even lower $\log\,\lambda_{\mathrm{sBHAR}} \lesssim -3$.  
    
    \item The median X-ray luminosity reveals a dependence on the host galaxy stellar mass for both star-forming and quiescent galaxies, while the median $\log\,\lambda_{\mathrm{sBHAR}}$ shows a slight increase with stellar mass only for SFG. Additionally, the star-forming galaxies accrete more at fixed stellar mass than quiescent galaxies.
    
    \item We observe a significant correlation between  $\log\,\lambda_{\mathrm{sBHAR}}$ and $\log$\,SFR in almost all stellar mass ranges,  where quiescent galaxies have a systematically lower level of $\lambda_{\mathrm{sBHAR}}$ than star-forming systems.   
\end{enumerate}

Our results support a picture where the local AGN population is dominated by very low-to-moderate luminosity systems, i.e. inefficiently-accreting SMBH. Our findings show that AGN activity in star-forming galaxies is enhanced with respect to quiescent systems, this may be caused by the different amounts of accreting material present in these types of galaxies or may also indicate a possible difference of the physical mechanisms responsible for the triggering and fuelling of AGN described in the literature (e.g. the stochastic fuelling by cold gas in SFGs and the stellar mass loss or cooling flows in quiescent galaxies).

\section*{Acknowledgements}

Authors gratefully thank Dr.~Duncan~Law-Green and Dr.~Clive~Page for their assistance in data acquisition from Flux Limits from Images from XMM-Newton with DR7 data (FLIX). We thank Jarle Brinchmann for his help with understanding the data presented in \textit{galSpec} catalogue.

OT and MP acknowledge financial contribution from the agreement ASI-INAF n.2017-14-H.O.
FJC acknowledges financial support from the Spanish Ministry MCIU under project RTI2018-096686-B-C21 (MCIU/AEI/FEDER/UE), cofunded by FEDER funds and from the Agencia Estatal de Investigaci\'{o}n, Unidad de Excelencia Mar\`{i}a de Maeztu, ref. MDM-2017-0765.
SC acknowledges financial support from FFABR 2017 (Fondo di Finanziamento per le Attivit\`{a} Base di Ricerca).

Funding for SDSS-III has been provided by the Alfred P. Sloan Foundation, the Participating Institutions, the National Science Foundation, and the U.S. Department of Energy Office of Science. The SDSS-III web site is \url{http://www.sdss3.org/}. 

This research has made use of data obtained from the 3XMM XMM-Newton serendipitous source catalogue compiled by the 10 institutes of the XMM-Newton Survey Science Centre selected by ESA.

This publication makes use of data products from the Two Micron All Sky Survey, which is a joint project of the University of Massachusetts and the Infrared Processing and Analysis Center/California Institute of Technology, funded by the National Aeronautics and Space Administration and the National Science Foundation.

\section*{Data availability}

All the data presented/used in this work are publicly available. The SDSS data (\textit{galSpec} catalogue) are accessible through the online service \texttt{CasJobs} SDSS SkyServer or the web-page \url{https://www.sdss.org/dr12/spectro/galaxy\_mpajhu/}. The \textit{XMM-Newton} data are available in the \textit{XMM-Newton} Survey Science Centre (\url{http://xmmssc.irap.omp.eu/Catalogue/3XMM-DR8/3XMM_DR8.html}). The 2MASS data can be found through the Infrared Processing \& Analysis Center (\url{https://old.ipac.caltech.edu/2mass/}). The count/flux upper limits for XMM data are obtained from the XMM FLIX at \url{https://www.ledas.ac.uk/flix/flix.html}.


\bibliographystyle{mnras}
\bibliography{references}

\begin{thebibliography}{}
\makeatletter
\relax
\def\mn@urlcharsother{\let\do\@makeother \do\$\do\&\do\#\do\^\do\_\do\%\do\~}
\def\mn@doi{\begingroup\mn@urlcharsother \@ifnextchar [ {\mn@doi@}
  {\mn@doi@[]}}
\def\mn@doi@[#1]#2{\def\@tempa{#1}\ifx\@tempa\@empty \href
  {http://dx.doi.org/#2} {doi:#2}\else \href {http://dx.doi.org/#2} {#1}\fi
  \endgroup}
\def\mn@eprint#1#2{\mn@eprint@#1:#2::\@nil}
\def\mn@eprint@arXiv#1{\href {http://arxiv.org/abs/#1} {{\tt arXiv:#1}}}
\def\mn@eprint@dblp#1{\href {http://dblp.uni-trier.de/rec/bibtex/#1.xml}
  {dblp:#1}}
\def\mn@eprint@#1:#2:#3:#4\@nil{\def\@tempa {#1}\def\@tempb {#2}\def\@tempc
  {#3}\ifx \@tempc \@empty \let \@tempc \@tempb \let \@tempb \@tempa \fi \ifx
  \@tempb \@empty \def\@tempb {arXiv}\fi \@ifundefined
  {mn@eprint@\@tempb}{\@tempb:\@tempc}{\expandafter \expandafter \csname
  mn@eprint@\@tempb\endcsname \expandafter{\@tempc}}}

\bibitem[\protect\citeauthoryear{{Aihara} et~al.,}{{Aihara}
  et~al.}{2011}]{Aihara:11}
{Aihara} H.,  et~al., 2011, \mn@doi [\apjs] {10.1088/0067-0049/193/2/29}, \href
  {https://ui.adsabs.harvard.edu/abs/2011ApJS..193...29A} {193, 29}

\bibitem[\protect\citeauthoryear{{Aird} et~al.,}{{Aird} et~al.}{2010}]{Aird:10}
{Aird} J.,  et~al., 2010, \mn@doi [\mnras] {10.1111/j.1365-2966.2009.15829.x},
  \href {https://ui.adsabs.harvard.edu/abs/2010MNRAS.401.2531A} {401, 2531}

\bibitem[\protect\citeauthoryear{{Aird} et~al.,}{{Aird} et~al.}{2012}]{Aird:12}
{Aird} J.,  et~al., 2012, \mn@doi [\apj] {10.1088/0004-637X/746/1/90}, \href
  {https://ui.adsabs.harvard.edu/abs/2012ApJ...746...90A} {746, 90}

\bibitem[\protect\citeauthoryear{{Aird} et~al.,}{{Aird} et~al.}{2013}]{Aird:13}
{Aird} J.,  et~al., 2013, \mn@doi [\apj] {10.1088/0004-637X/775/1/41}, \href
  {https://ui.adsabs.harvard.edu/abs/2013ApJ...775...41A} {775, 41}

\bibitem[\protect\citeauthoryear{{Aird}, {Coil}, {Georgakakis}, {Nandra},
  {Barro}  \& {P{\'e}rez-Gonz{\'a}lez}}{{Aird} et~al.}{2015}]{Aird:15}
{Aird} J.,  {Coil} A.~L.,  {Georgakakis} A.,  {Nandra} K.,  {Barro} G.,
  {P{\'e}rez-Gonz{\'a}lez} P.~G.,  2015, \mn@doi [\mnras]
  {10.1093/mnras/stv1062}, \href
  {https://ui.adsabs.harvard.edu/abs/2015MNRAS.451.1892A} {451, 1892}

\bibitem[\protect\citeauthoryear{{Aird}, {Coil}  \& {Georgakakis}}{{Aird}
  et~al.}{2017}]{Aird:17}
{Aird} J.,  {Coil} A.~L.,   {Georgakakis} A.,  2017, \mn@doi [\mnras]
  {10.1093/mnras/stw2932}, \href
  {https://ui.adsabs.harvard.edu/abs/2017MNRAS.465.3390A} {465, 3390}

\bibitem[\protect\citeauthoryear{{Aird}, {Coil}  \& {Georgakakis}}{{Aird}
  et~al.}{2018}]{Aird:18}
{Aird} J.,  {Coil} A.~L.,   {Georgakakis} A.,  2018, \mn@doi [\mnras]
  {10.1093/mnras/stx2700}, \href
  {https://ui.adsabs.harvard.edu/abs/2018MNRAS.474.1225A} {474, 1225}

\bibitem[\protect\citeauthoryear{{Aird}, {Coil}  \& {Georgakakis}}{{Aird}
  et~al.}{2019}]{Aird:19}
{Aird} J.,  {Coil} A.~L.,   {Georgakakis} A.,  2019, \mn@doi [\mnras]
  {10.1093/mnras/stz125}, \href
  {https://ui.adsabs.harvard.edu/abs/2019MNRAS.484.4360A} {484, 4360}

\bibitem[\protect\citeauthoryear{{Alexander} \& {Hickox}}{{Alexander} \&
  {Hickox}}{2012}]{Alexander:12}
{Alexander} D.~M.,  {Hickox} R.~C.,  2012, \mn@doi [\nar]
  {10.1016/j.newar.2011.11.003}, \href
  {https://ui.adsabs.harvard.edu/abs/2012NewAR..56...93A} {56, 93}

\bibitem[\protect\citeauthoryear{{Allen}, {Dunn}, {Fabian}, {Taylor}  \&
  {Reynolds}}{{Allen} et~al.}{2006}]{Allen:06}
{Allen} S.~W.,  {Dunn} R.~J.~H.,  {Fabian} A.~C.,  {Taylor} G.~B.,   {Reynolds}
  C.~S.,  2006, \mn@doi [\mnras] {10.1111/j.1365-2966.2006.10778.x}, \href
  {https://ui.adsabs.harvard.edu/abs/2006MNRAS.372...21A} {372, 21}

\bibitem[\protect\citeauthoryear{{Ananna} et~al.,}{{Ananna}
  et~al.}{2017}]{Ananna:17}
{Ananna} T.~T.,  et~al., 2017, \mn@doi [\apj] {10.3847/1538-4357/aa937d}, \href
  {https://ui.adsabs.harvard.edu/abs/2017ApJ...850...66A} {850, 66}

\bibitem[\protect\citeauthoryear{{Azadi} et~al.,}{{Azadi}
  et~al.}{2015}]{Azadi:15}
{Azadi} M.,  et~al., 2015, \mn@doi [\apj] {10.1088/0004-637X/806/2/187}, \href
  {https://ui.adsabs.harvard.edu/abs/2015ApJ...806..187A} {806, 187}

\bibitem[\protect\citeauthoryear{{Baldwin}, {Phillips}  \&
  {Terlevich}}{{Baldwin} et~al.}{1981}]{Baldwin:81}
{Baldwin} J.~A.,  {Phillips} M.~M.,   {Terlevich} R.,  1981, \mn@doi [\pasp]
  {10.1086/130766}, \href
  {https://ui.adsabs.harvard.edu/abs/1981PASP...93....5B} {93, 5}

\bibitem[\protect\citeauthoryear{{Bernardi}, {Sheth}, {Tundo}  \&
  {Hyde}}{{Bernardi} et~al.}{2007}]{Bernardi:07}
{Bernardi} M.,  {Sheth} R.~K.,  {Tundo} E.,   {Hyde} J.~B.,  2007, \mn@doi
  [\apj] {10.1086/512719}, \href
  {https://ui.adsabs.harvard.edu/abs/2007ApJ...660..267B} {660, 267}

\bibitem[\protect\citeauthoryear{{Best} \& {Heckman}}{{Best} \&
  {Heckman}}{2012}]{Best:12}
{Best} P.~N.,  {Heckman} T.~M.,  2012, \mn@doi [\mnras]
  {10.1111/j.1365-2966.2012.20414.x}, \href
  {https://ui.adsabs.harvard.edu/abs/2012MNRAS.421.1569B} {421, 1569}

\bibitem[\protect\citeauthoryear{{Birchall}, {Watson}  \& {Aird}}{{Birchall}
  et~al.}{2020}]{Birchall:20}
{Birchall} K.~L.,  {Watson} M.~G.,   {Aird} J.,  2020, \mn@doi [\mnras]
  {10.1093/mnras/staa040}, \href
  {https://ui.adsabs.harvard.edu/abs/2020MNRAS.492.2268B} {492, 2268}

\bibitem[\protect\citeauthoryear{{Blanton} \& {Moustakas}}{{Blanton} \&
  {Moustakas}}{2009}]{Blanton:09}
{Blanton} M.~R.,  {Moustakas} J.,  2009, \mn@doi [\araa]
  {10.1146/annurev-astro-082708-101734}, \href
  {https://ui.adsabs.harvard.edu/abs/2009ARA&A..47..159B} {47, 159}

\bibitem[\protect\citeauthoryear{{Bluck} et~al.,}{{Bluck}
  et~al.}{2020}]{Bluck:20}
{Bluck} A. F.~L.,  et~al., 2020, \mn@doi [\mnras] {10.1093/mnras/staa2806},
  \href {https://ui.adsabs.harvard.edu/abs/2020MNRAS.499..230B} {499, 230}

\bibitem[\protect\citeauthoryear{{Bongiorno} et~al.,}{{Bongiorno}
  et~al.}{2007}]{Bongiorno:07}
{Bongiorno} A.,  et~al., 2007, \mn@doi [\aap] {10.1051/0004-6361:20077611},
  \href {https://ui.adsabs.harvard.edu/abs/2007A&A...472..443B} {472, 443}

\bibitem[\protect\citeauthoryear{{Bongiorno} et~al.,}{{Bongiorno}
  et~al.}{2012}]{Bongiorno:12}
{Bongiorno} A.,  et~al., 2012, \mn@doi [\mnras]
  {10.1111/j.1365-2966.2012.22089.x}, \href
  {https://ui.adsabs.harvard.edu/abs/2012MNRAS.427.3103B} {427, 3103}

\bibitem[\protect\citeauthoryear{{Bongiorno} et~al.,}{{Bongiorno}
  et~al.}{2016}]{Bongiorno:16}
{Bongiorno} A.,  et~al., 2016, \mn@doi [\aap] {10.1051/0004-6361/201527436},
  \href {https://ui.adsabs.harvard.edu/abs/2016A&A...588A..78B} {588, A78}

\bibitem[\protect\citeauthoryear{{Boroson}, {Kim}  \& {Fabbiano}}{{Boroson}
  et~al.}{2011}]{Boroson:11}
{Boroson} B.,  {Kim} D.-W.,   {Fabbiano} G.,  2011, \mn@doi [\apj]
  {10.1088/0004-637X/729/1/12}, \href
  {https://ui.adsabs.harvard.edu/abs/2011ApJ...729...12B} {729, 12}

\bibitem[\protect\citeauthoryear{{Boyle} \& {Terlevich}}{{Boyle} \&
  {Terlevich}}{1998}]{Boyle:98}
{Boyle} B.~J.,  {Terlevich} R.~J.,  1998, \mn@doi [\mnras]
  {10.1046/j.1365-8711.1998.01264.x}, \href
  {https://ui.adsabs.harvard.edu/abs/1998MNRAS.293L..49B} {293, L49}

\bibitem[\protect\citeauthoryear{Brandt \& Hasinger}{Brandt \&
  Hasinger}{2005}]{Brandt:05}
Brandt W.,  Hasinger G.,  2005, \mn@doi [Annual Review of Astronomy and
  Astrophysics] {10.1146/annurev.astro.43.051804.102213}, 43, 827

\bibitem[\protect\citeauthoryear{{Brinchmann}, {Charlot}, {White}, {Tremonti},
  {Kauffmann}, {Heckman}  \& {Brinkmann}}{{Brinchmann}
  et~al.}{2004}]{Brinchmann:04}
{Brinchmann} J.,  {Charlot} S.,  {White} S.~D.~M.,  {Tremonti} C.,  {Kauffmann}
  G.,  {Heckman} T.,   {Brinkmann} J.,  2004, \mn@doi [\mnras]
  {10.1111/j.1365-2966.2004.07881.x}, \href
  {https://ui.adsabs.harvard.edu/abs/2004MNRAS.351.1151B} {351, 1151}

\bibitem[\protect\citeauthoryear{{Carrera} et~al.,}{{Carrera}
  et~al.}{2007}]{Carrera:07}
{Carrera} F.~J.,  et~al., 2007, \mn@doi [\aap] {10.1051/0004-6361:20066271},
  \href {https://ui.adsabs.harvard.edu/abs/2007A&A...469...27C} {469, 27}

\bibitem[\protect\citeauthoryear{{Chabrier}}{{Chabrier}}{2003}]{Chabrier:03}
{Chabrier} G.,  2003, \mn@doi [\pasp] {10.1086/376392}, \href
  {https://ui.adsabs.harvard.edu/abs/2003PASP..115..763C} {115, 763}

\bibitem[\protect\citeauthoryear{{Chen} et~al.,}{{Chen} et~al.}{2013}]{Chen:13}
{Chen} C.-T.~J.,  et~al., 2013, \mn@doi [\apj] {10.1088/0004-637X/773/1/3},
  \href {https://ui.adsabs.harvard.edu/abs/2013ApJ...773....3C} {773, 3}

\bibitem[\protect\citeauthoryear{{Churazov}, {Sazonov}, {Sunyaev}, {Forman},
  {Jones}  \& {B{\"o}hringer}}{{Churazov} et~al.}{2005}]{Churazov:05}
{Churazov} E.,  {Sazonov} S.,  {Sunyaev} R.,  {Forman} W.,  {Jones} C.,
  {B{\"o}hringer} H.,  2005, \mn@doi [\mnras]
  {10.1111/j.1745-3933.2005.00093.x}, \href
  {https://ui.adsabs.harvard.edu/abs/2005MNRAS.363L..91C} {363, L91}

\bibitem[\protect\citeauthoryear{{Civano}, {Fabbiano}, {Pellegrini}, {Kim},
  {Paggi}, {Feder}  \& {Elvis}}{{Civano} et~al.}{2014}]{Civano:14}
{Civano} F.,  {Fabbiano} G.,  {Pellegrini} S.,  {Kim} D.~W.,  {Paggi} A.,
  {Feder} R.,   {Elvis} M.,  2014, \mn@doi [\apj] {10.1088/0004-637X/790/1/16},
  \href {https://ui.adsabs.harvard.edu/abs/2014ApJ...790...16C} {790, 16}

\bibitem[\protect\citeauthoryear{{Combes}}{{Combes}}{2017}]{Combes:17}
{Combes} F.,  2017, \mn@doi [Frontiers in Astronomy and Space Sciences]
  {10.3389/fspas.2017.00010}, \href
  {https://ui.adsabs.harvard.edu/abs/2017FrASS...4...10C} {4, 10}

\bibitem[\protect\citeauthoryear{{Cowie}, {Songaila}, {Hu}  \& {Cohen}}{{Cowie}
  et~al.}{1996}]{Cowie:96}
{Cowie} L.~L.,  {Songaila} A.,  {Hu} E.~M.,   {Cohen} J.~G.,  1996, \mn@doi
  [\aj] {10.1086/118058}, \href
  {https://ui.adsabs.harvard.edu/abs/1996AJ....112..839C} {112, 839}

\bibitem[\protect\citeauthoryear{{Crocker}, {Bureau}, {Young}  \&
  {Combes}}{{Crocker} et~al.}{2011}]{Crocker:11}
{Crocker} A.~F.,  {Bureau} M.,  {Young} L.~M.,   {Combes} F.,  2011, \mn@doi
  [\mnras] {10.1111/j.1365-2966.2010.17537.x}, \href
  {https://ui.adsabs.harvard.edu/abs/2011MNRAS.410.1197C} {410, 1197}

\bibitem[\protect\citeauthoryear{{Croton} et~al.,}{{Croton}
  et~al.}{2006}]{Croton:06}
{Croton} D.~J.,  et~al., 2006, \mn@doi [\mnras]
  {10.1111/j.1365-2966.2005.09675.x}, \href
  {https://ui.adsabs.harvard.edu/abs/2006MNRAS.365...11C} {365, 11}

\bibitem[\protect\citeauthoryear{{Davis} et~al.,}{{Davis}
  et~al.}{2011}]{Davis:11}
{Davis} T.~A.,  et~al., 2011, \mn@doi [\mnras]
  {10.1111/j.1365-2966.2011.19355.x}, \href
  {https://ui.adsabs.harvard.edu/abs/2011MNRAS.417..882D} {417, 882}

\bibitem[\protect\citeauthoryear{{Delvecchio} et~al.,}{{Delvecchio}
  et~al.}{2014}]{Delvecchio:14}
{Delvecchio} I.,  et~al., 2014, \mn@doi [\mnras] {10.1093/mnras/stu130}, \href
  {https://ui.adsabs.harvard.edu/abs/2014MNRAS.439.2736D} {439, 2736}

\bibitem[\protect\citeauthoryear{{Delvecchio} et~al.,}{{Delvecchio}
  et~al.}{2015}]{Delvecchio:15}
{Delvecchio} I.,  et~al., 2015, \mn@doi [\mnras] {10.1093/mnras/stv213}, \href
  {https://ui.adsabs.harvard.edu/abs/2015MNRAS.449..373D} {449, 373}

\bibitem[\protect\citeauthoryear{{Elbaz} et~al.,}{{Elbaz}
  et~al.}{2011}]{Elbaz:11}
{Elbaz} D.,  et~al., 2011, \mn@doi [\aap] {10.1051/0004-6361/201117239}, \href
  {https://ui.adsabs.harvard.edu/abs/2011A&A...533A.119E} {533, A119}

\bibitem[\protect\citeauthoryear{{Fabbiano}}{{Fabbiano}}{1989}]{Fabbiano:89}
{Fabbiano} G.,  1989, \mn@doi [\araa] {10.1146/annurev.aa.27.090189.000511},
  \href {https://ui.adsabs.harvard.edu/abs/1989ARA&A..27...87F} {27, 87}

\bibitem[\protect\citeauthoryear{{Faber} et~al.,}{{Faber}
  et~al.}{2007}]{Faber:07}
{Faber} S.~M.,  et~al., 2007, \mn@doi [\apj] {10.1086/519294}, \href
  {https://ui.adsabs.harvard.edu/abs/2007ApJ...665..265F} {665, 265}

\bibitem[\protect\citeauthoryear{{Fabian}}{{Fabian}}{2012}]{Fabian:12}
{Fabian} A.~C.,  2012, \mn@doi [\araa] {10.1146/annurev-astro-081811-125521},
  \href {https://ui.adsabs.harvard.edu/abs/2012ARA&A..50..455F} {50, 455}

\bibitem[\protect\citeauthoryear{{Falcke}, {K{\"o}rding}  \&
  {Markoff}}{{Falcke} et~al.}{2004}]{Falcke:04}
{Falcke} H.,  {K{\"o}rding} E.,   {Markoff} S.,  2004, \mn@doi [\aap]
  {10.1051/0004-6361:20031683}, \href
  {https://ui.adsabs.harvard.edu/abs/2004A&A...414..895F} {414, 895}

\bibitem[\protect\citeauthoryear{{Fathi}, {Storchi-Bergmann}, {Riffel},
  {Winge}, {Axon}, {Robinson}, {Capetti}  \& {Marconi}}{{Fathi}
  et~al.}{2006}]{Fathi:06}
{Fathi} K.,  {Storchi-Bergmann} T.,  {Riffel} R.~A.,  {Winge} C.,  {Axon}
  D.~J.,  {Robinson} A.,  {Capetti} A.,   {Marconi} A.,  2006, \mn@doi [\apjl]
  {10.1086/503832}, \href
  {https://ui.adsabs.harvard.edu/abs/2006ApJ...641L..25F} {641, L25}

\bibitem[\protect\citeauthoryear{{Ferrarese} \& {Merritt}}{{Ferrarese} \&
  {Merritt}}{2000}]{Ferrarese:00}
{Ferrarese} L.,  {Merritt} D.,  2000, \mn@doi [\apjl] {10.1086/312838}, \href
  {https://ui.adsabs.harvard.edu/abs/2000ApJ...539L...9F} {539, L9}

\bibitem[\protect\citeauthoryear{{Fischer}, {Crenshaw}, {Kraemer}, {Schmitt},
  {Storchi-Bergmann}  \& {Riffel}}{{Fischer} et~al.}{2015}]{Fischer:15}
{Fischer} T.~C.,  {Crenshaw} D.~M.,  {Kraemer} S.~B.,  {Schmitt} H.~R.,
  {Storchi-Bergmann} T.,   {Riffel} R.~A.,  2015, \mn@doi [\apj]
  {10.1088/0004-637X/799/2/234}, \href
  {https://ui.adsabs.harvard.edu/abs/2015ApJ...799..234F} {799, 234}

\bibitem[\protect\citeauthoryear{{Fragos} et~al.,}{{Fragos}
  et~al.}{2013}]{Fragos:13}
{Fragos} T.,  et~al., 2013, \mn@doi [\apj] {10.1088/0004-637X/764/1/41}, \href
  {https://ui.adsabs.harvard.edu/abs/2013ApJ...764...41F} {764, 41}

\bibitem[\protect\citeauthoryear{{Gebhardt} et~al.,}{{Gebhardt}
  et~al.}{2000}]{Gebhardt:00}
{Gebhardt} K.,  et~al., 2000, \mn@doi [\apjl] {10.1086/312840}, \href
  {https://ui.adsabs.harvard.edu/abs/2000ApJ...539L..13G} {539, L13}

\bibitem[\protect\citeauthoryear{{Georgakakis} et~al.,}{{Georgakakis}
  et~al.}{2014}]{Georgakakis:14}
{Georgakakis} A.,  et~al., 2014, \mn@doi [\mnras] {10.1093/mnras/stu236}, \href
  {https://ui.adsabs.harvard.edu/abs/2014MNRAS.440..339G} {440, 339}

\bibitem[\protect\citeauthoryear{{Goulding} et~al.,}{{Goulding}
  et~al.}{2014}]{Goulding:14}
{Goulding} A.~D.,  et~al., 2014, \mn@doi [\apj] {10.1088/0004-637X/783/1/40},
  \href {https://ui.adsabs.harvard.edu/abs/2014ApJ...783...40G} {783, 40}

\bibitem[\protect\citeauthoryear{{G{\"u}ltekin} et~al.,}{{G{\"u}ltekin}
  et~al.}{2009}]{Gultekin:09}
{G{\"u}ltekin} K.,  et~al., 2009, \mn@doi [\apj] {10.1088/0004-637X/698/1/198},
  \href {https://ui.adsabs.harvard.edu/abs/2009ApJ...698..198G} {698, 198}

\bibitem[\protect\citeauthoryear{{Hardcastle}, {Evans}  \&
  {Croston}}{{Hardcastle} et~al.}{2007}]{Hardcastle:07}
{Hardcastle} M.~J.,  {Evans} D.~A.,   {Croston} J.~H.,  2007, \mn@doi [\mnras]
  {10.1111/j.1365-2966.2007.11572.x}, \href
  {https://ui.adsabs.harvard.edu/abs/2007MNRAS.376.1849H} {376, 1849}

\bibitem[\protect\citeauthoryear{{H{\"a}ring} \& {Rix}}{{H{\"a}ring} \&
  {Rix}}{2004}]{Haring:04}
{H{\"a}ring} N.,  {Rix} H.-W.,  2004, \mn@doi [\apjl] {10.1086/383567}, \href
  {https://ui.adsabs.harvard.edu/abs/2004ApJ...604L..89H} {604, L89}

\bibitem[\protect\citeauthoryear{{Hasinger}, {Miyaji}  \& {Schmidt}}{{Hasinger}
  et~al.}{2005}]{Hasinger:05}
{Hasinger} G.,  {Miyaji} T.,   {Schmidt} M.,  2005, \mn@doi [\aap]
  {10.1051/0004-6361:20042134}, \href
  {https://ui.adsabs.harvard.edu/abs/2005A&A...441..417H} {441, 417}

\bibitem[\protect\citeauthoryear{{Heckman} \& {Best}}{{Heckman} \&
  {Best}}{2014}]{Heckman:14}
{Heckman} T.~M.,  {Best} P.~N.,  2014, \mn@doi [\araa]
  {10.1146/annurev-astro-081913-035722}, \href
  {https://ui.adsabs.harvard.edu/abs/2014ARA&A..52..589H} {52, 589}

\bibitem[\protect\citeauthoryear{{Heinis} et~al.,}{{Heinis}
  et~al.}{2016}]{Heinis:16}
{Heinis} S.,  et~al., 2016, \mn@doi [\apj] {10.3847/0004-637X/826/1/62}, \href
  {https://ui.adsabs.harvard.edu/abs/2016ApJ...826...62H} {826, 62}

\bibitem[\protect\citeauthoryear{{Hickox} et~al.,}{{Hickox}
  et~al.}{2009}]{Hickox:09}
{Hickox} R.~C.,  et~al., 2009, \mn@doi [\apj] {10.1088/0004-637X/696/1/891},
  \href {https://ui.adsabs.harvard.edu/abs/2009ApJ...696..891H} {696, 891}

\bibitem[\protect\citeauthoryear{{Hickox}, {Mullaney}, {Alexander}, {Chen},
  {Civano}, {Goulding}  \& {Hainline}}{{Hickox} et~al.}{2014}]{Hickox:14}
{Hickox} R.~C.,  {Mullaney} J.~R.,  {Alexander} D.~M.,  {Chen} C.-T.~J.,
  {Civano} F.~M.,  {Goulding} A.~D.,   {Hainline} K.~N.,  2014, \mn@doi [\apj]
  {10.1088/0004-637X/782/1/9}, \href
  {https://ui.adsabs.harvard.edu/abs/2014ApJ...782....9H} {782, 9}

\bibitem[\protect\citeauthoryear{{Ho}}{{Ho}}{2008}]{Ho:08}
{Ho} L.~C.,  2008, \mn@doi [\araa] {10.1146/annurev.astro.45.051806.110546},
  \href {https://ui.adsabs.harvard.edu/abs/2008ARA&A..46..475H} {46, 475}

\bibitem[\protect\citeauthoryear{{Hopkins} \& {Quataert}}{{Hopkins} \&
  {Quataert}}{2010}]{Hopkins:10}
{Hopkins} P.~F.,  {Quataert} E.,  2010, \mn@doi [\mnras]
  {10.1111/j.1365-2966.2010.17064.x}, \href
  {https://ui.adsabs.harvard.edu/abs/2010MNRAS.407.1529H} {407, 1529}

\bibitem[\protect\citeauthoryear{{Hopkins}, {Hernquist}, {Cox}  \&
  {Kere{\v{s}}}}{{Hopkins} et~al.}{2008}]{Hopkins:08}
{Hopkins} P.~F.,  {Hernquist} L.,  {Cox} T.~J.,   {Kere{\v{s}}} D.,  2008,
  \mn@doi [\apjs] {10.1086/524362}, \href
  {https://ui.adsabs.harvard.edu/abs/2008ApJS..175..356H} {175, 356}

\bibitem[\protect\citeauthoryear{{Hopkins}, {Torrey}, {Faucher-Gigu{\`e}re},
  {Quataert}  \& {Murray}}{{Hopkins} et~al.}{2016}]{Hopkins:16}
{Hopkins} P.~F.,  {Torrey} P.,  {Faucher-Gigu{\`e}re} C.-A.,  {Quataert} E.,
  {Murray} N.,  2016, \mn@doi [\mnras] {10.1093/mnras/stw289}, \href
  {https://ui.adsabs.harvard.edu/abs/2016MNRAS.458..816H} {458, 816}

\bibitem[\protect\citeauthoryear{{Ishibashi} \& {Fabian}}{{Ishibashi} \&
  {Fabian}}{2012}]{Ishibashi:12}
{Ishibashi} W.,  {Fabian} A.~C.,  2012, \mn@doi [\mnras]
  {10.1111/j.1365-2966.2012.22074.x}, \href
  {https://ui.adsabs.harvard.edu/abs/2012MNRAS.427.2998I} {427, 2998}

\bibitem[\protect\citeauthoryear{{Kauffmann} \& {Haehnelt}}{{Kauffmann} \&
  {Haehnelt}}{2000}]{Kauffmann:00}
{Kauffmann} G.,  {Haehnelt} M.,  2000, \mn@doi [\mnras]
  {10.1046/j.1365-8711.2000.03077.x}, \href
  {https://ui.adsabs.harvard.edu/abs/2000MNRAS.311..576K} {311, 576}

\bibitem[\protect\citeauthoryear{{Kauffmann} \& {Heckman}}{{Kauffmann} \&
  {Heckman}}{2009}]{Kauffmann:09}
{Kauffmann} G.,  {Heckman} T.~M.,  2009, \mn@doi [\mnras]
  {10.1111/j.1365-2966.2009.14960.x}, \href
  {https://ui.adsabs.harvard.edu/abs/2009MNRAS.397..135K} {397, 135}

\bibitem[\protect\citeauthoryear{{Kauffmann} et~al.,}{{Kauffmann}
  et~al.}{2003a}]{Kauffmann:03a}
{Kauffmann} G.,  et~al., 2003a, \mn@doi [\mnras]
  {10.1046/j.1365-8711.2003.06291.x}, \href
  {https://ui.adsabs.harvard.edu/abs/2003MNRAS.341...33K} {341, 33}

\bibitem[\protect\citeauthoryear{{Kauffmann} et~al.,}{{Kauffmann}
  et~al.}{2003b}]{Kauffmann:03c}
{Kauffmann} G.,  et~al., 2003b, \mn@doi [\mnras]
  {10.1111/j.1365-2966.2003.07154.x}, \href
  {https://ui.adsabs.harvard.edu/abs/2003MNRAS.346.1055K} {346, 1055}

\bibitem[\protect\citeauthoryear{{Kewley}, {Groves}, {Kauffmann}  \&
  {Heckman}}{{Kewley} et~al.}{2006}]{Kewley:06}
{Kewley} L.~J.,  {Groves} B.,  {Kauffmann} G.,   {Heckman} T.,  2006, \mn@doi
  [\mnras] {10.1111/j.1365-2966.2006.10859.x}, \href
  {https://ui.adsabs.harvard.edu/abs/2006MNRAS.372..961K} {372, 961}

\bibitem[\protect\citeauthoryear{{Kim} \& {Fabbiano}}{{Kim} \&
  {Fabbiano}}{2013}]{Kim:13}
{Kim} D.-W.,  {Fabbiano} G.,  2013, \mn@doi [\apj]
  {10.1088/0004-637X/776/2/116}, \href
  {https://ui.adsabs.harvard.edu/abs/2013ApJ...776..116K} {776, 116}

\bibitem[\protect\citeauthoryear{{Kormendy} \& {Ho}}{{Kormendy} \&
  {Ho}}{2013}]{Kormendy:13}
{Kormendy} J.,  {Ho} L.~C.,  2013, \mn@doi [\araa]
  {10.1146/annurev-astro-082708-101811}, \href
  {https://ui.adsabs.harvard.edu/abs/2013ARA&A..51..511K} {51, 511}

\bibitem[\protect\citeauthoryear{{Kroupa}}{{Kroupa}}{2001}]{Kroupa:01}
{Kroupa} P.,  2001, \mn@doi [\mnras] {10.1046/j.1365-8711.2001.04022.x}, \href
  {https://ui.adsabs.harvard.edu/abs/2001MNRAS.322..231K} {322, 231}

\bibitem[\protect\citeauthoryear{{Lehmer} et~al.,}{{Lehmer}
  et~al.}{2016}]{Lehmer:16}
{Lehmer} B.~D.,  et~al., 2016, \mn@doi [\apj] {10.3847/0004-637X/825/1/7},
  \href {https://ui.adsabs.harvard.edu/abs/2016ApJ...825....7L} {825, 7}

\bibitem[\protect\citeauthoryear{{Liu} et~al.,}{{Liu} et~al.}{2017}]{Liu:17}
{Liu} T.,  et~al., 2017, \mn@doi [\apjs] {10.3847/1538-4365/aa7847}, \href
  {https://ui.adsabs.harvard.edu/abs/2017ApJS..232....8L} {232, 8}

\bibitem[\protect\citeauthoryear{{Luo} et~al.,}{{Luo} et~al.}{2017}]{Luo:17}
{Luo} B.,  et~al., 2017, \mn@doi [\apjs] {10.3847/1538-4365/228/1/2}, \href
  {https://ui.adsabs.harvard.edu/abs/2017ApJS..228....2L} {228, 2}

\bibitem[\protect\citeauthoryear{{Lusso} et~al.,}{{Lusso}
  et~al.}{2012}]{Lusso:12}
{Lusso} E.,  et~al., 2012, \mn@doi [\mnras] {10.1111/j.1365-2966.2012.21513.x},
  \href {https://ui.adsabs.harvard.edu/abs/2012MNRAS.425..623L} {425, 623}

\bibitem[\protect\citeauthoryear{{Lutz} et~al.,}{{Lutz} et~al.}{2010}]{Lutz:10}
{Lutz} D.,  et~al., 2010, \mn@doi [\apj] {10.1088/0004-637X/712/2/1287}, \href
  {https://ui.adsabs.harvard.edu/abs/2010ApJ...712.1287L} {712, 1287}

\bibitem[\protect\citeauthoryear{{Madau} \& {Dickinson}}{{Madau} \&
  {Dickinson}}{2014}]{Madau:14}
{Madau} P.,  {Dickinson} M.,  2014, \mn@doi [\araa]
  {10.1146/annurev-astro-081811-125615}, \href
  {https://ui.adsabs.harvard.edu/abs/2014ARA&A..52..415M} {52, 415}

\bibitem[\protect\citeauthoryear{{Marconi} \& {Hunt}}{{Marconi} \&
  {Hunt}}{2003}]{Marconi:03}
{Marconi} A.,  {Hunt} L.~K.,  2003, \mn@doi [\apjl] {10.1086/375804}, \href
  {https://ui.adsabs.harvard.edu/abs/2003ApJ...589L..21M} {589, L21}

\bibitem[\protect\citeauthoryear{{Marconi}, {Risaliti}, {Gilli}, {Hunt},
  {Maiolino}  \& {Salvati}}{{Marconi} et~al.}{2004}]{Marconi:04}
{Marconi} A.,  {Risaliti} G.,  {Gilli} R.,  {Hunt} L.~K.,  {Maiolino} R.,
  {Salvati} M.,  2004, \mn@doi [\mnras] {10.1111/j.1365-2966.2004.07765.x},
  \href {https://ui.adsabs.harvard.edu/abs/2004MNRAS.351..169M} {351, 169}

\bibitem[\protect\citeauthoryear{{Masoura}, {Mountrichas}, {Georgantopoulos},
  {Ruiz}, {Magdis}  \& {Plionis}}{{Masoura} et~al.}{2018}]{Masoura:18}
{Masoura} V.~A.,  {Mountrichas} G.,  {Georgantopoulos} I.,  {Ruiz} A.,
  {Magdis} G.,   {Plionis} M.,  2018, \mn@doi [\aap]
  {10.1051/0004-6361/201833397}, \href
  {https://ui.adsabs.harvard.edu/abs/2018A&A...618A..31M} {618, A31}

\bibitem[\protect\citeauthoryear{{McConnell} \& {Ma}}{{McConnell} \&
  {Ma}}{2013}]{McConnel:13}
{McConnell} N.~J.,  {Ma} C.-P.,  2013, \mn@doi [\apj]
  {10.1088/0004-637X/764/2/184}, \href
  {https://ui.adsabs.harvard.edu/abs/2013ApJ...764..184M} {764, 184}

\bibitem[\protect\citeauthoryear{{Mendel}, {Simard}, {Palmer}, {Ellison}  \&
  {Patton}}{{Mendel} et~al.}{2014}]{Mendel:14}
{Mendel} J.~T.,  {Simard} L.,  {Palmer} M.,  {Ellison} S.~L.,   {Patton} D.~R.,
   2014, \mn@doi [\apjs] {10.1088/0067-0049/210/1/3}, \href
  {https://ui.adsabs.harvard.edu/abs/2014ApJS..210....3M} {210, 3}

\bibitem[\protect\citeauthoryear{{Mendez} et~al.,}{{Mendez}
  et~al.}{2013}]{Mendez:13}
{Mendez} A.~J.,  et~al., 2013, \mn@doi [\apj] {10.1088/0004-637X/770/1/40},
  \href {https://ui.adsabs.harvard.edu/abs/2013ApJ...770...40M} {770, 40}

\bibitem[\protect\citeauthoryear{{Merloni} et~al.,}{{Merloni}
  et~al.}{2010}]{Merloni:10}
{Merloni} A.,  et~al., 2010, \mn@doi [\apj] {10.1088/0004-637X/708/1/137},
  \href {https://ui.adsabs.harvard.edu/abs/2010ApJ...708..137M} {708, 137}

\bibitem[\protect\citeauthoryear{{Mineo}, {Gilfanov}  \& {Sunyaev}}{{Mineo}
  et~al.}{2012}]{Mineo:12}
{Mineo} S.,  {Gilfanov} M.,   {Sunyaev} R.,  2012, \mn@doi [\mnras]
  {10.1111/j.1365-2966.2011.19862.x}, \href
  {https://ui.adsabs.harvard.edu/abs/2012MNRAS.419.2095M} {419, 2095}

\bibitem[\protect\citeauthoryear{{Mullaney} et~al.,}{{Mullaney}
  et~al.}{2012a}]{Mullaney:12a}
{Mullaney} J.~R.,  et~al., 2012a, \mn@doi [\mnras]
  {10.1111/j.1365-2966.2011.19675.x}, \href
  {https://ui.adsabs.harvard.edu/abs/2012MNRAS.419...95M} {419, 95}

\bibitem[\protect\citeauthoryear{{Mullaney} et~al.,}{{Mullaney}
  et~al.}{2012b}]{Mullaney:12b}
{Mullaney} J.~R.,  et~al., 2012b, \mn@doi [\apjl]
  {10.1088/2041-8205/753/2/L30}, \href
  {https://ui.adsabs.harvard.edu/abs/2012ApJ...753L..30M} {753, L30}

\bibitem[\protect\citeauthoryear{{Narayan}, {Garcia}  \&
  {McClintock}}{{Narayan} et~al.}{1997}]{Narayan:97}
{Narayan} R.,  {Garcia} M.~R.,   {McClintock} J.~E.,  1997, \mn@doi [\apjl]
  {10.1086/310554}, \href
  {https://ui.adsabs.harvard.edu/abs/1997ApJ...478L..79N} {478, L79}

\bibitem[\protect\citeauthoryear{{Noeske} et~al.,}{{Noeske}
  et~al.}{2007}]{Noeske:07}
{Noeske} K.~G.,  et~al., 2007, \mn@doi [\apjl] {10.1086/517926}, \href
  {https://ui.adsabs.harvard.edu/abs/2007ApJ...660L..43N} {660, L43}

\bibitem[\protect\citeauthoryear{{Oemler}, {Abramson}, {Gladders}, {Dressler},
  {Poggianti}  \& {Vulcani}}{{Oemler} et~al.}{2017}]{Oemler:17}
{Oemler} Augustus J.,  {Abramson} L.~E.,  {Gladders} M.~D.,  {Dressler} A.,
  {Poggianti} B.~M.,   {Vulcani} B.,  2017, \mn@doi [\apj]
  {10.3847/1538-4357/aa789e}, \href
  {https://ui.adsabs.harvard.edu/abs/2017ApJ...844...45O} {844, 45}

\bibitem[\protect\citeauthoryear{{Paolillo} et~al.,}{{Paolillo}
  et~al.}{2017}]{Paolillo:17}
{Paolillo} M.,  et~al., 2017, \mn@doi [\mnras] {10.1093/mnras/stx1761}, \href
  {https://ui.adsabs.harvard.edu/abs/2017MNRAS.471.4398P} {471, 4398}

\bibitem[\protect\citeauthoryear{{Popesso} et~al.,}{{Popesso}
  et~al.}{2019}]{Popesso:19}
{Popesso} P.,  et~al., 2019, \mn@doi [\mnras] {10.1093/mnras/sty3210}, \href
  {https://ui.adsabs.harvard.edu/abs/2019MNRAS.483.3213P} {483, 3213}

\bibitem[\protect\citeauthoryear{{Ranalli}, {Comastri}  \& {Setti}}{{Ranalli}
  et~al.}{2003}]{Ranalli:03}
{Ranalli} P.,  {Comastri} A.,   {Setti} G.,  2003, \mn@doi [\aap]
  {10.1051/0004-6361:20021600}, \href
  {https://ui.adsabs.harvard.edu/abs/2003A&A...399...39R} {399, 39}

\bibitem[\protect\citeauthoryear{{Reines} \& {Volonteri}}{{Reines} \&
  {Volonteri}}{2015}]{Reines:15}
{Reines} A.~E.,  {Volonteri} M.,  2015, \mn@doi [\apj]
  {10.1088/0004-637X/813/2/82}, \href
  {https://ui.adsabs.harvard.edu/abs/2015ApJ...813...82R} {813, 82}

\bibitem[\protect\citeauthoryear{{Rodighiero} et~al.,}{{Rodighiero}
  et~al.}{2015}]{Rodighiero:15}
{Rodighiero} G.,  et~al., 2015, \mn@doi [\apjl] {10.1088/2041-8205/800/1/L10},
  \href {https://ui.adsabs.harvard.edu/abs/2015ApJ...800L..10R} {800, L10}

\bibitem[\protect\citeauthoryear{{Rosario} et~al.,}{{Rosario}
  et~al.}{2013}]{Rosario:13}
{Rosario} D.~J.,  et~al., 2013, \mn@doi [\apj] {10.1088/0004-637X/771/1/63},
  \href {https://ui.adsabs.harvard.edu/abs/2013ApJ...771...63R} {771, 63}

\bibitem[\protect\citeauthoryear{{Rosen} et~al.,}{{Rosen}
  et~al.}{2016}]{Rosen:16}
{Rosen} S.~R.,  et~al., 2016, \mn@doi [\aap] {10.1051/0004-6361/201526416},
  \href {https://ui.adsabs.harvard.edu/abs/2016A&A...590A...1R} {590, A1}

\bibitem[\protect\citeauthoryear{{Rovilos} et~al.,}{{Rovilos}
  et~al.}{2012}]{Rovilos:12}
{Rovilos} E.,  et~al., 2012, \mn@doi [\aap] {10.1051/0004-6361/201218952},
  \href {https://ui.adsabs.harvard.edu/abs/2012A&A...546A..58R} {546, A58}

\bibitem[\protect\citeauthoryear{{Salim} et~al.,}{{Salim}
  et~al.}{2007}]{Salim:07}
{Salim} S.,  et~al., 2007, \mn@doi [\apjs] {10.1086/519218}, \href
  {https://ui.adsabs.harvard.edu/abs/2007ApJS..173..267S} {173, 267}

\bibitem[\protect\citeauthoryear{{Salim} et~al.,}{{Salim}
  et~al.}{2016}]{Salim:16}
{Salim} S.,  et~al., 2016, \mn@doi [\apjs] {10.3847/0067-0049/227/1/2}, \href
  {https://ui.adsabs.harvard.edu/abs/2016ApJS..227....2S} {227, 2}

\bibitem[\protect\citeauthoryear{{Savorgnan}, {Graham}, {Marconi}  \&
  {Sani}}{{Savorgnan} et~al.}{2016}]{Savorgnan:16}
{Savorgnan} G. A.~D.,  {Graham} A.~W.,  {Marconi} A.,   {Sani} E.,  2016,
  \mn@doi [\apj] {10.3847/0004-637X/817/1/21}, \href
  {https://ui.adsabs.harvard.edu/abs/2016ApJ...817...21S} {817, 21}

\bibitem[\protect\citeauthoryear{{Schartmann}, {Meisenheimer}, {Klahr},
  {Camenzind}, {Wolf}  \& {Henning}}{{Schartmann} et~al.}{2009}]{Schartmann:09}
{Schartmann} M.,  {Meisenheimer} K.,  {Klahr} H.,  {Camenzind} M.,  {Wolf} S.,
   {Henning} T.,  2009, \mn@doi [\mnras] {10.1111/j.1365-2966.2008.14220.x},
  \href {https://ui.adsabs.harvard.edu/abs/2009MNRAS.393..759S} {393, 759}

\bibitem[\protect\citeauthoryear{{Serra} et~al.,}{{Serra}
  et~al.}{2012}]{Serra:12}
{Serra} P.,  et~al., 2012, \mn@doi [\mnras] {10.1111/j.1365-2966.2012.20219.x},
  \href {https://ui.adsabs.harvard.edu/abs/2012MNRAS.422.1835S} {422, 1835}

\bibitem[\protect\citeauthoryear{{Shankar} et~al.,}{{Shankar}
  et~al.}{2016}]{Shankar:16}
{Shankar} F.,  et~al., 2016, \mn@doi [\mnras] {10.1093/mnras/stw678}, \href
  {https://ui.adsabs.harvard.edu/abs/2016MNRAS.460.3119S} {460, 3119}

\bibitem[\protect\citeauthoryear{{Shankar}, {Bernardi}  \& {Sheth}}{{Shankar}
  et~al.}{2017}]{Shankar:17}
{Shankar} F.,  {Bernardi} M.,   {Sheth} R.~K.,  2017, \mn@doi [\mnras]
  {10.1093/mnras/stw3368}, \href
  {https://ui.adsabs.harvard.edu/abs/2017MNRAS.466.4029S} {466, 4029}

\bibitem[\protect\citeauthoryear{{Shankar} et~al.,}{{Shankar}
  et~al.}{2020}]{Shankar:20}
{Shankar} F.,  et~al., 2020, \mn@doi [Nature Astronomy]
  {10.1038/s41550-019-0949-y}, \href
  {https://ui.adsabs.harvard.edu/abs/2020NatAs...4..282S} {4, 282}

\bibitem[\protect\citeauthoryear{{Shen} et~al.,}{{Shen} et~al.}{2011}]{Shen:11}
{Shen} Y.,  et~al., 2011, \mn@doi [\apjs] {10.1088/0067-0049/194/2/45}, \href
  {https://ui.adsabs.harvard.edu/abs/2011ApJS..194...45S} {194, 45}

\bibitem[\protect\citeauthoryear{{Shimizu}, {Mushotzky}, {Mel{\'e}ndez}, {Koss}
   \& {Rosario}}{{Shimizu} et~al.}{2015}]{Shimizu:15}
{Shimizu} T.~T.,  {Mushotzky} R.~F.,  {Mel{\'e}ndez} M.,  {Koss} M.,
  {Rosario} D.~J.,  2015, \mn@doi [\mnras] {10.1093/mnras/stv1407}, \href
  {https://ui.adsabs.harvard.edu/abs/2015MNRAS.452.1841S} {452, 1841}

\bibitem[\protect\citeauthoryear{{Skrutskie} et~al.,}{{Skrutskie}
  et~al.}{2006}]{Skrutskie:06}
{Skrutskie} M.~F.,  et~al., 2006, \mn@doi [\aj] {10.1086/498708}, \href
  {https://ui.adsabs.harvard.edu/abs/2006AJ....131.1163S} {131, 1163}

\bibitem[\protect\citeauthoryear{{Stemo}, {Comerford}, {Barrows}, {Stern},
  {Assef}  \& {Griffith}}{{Stemo} et~al.}{2020}]{Stemo:20}
{Stemo} A.,  {Comerford} J.~M.,  {Barrows} R.~S.,  {Stern} D.,  {Assef} R.~J.,
   {Griffith} R.~L.,  2020, \mn@doi [\apj] {10.3847/1538-4357/ab5f66}, \href
  {https://ui.adsabs.harvard.edu/abs/2020ApJ...888...78S} {888, 78}

\bibitem[\protect\citeauthoryear{{Symeonidis} et~al.,}{{Symeonidis}
  et~al.}{2014}]{Symeonidis:14}
{Symeonidis} M.,  et~al., 2014, \mn@doi [\mnras] {10.1093/mnras/stu1441}, \href
  {https://ui.adsabs.harvard.edu/abs/2014MNRAS.443.3728S} {443, 3728}

\bibitem[\protect\citeauthoryear{{Thom} et~al.,}{{Thom} et~al.}{2012}]{Thom:12}
{Thom} C.,  et~al., 2012, \mn@doi [\apjl] {10.1088/2041-8205/758/2/L41}, \href
  {https://ui.adsabs.harvard.edu/abs/2012ApJ...758L..41T} {758, L41}

\bibitem[\protect\citeauthoryear{{Tozzi} et~al.,}{{Tozzi}
  et~al.}{2006}]{Tozzi:06}
{Tozzi} P.,  et~al., 2006, \mn@doi [\aap] {10.1051/0004-6361:20042592}, \href
  {https://ui.adsabs.harvard.edu/abs/2006A&A...451..457T} {451, 457}

\bibitem[\protect\citeauthoryear{{Tremonti} et~al.,}{{Tremonti}
  et~al.}{2004}]{Tremonti:04}
{Tremonti} C.~A.,  et~al., 2004, \mn@doi [\apj] {10.1086/423264}, \href
  {https://ui.adsabs.harvard.edu/abs/2004ApJ...613..898T} {613, 898}

\bibitem[\protect\citeauthoryear{{Ueda}, {Akiyama}, {Ohta}  \& {Miyaji}}{{Ueda}
  et~al.}{2003}]{Ueda:03}
{Ueda} Y.,  {Akiyama} M.,  {Ohta} K.,   {Miyaji} T.,  2003, \mn@doi [\apj]
  {10.1086/378940}, \href
  {https://ui.adsabs.harvard.edu/abs/2003ApJ...598..886U} {598, 886}

\bibitem[\protect\citeauthoryear{{Ueda}, {Akiyama}, {Hasinger}, {Miyaji}  \&
  {Watson}}{{Ueda} et~al.}{2014}]{Ueda:14}
{Ueda} Y.,  {Akiyama} M.,  {Hasinger} G.,  {Miyaji} T.,   {Watson} M.~G.,
  2014, \mn@doi [\apj] {10.1088/0004-637X/786/2/104}, \href
  {https://ui.adsabs.harvard.edu/abs/2014ApJ...786..104U} {786, 104}

\bibitem[\protect\citeauthoryear{{Vattakunnel} et~al.,}{{Vattakunnel}
  et~al.}{2012}]{Vattakunnel:12}
{Vattakunnel} S.,  et~al., 2012, \mn@doi [\mnras]
  {10.1111/j.1365-2966.2011.20185.x}, \href
  {https://ui.adsabs.harvard.edu/abs/2012MNRAS.420.2190V} {420, 2190}

\bibitem[\protect\citeauthoryear{{Veilleux} \& {Osterbrock}}{{Veilleux} \&
  {Osterbrock}}{1987}]{Veilleux:87}
{Veilleux} S.,  {Osterbrock} D.~E.,  1987, \mn@doi [\apjs] {10.1086/191166},
  \href {https://ui.adsabs.harvard.edu/abs/1987ApJS...63..295V} {63, 295}

\bibitem[\protect\citeauthoryear{{Yang} et~al.,}{{Yang} et~al.}{2018}]{Yang:18}
{Yang} G.,  et~al., 2018, \mn@doi [\mnras] {10.1093/mnras/stx2805}, \href
  {https://ui.adsabs.harvard.edu/abs/2018MNRAS.475.1887Y} {475, 1887}

\bibitem[\protect\citeauthoryear{{Young} et~al.,}{{Young}
  et~al.}{2011}]{Young:11}
{Young} L.~M.,  et~al., 2011, \mn@doi [\mnras]
  {10.1111/j.1365-2966.2011.18561.x}, \href
  {https://ui.adsabs.harvard.edu/abs/2011MNRAS.414..940Y} {414, 940}

\bibitem[\protect\citeauthoryear{{Zubovas}, {Nayakshin}, {King}  \&
  {Wilkinson}}{{Zubovas} et~al.}{2013}]{Zubovas:13}
{Zubovas} K.,  {Nayakshin} S.,  {King} A.,   {Wilkinson} M.,  2013, \mn@doi
  [\mnras] {10.1093/mnras/stt952}, \href
  {https://ui.adsabs.harvard.edu/abs/2013MNRAS.433.3079Z} {433, 3079}

\bibitem[\protect\citeauthoryear{{de Nicola}, {Marconi}  \& {Longo}}{{de
  Nicola} et~al.}{2019}]{deNicola:19}
{de Nicola} S.,  {Marconi} A.,   {Longo} G.,  2019, \mn@doi [\mnras]
  {10.1093/mnras/stz2472}, \href
  {https://ui.adsabs.harvard.edu/abs/2019MNRAS.490..600D} {490, 600}

\makeatother
\end{thebibliography}



\bsp	
\label{lastpage}
\end{document}